\pdfoutput=1
\documentclass[aps,nofootinbib,11pt,notitlepage]{revtex4-1} 
\usepackage[letterpaper,hmargin=1in,vmargin=1in]{geometry}
\usepackage{graphicx,epstopdf,amsmath,amsfonts,amssymb,color,cancel,subfigure,comment}
\usepackage[utf8x]{inputenc}
\usepackage[fixamsmath]{mathtools}
\usepackage{textcomp,subfigure,natbib,hyperref}
\usepackage[cm]{fullpage}

\def\be{\begin{equation}}
\def\ee{\end{equation}}
\def\ba{\begin{eqnarray}}
\def\ea{\end{eqnarray}}
\def\ge{\mathrel{\raise.3ex\hbox{$>$\kern-.75em\lower1ex\hbox{$\sim$}}}}
\def\la{\mathrel{\raise.3ex\hbox{$<$\kern-.75em\lower1ex\hbox{$\sim$}}}}

\def\simgt{\mathrel{\raise.3ex\hbox{$>$\kern-.75em\lower1ex\hbox{$\sim$}}}}
\def\simlt{\mathrel{\raise.3ex\hbox{$<$\kern-.75em\lower1ex\hbox{$\sim$}}}}

\newcommand{\nc}{\newcommand}

\newcommand{\MM}{\mathcal{M}}
\newcommand{\Int}{\int\frac{d^4l}{(2\pi)^4}}
\newcommand{\Leq}{\leqslant}
\newcommand{\Geq}{\geqslant}

\newcommand{\gev}{\text{GeV}}
\newcommand{\tev}{\text{TeV}}

\nc{\gone}{\bar g_{\pi NN}^{(1)}}
\nc{\gzero}{\bar g_{\pi NN}^{(0)}}
\nc{\al}{\alpha}
\nc{\ga}{\gamma}
\nc{\de}{\delta}
\nc{\ep}{\epsilon}
\nc{\ze}{\zeta}
\nc{\et}{\eta}
\nc{\ka}{\kappa}
\nc{\rh}{\rho}
\nc{\si}{\sigma}
\nc{\ta}{\tau}
\nc{\up}{\upsilon}
\nc{\ph}{\phi}
\nc{\ch}{\chi}
\nc{\ps}{\psi}
\nc{\om}{\omega}
\nc{\Ga}{\Gamma}
\nc{\De}{\Delta}
\nc{\La}{\Lambda}
\nc{\Si}{\Sigma}
\nc{\Up}{\Upsilon}
\nc{\Ph}{\Phi}
\nc{\Ps}{\Psi}
\nc{\Om}{\Omega}
\nc{\ptl}{\partial}
\nc{\del}{\nabla}
\nc{\ov}{\overline}
\nc{\newcaption}[1]{\centerline{\parbox{15cm}{\caption{#1}}}}
\nc{\us}{U(1)$_S$}

\def\beq{\begin{equation}}
\def\eeq{\end{equation}}
\def\bmat{\begin{displaymath}}
\def\emat{\end{displaymath}}
\def\bear{\begin{eqnarray}}
\def\eear{\end{eqnarray}}
\def\ba{\begin{eqnarray}}
\def\ea{\end{eqnarray}}
\def\bery{\begin{array}}
\def\ery{\end{array}}
\def\bit{\begin{itemize}}
\def\eit{\end{itemize}}
\def\ben{\begin{enumerate}}
\def\een{\end{enumerate}}
\def\btab{\begin{tabular}}
\def\etab{\end{tabular}}
\def\btbl{\begin{table}}
\def\etbl{\end{table}}
\def\bfig{\begin{figure}[htb]}
\def\efig{\end{figure}}
\def\bpic{\begin{picture}}
\def\epic{\end{picture}}

\def\ga{\mathrel{\raise.3ex\hbox{$>$\kern-.75em\lower1ex\hbox{$\sim$}}}}
\def\la{\mathrel{\raise.3ex\hbox{$<$\kern-.75em\lower1ex\hbox{$\sim$}}}}
\def\gappeq{\mathrel{\rlap {\raise.5ex\hbox{$>$}}
{\lower.5ex\hbox{$\sim$}}}}
\def\lappeq{\mathrel{\rlap{\raise.5ex\hbox{$<$}}
{\lower.5ex\hbox{$\sim$}}}}

\def\gyr{{\rm \, G\kern-0.125em yr}}
\def\mev{{\rm \, Me\kern-0.125em V}}
\def\gev{{\rm \, Ge\kern-0.125em V}}
\def\tev{{\rm \, Te\kern-0.125em V}}

\begin{document}
 
\title{Leptogenesis and the Higgs Portal}

\author{Matthias Le Dall and Adam Ritz}
\affiliation{Department of Physics and Astronomy, University of Victoria, 
     Victoria, BC, V8P 5C2 Canada}

\date{August 2014}

\begin{abstract}
\noindent
We study the impact on leptogenesis of Higgs portal couplings to a new scalar
singlet. These couplings open up additional $CP$-violating decay channels for the
higher mass singlet neutrinos $N_2$ and $N_3$. We analyze the simplest case of
two-level $N_1-N_2$ leptogenesis, including significant mass hierarchies, in which the $CP$ asymmetry 
is generated in part by singlet-mediated decays of $N_2$. For these models, provided the lightest singlet neutrino $N_1$ 
is sufficiently weakly coupled to avoid excessive washout, its mass scale is not directly constrained by the
Davidson-Ibarra bound.
\end{abstract}
\maketitle

\tableofcontents

\section{Introduction}\label{sec: Introduction}

The discovery of neutrino oscillations 
\cite{Mohapatra:2005wg,*deGouvea:2004gd,*GonzalezGarcia:2002dz}, and thus small neutrino masses 
 \cite{Capozzi:2013csa,*Fogli:2012ua,*GonzalezGarcia:2012sz,*Tortola:2012te}, provides motivation for leptogenesis 
\cite{Fukugita:1986hr,Luty:1992un,Buchmuller:2004nz,Davidson:2008bu,Blanchet:2012bk} as a simple, and seemingly 
generic,
mechanism
for producing the observed baryon asymmetry in the universe. The simplest UV
completions of the dimension five Weinberg operator $HLHL$ \cite{Weinberg:1979sa}, which contribute to
neutrino masses, naturally incorporate heavy degrees of freedom with allowed
$CP$-violating couplings, whose out-of-equilibrium decay can generate a lepton
asymmetry. Standard Model (SM) sphaleron processes can then equilibrate $B-L$
above the weak scale, resulting in the required late time asymmetry in baryon
number \cite{Klinkhamer:1984di,*Kuzmin:1985mm,*Arnold:1987mh,*Arnold:1987zg,*Harvey:1990qw}.

Beyond the clear possibility to test the Majorana nature of neutrinos through
neutrinoless double beta decay \cite{Racah:1937qq,*Furry:1939qr,*Vergados:2012xy,*Bilenky:2012qi}, the high-scale nature of leptogenesis - characterized for
example by the Davidson-Ibarra bound \cite{Davidson:2002qv} -  and the lack of model-independent links
between the high-scale and low-scale manifestations of $CP$-violation, renders
the mechanism feasible but hard to test \cite{Branco:2011zb}. This has motivated continuing study of
variations of this general framework which may be placed under further
experimental scrutiny, particularly those that allow a lowering of the scale (see e.g.
\cite{Hambye:2001eu,*Fong:2013gaa,*Tsuyuki:2014aia, *Ma:2006ci}).
This turns out to be quite difficult for the basic reason that the asymmetries
generated by loop-level decays are counteracted by similar washout processes
from two-to-two scattering.
Lowering the scale at
which the asymmetry is generated means a lower Hubble expansion rate, and thus
more scattering processes will be in equilibrium and able to efficiently cancel the asymmetry. The conclusion being that it
becomes increasingly difficult to find viable scenarios which operate at or close to the
weak scale; an exception is the case of resonant decays \cite{Pilaftsis:2003gt,*Pilaftsis:2005rv}. 

From a theoretical perspective, the simplest realization of leptogenesis, with
heavy right-handed (RH), i.e. singlet, Majorana neutrinos, involves one of the few renormalizable
interactions between the SM and a neutral hidden (or dark) sector. The RH
neutrino coupling is described by the Langrangian,
\be
  {\cal L} = {\cal  L}_{\rm SM} + i \overline{N}_i \gamma^\mu \ptl_\mu N_i - 
\lambda_{ji} \overline{N}_i P_L L_j\cdot H  - \frac{1}{2}M_i \overline{N}^c_i N_i + h.c. 
\ee
The class of UV complete relevant or marginal interactions of the type characterized by $\lambda_{ji}$, known
as portals, is very small. If we require no additional states, the list contains
this right-handed neutrino coupling, $\lambda \bar{N}L\cdot H$, the coupling of a
scalar singlet to the Higgs, $(\beta S + \lambda S^2)H^\dagger H$, and kinetic
mixing of a U(1) vector with hypercharge, $\kappa V_{\mu\nu} B^{\mu\nu}$. Since
the coefficients of these operators are unsuppressed by any heavy new physics
scale, they are a natural place to look for signs of new short-distance physics. Couplings to a hidden
sector are also motivated by our other primary piece of empirical evidence for
new physics, namely dark matter, and thus these portals have been the focus of
considerable recent attention \cite{Essig:2013lka}.

In this paper, we explore the minimal extension of `standard' leptogenesis
incorporating the Higgs portal coupling, which is now subject to direct
experimental probes at the LHC. The relevant and marginal interactions then
include,
\begin{align}
 {\cal L} &= {\cal  L}_{\rm SM} + i \overline{N}_i \gamma^\mu \ptl_\mu N_i -  \lambda_{ji}
\overline{N}_i P_L L_j\cdot H  - \frac{1}{2} M_i \overline{N}^c_i N_i  \nonumber \\
 & \qquad\qquad - \beta S H^\dagger H - \al_{ij} S
\overline{N}^c_i P_L N_j + h.c. + \cdots \label{L}
\end{align}
We have not shown the quartic Higgs portal coupling here as it will not play a
significant role, other than for the full scalar potential, but have added
the allowed Majoron coupling $\al_{ij}$ between the scalar $S$ and $N_i$. The ensuing scenario for baryogenesis will be referred to as Higgs Portal Leptogenesis (HPL).

As we discuss below, this minimal extension is sufficient to open up new $CP$-violating
decay channels for the next-to-lightest singlet neutrino $N_2$ (and $N_3$). These new channels also decouple the source of the $CP$-asymmetry from the seesaw contribution 
to the light neutrino mass. Consequently, leptogenesis can be viable in a wider mass range, and in particular when the lightest singlet neutrino is very light, 
e.g. 
below the sphaleron threshold. To analyze the impact of the new decay channels, we study the case of two-level $N_1-N_2$ leptogenesis in detail, paying 
attention to the washout induced by scattering associated with $N_1$ and the additional scalar. In minimal leptogenesis, these two-to-two processes are generally negligible as the 
neutrino abundance is Boltzmann suppressed. This is not necessarily the case for $N_{2,1}$ scattering mediated by the hidden sector, and washout would 
be 
problematic if $N_1$ were in equilibrium for an extended period. Nonetheless, we find 
that there are viable regions of 
parameter space in which $N_1$ is parametrically light, but also very weakly coupled, in which the lepton asymmetry generated by $N_2$ decays survives to 
provide the observed baryon abundance (see e.g.~\cite{Engelhard:2006yg,*Vives:2005ra,*DiBari:2005st,*Blanchet:2006dq}). The possibility of effectively decoupling $N_1$ from 
its normally dominant role in leptogenesis, and having it be 
parametrically light and experimentally accessible \cite{Hambye:2001eu,Engelhard:2006yg}, is one of the interesting features of HPL.

The rest of the paper is organized as follows. In Section~2, we outline the model and determine the additional contributions to the $CP$ asymmetry from the 
hidden sector. Since the hidden sector contributions enter only through the decays of the next-to-lightest singlet neutrinos, in Section~3 we turn to the 
Boltzmann evolution of the coupled $N_1-N_2$ system, and study in some detail the important role of $N_1$-mediated two-to-two scattering in washing out the 
$N_2$-generated asymmetry. Section~4 presents a number of results for the final lepton asymmetry in different mass regimes. The new hidden sector contributions 
to the $CP$ asymmetry are crucial in allowing a lowering of the overall mass scale without violating the Davidson-Ibarra bound. We conclude in Section~5, while 
a series of Appendices contain further technical details.

\section{Higgs Portal and the $CP$-asymmetry}\label{sec: HPL CP-asymmetry}

\subsection{The Model}

Our focus in this paper will be a minimal extension of conventional leptogenesis, which adds a scalar singlet $S$
along with the right-handed singlet neutrinos $N_i$. This opens up the Higgs and neutrino portals and, significantly, allows 
for a new $CP$-odd source in the hidden sector. $CP$-violation is of course a central ingredient in leptogenesis, as it is in 
any theory of matter genesis according to the Sakharov conditions \cite{Sakharov:1967dj}. The portal couplings in the Lagrangian (\ref{L}) include
\be
{\cal L}_{\rm portals} =  -  \lambda_{ji}\overline{N}_i P_L L_j\cdot H - \beta S H^\dagger H - \left(\frac{1}{2}M_i \de_{ij}+ \al_{ij} S\right)\overline{N}^c_i 
P_L N_j + h.c. + \cdots \label{Lp}
\ee
The Higgs portal coupling $\beta$ is one part of the full scalar potential $V(H,S)$, and necessarily breaks any $S\rightarrow -S$ symmetry. Thus, determining 
the vacuum structure requires a separate analysis incorporating thermal corrections. This potential has been studied in detail elsewhere 
\cite{Profumo:2007wc,*Ahriche:2007jp,*Biswas:2011td}, 
and here we simply assume that the parameters are 
chosen to ensure viable electroweak symmetry breaking, and importantly that $\langle S\rangle =0$. The possibility of a more complex behaviour of $\langle 
S\rangle$, which modifies the effective RHN mass is nonetheless interesting, and will be discussed further in the concluding section.

In minimal leptogenesis, the Yukawa couplings $\lambda$ determine both the light neutrino mass spectrum and the $CP$-asymmetry generated in RH neutrino decays 
\cite{Covi:1996wh,*Roulet:1997xa,*Buchmuller:1997yu,*Pilaftsis:1997jf}. 
Opening the Higgs portal allows these two physical phenomena to be decoupled, with the Majoron coupling $\al_{ij}$ providing a new $CP$-odd source that is 
unconstrained (for $\langle S\rangle =0$) by the light neutrino mass spectrum. In the Lagrangian (\ref{Lp}), a unitary rotation has been used to diagonalize the RH neutrino mass 
matrix 
$M_{ij} \rightarrow M_i\de_{ij}$. The Majorana 
nature of $N_i=N_i^c$ ensures that $M_{ij}$ is symmetric,
and for $n$ flavors the diagonalization leaves $M_i$ as $n$ real mass 
eigenvalues \cite{Bigi:2000yz,*Doi:1985dx}. In general, the corresponding rotation simply rearranges the $n(n+1)/2$ phases in the 
symmetric matrix 
$\al_{ij}$, which is thus a physical $CP$-odd source in addition to the neutrino Yukawa $\lambda_{ij}$.\footnote{The coupling $\al_{ij} S\overline{N}^c_i P_L 
N_j+h.c$ is more commonly used to generate the right-handed neutrino masses $M_i$ by having $S$ develop a vev $\langle 
S\rangle$, spontaneously breaking a global lepton number symmetry. In such cases where an explicit mass term $M_{ij}$ is forbidden, the matrix $\alpha_{ij}$ can 
be be made real and diagonal.}

\subsection{CP Asymmetry}

\begin{figure}[t]
\centerline{\includegraphics[width=15cm]{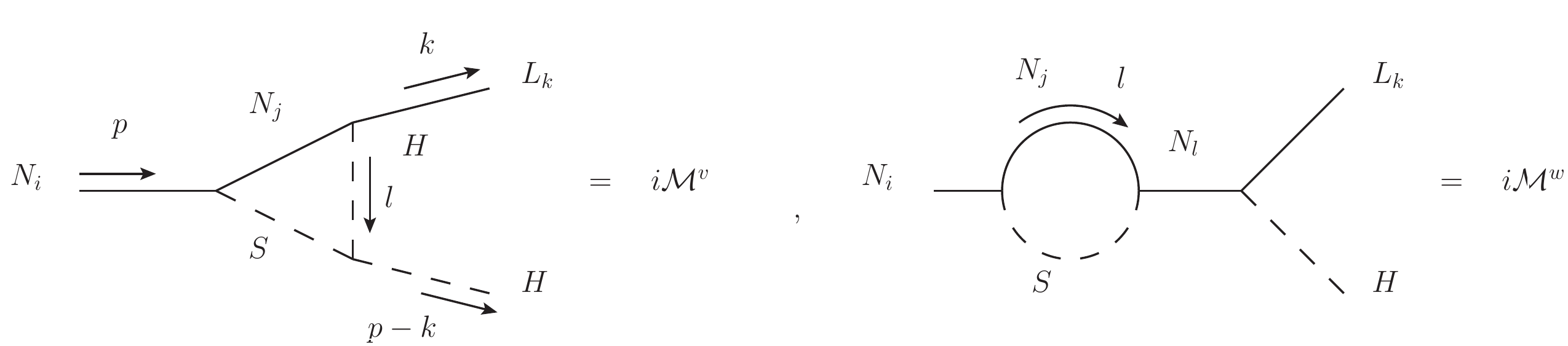}}
\caption{\footnotesize Two hidden sector decay channels for RHN that contribute to the $CP$ asymmetry. The superscripts `v', and `w' stand for the vertex and 
wave function diagrams.}
\label{fig: Hidden sector asymmetry channels}
\end{figure}

In leptogenesis, the $CP$-asymmetry arises from RH neutrino (RHN) decays to leptons, $N\rightarrow LH$ and $N\rightarrow \overline{L}\overline{H}$, and is measured by
$\epsilon_i$,
\beq\label{eq: asymmetry definition}
 \epsilon_i\equiv\frac{\Gamma(N_i\rightarrow LH)-\Gamma(N_i\rightarrow\overline L\overline H)}{\Gamma_i}.
\eeq
In the denominator, the $N_i$ decay rate is calculated at tree level, and reads
\beq\label{eq: total decay rate}
 \Gamma_i\equiv\sum_{k,\alpha,\beta}\left(\Gamma(N_i\rightarrow
L^\alpha_kH^\beta)+\Gamma(N_i\rightarrow
\overline{L_k^\alpha H^\beta})\right)=\frac{(\lambda^\dagger\lambda)_{ii}}{8\pi}M_i,
\eeq
where the lepton family index, $k=1,2,3$ stands for the electron, muon and tau families respectively. The
$\alpha,\beta=1,2$ indices denote the components of the SU(2) lepton and Higgs doublets $L=(\nu_l\quad e_l)^T$ and $H=(H^+\quad
H^0)^T$. If we schematically write the decay amplitude as 
$i\MM=\Gamma_{0}+\Gamma_{1}I$, with $\Gamma_{0,1}$ the tree and loop level combinations of coupling constants, and $I$ the loop 
function, then the decay amplitude for the antiparticle is $i\overline\MM=\Gamma_{0}^*+\Gamma_{1}^*I$, while the decay rates 
$\Gamma(N_i\rightarrow LH)$ and $\Gamma(N_i\rightarrow \overline{LH})$ are proportional to $|i\MM|^2$ and $|i\overline{\MM}|^2$ respectively. At tree level, the 
difference vanishes, but at the loop level, the $CP$-asymmetry takes the schematic form
\beq
 \epsilon_i\sim2\frac{\text{Im}\{\Gamma_0\Gamma_1^*\}}{|\Gamma_0|^2}\text{Im}\{I\}.
\eeq
Thus, at least for two-body decays, the $CP$-asymmetry requires loops, and a phase in the loop function itself. In standard leptogenesis, only 
the Yukawa $\lambda_{ij}$ allows for this decay channel, and can accommodate CP-violation. In
Higgs Portal Leptogenesis, additional lepton number violating and $CP$-violating sources are present in the theory, specifically the $\al_{ij}$ coupling as 
discussed above. 
As a result, additional loop-induced decay channels open up, as displayed in Fig.~\ref{fig: Hidden sector asymmetry channels}.

The corresponding $CP$-asymmetries will be discussed in the following subsections. We utilize the Majorana Feynman rules 
\cite{Gates:1987ay,*Denner:1992me,*Denner:1992vza} for the RH neutrinos, and determine the imaginary parts of the loop functions using the standard Cutkosky 
rules \cite{Cutkosky:1960sp}.

\subsubsection*{Vertex Corrections: 2-body final states}\label{subsec: Vertex correction}

The contribution of the diagram $i\MM^{v}$ to the asymmetry is
\beq
 \epsilon_i^{v}=4\sum_j\left(\frac{\text{Im}\{(\lambda^\dagger\lambda)_{ji}\beta\alpha_{ij}\}}{(\lambda^\dagger\lambda)_{ii}M_i}
\text { Im}\{I_{jLL}\}+\frac{\text{Im}\{ (\lambda^\dagger\lambda)_{ji}
\beta\alpha_{ij}^*\}}{(\lambda^\dagger\lambda)_{ii}M_i}\text {Im}\{I_{jRL}\}\right),
\eeq
where $I_{jLL}$ and $I_{jRL}$ are loop function integrals.
The vertex contribution splits into two halves, proportional to $\text{Im}\{I_{jLL}\}$ and $\text{Im}\{I_{jRL}\}$, corresponding to the mixing of 
the left- and right- chiralities of the Majorana fermions along the fermion lines, effectively leading to the two chirality chains
\beq
\begin{split}
 &\nu_R(\text{RH})\longrightarrow\nu_R^C(\text{LH})\longrightarrow L(\text{LH}),\\
 &\nu_R^C(\text{LH})\longrightarrow\nu_R^C(\text{LH})\longrightarrow L(\text{LH}).
\end{split}
\eeq
Because the final leptons (assumed massless) have a definite chirality, the Yukawa coupling forces the next-to-last neutrino to be of the same chirality as
the final lepton, left-handed. The functions $I_{jLL}$ and $I_{jRL}$ correspond to the L-L-L and R-L-L chains respectively. The reason why the chirality 
chains do not combine, owes to the fact that $\alpha_{ij}$ is neither real nor diagonal. In the vertex contribution, there are 
three possible cuts which lead to an imaginary part: cuts along the $S/N$ lines, the $H/S$ lines and the 
$H/N$ lines, each of which contains the two chirality chain contributions. 
Thus, each chirality chain function, $\text{Im}\{I_{jLL}\}$ and 
$\text{Im}\{I_{jRL}\}$ is the sum,
\beq
 \text{Im}\{I\}=\text{Im}\{I\}^{S/N}+\text{Im}\{I\}^{H/S}+\text{Im}\{I\}^{H/N}.
\eeq

It will be convenient to graphically represent the interference terms in $|{\cal M}|^2$ which contribute to the imaginary part in the form of \textit{bubble} 
diagrams. For now we focus on the $N/S$ cut which is shown in Fig.~\ref{fig: kinematic constraints NS cut} (the full set of cuts is presented later in 
Fig.~\ref{fig: 
Imaginary_part_bubble_diagram_vertex}). The double line indicates external lines that are on-shell by definition in $|{\cal M}|^2$, in this case the final lepton 
and Higgs, while the single 
line shows the Cutkosky-cut. According to the Cutkosky rules, the $N/S$
cut is given by

\newsavebox\mybox
\savebox{\mybox}{\includegraphics[width=9cm]{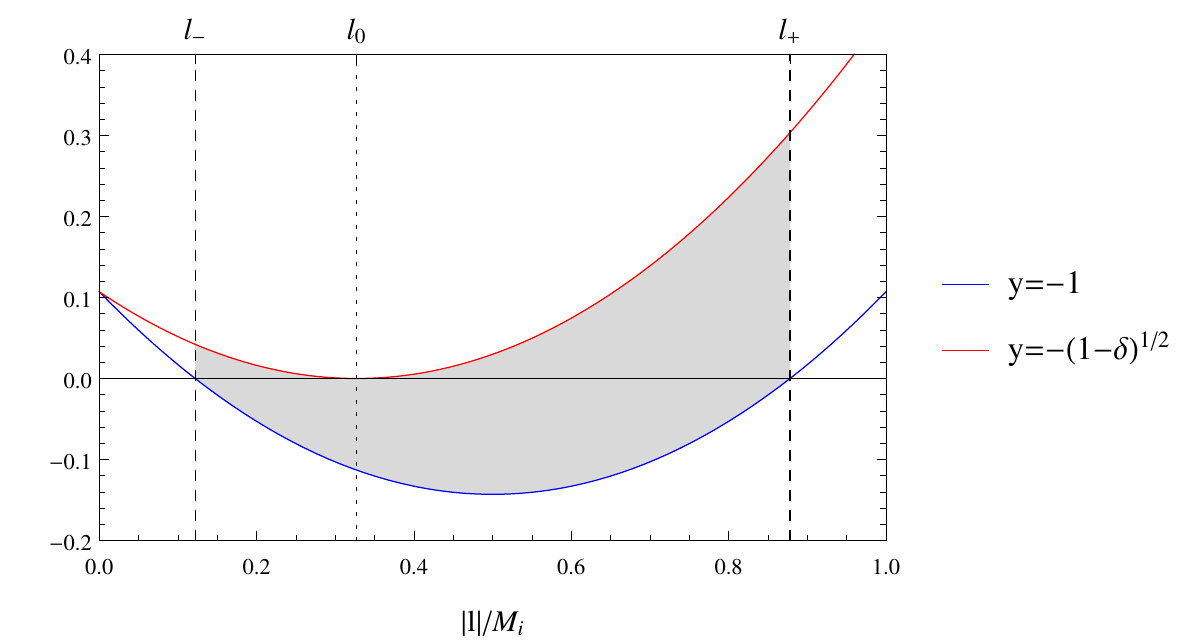}}
\begin{figure}
    \centering
      \begin{minipage}{0.45\textwidth}
        \centering
        \vbox to \ht\mybox{%
            \vfill
            \includegraphics[width=7cm]{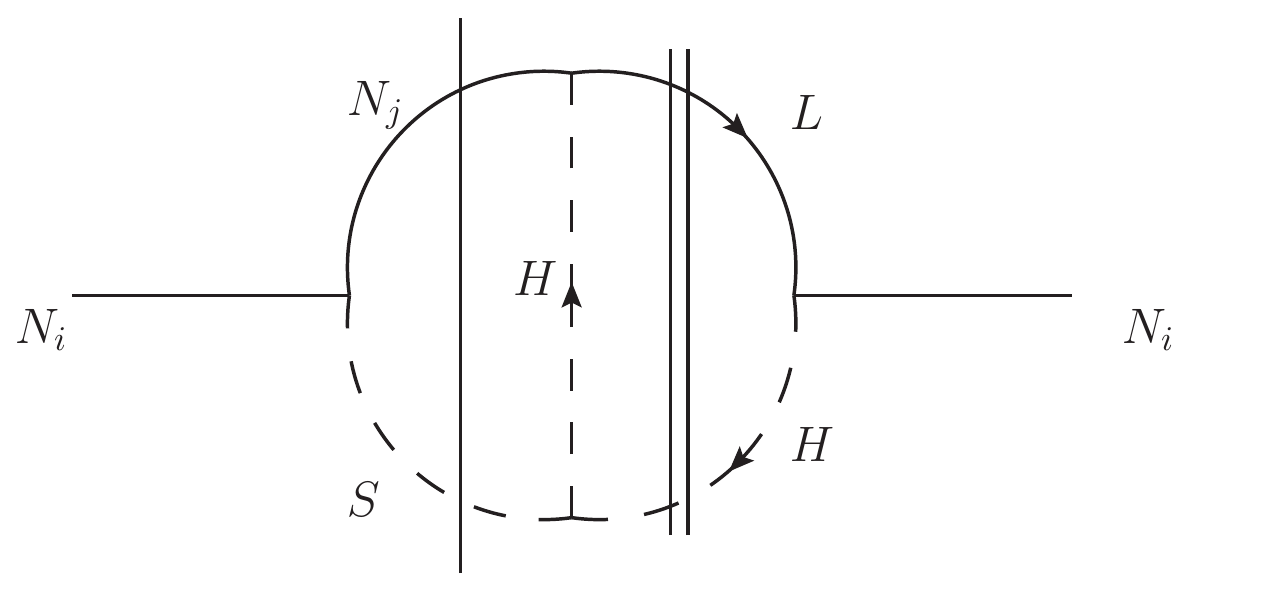}
            \vfill
        }
    \end{minipage}
    \begin{minipage}{0.45\textwidth}
        \centering
        \usebox{\mybox}
    \end{minipage}
    \caption{\footnotesize The left 
plot graphically represents the $N/S$ cut. The right figure exhibits the quadratic 
constraint in Eq.~\eqref{eq: conditions for imaginary part of vertex correction SN cut}. The vertical dashed lines represent the momentum $|\vec l|$ upper and 
lower bounds, $|\vec{l}|_{\pm}=M_i(1\pm\sqrt{\delta})/2$, which become $|\vec{l}|_{0}=M_i\sqrt{1-\delta}/2$ at 
$y=-\sqrt{1-\delta}$. The imaginary part of the vertex correction is non-zero contributions only from the gray area.}
\label{fig: kinematic constraints NS cut}
\end{figure}
\beq
 \begin{split}
 \text{Im}\{I_{LL}\}^{S/N}=2\pi^2\Int\frac{l\cdot
k}{l^2}&\delta((l+k)^2-M_j^2)\delta((l+k-p)^2-m_S^2)\\&\times\Theta(p^0-l^0-k^0)\Theta(k^0+l^0).
\end{split}
\eeq
The delta functions impose the on-shell condition for the $N/S$ lines, and the Heaviside functions $\Theta$ require these on-shell lines to be 
physical (timelike) processes. In other words, imposing positivity of the energies $p^0-l^0-k^0$ and $k^0+l^0$, requires that the cut diagram corresponds to
the decay $N_i\rightarrow N_jS$ followed by the scattering $N_jS\rightarrow LH$. Combining the energy constraints also restricts the individual energies 
$l^0=M_i(r_{ji}-\sigma_i)/2$, and $M_i/2>l^0>-M_i/2$. The on-shell conditions in turn, imply the quadratic 
constraint on the three momentum $\vec l$
\beq\label{eq: conditions for imaginary part of vertex correction SN cut}
 |\vec l|^2+M_i|\vec l|y-\frac{M_i^2}{4}\Big(\delta(1,r_{ji},\sigma_i)-1\Big)=0,
\eeq
where we have introduced the shorthand notations
\beq\label{eq: definition of mass ratio r and function delta}
 r_{ji}=\frac{M_j^2}{M_i^2},\quad\sigma_i=\frac{m_S^2}{M_i^2},\quad\delta(a,b,c)=(a-b-c)^2-4bc,
\quad y=\cos\theta,
\eeq
For simplicity below, $\delta$ without specified variables will implicitly be understood to mean $\delta(1,r_{ji},\sigma_i)$, 
unless stated otherwise. The angle $\theta$ lies between the 3-momenta $\vec{l}$
and $\vec{k}$. The combined constraints on $l^0$ given above imply the equivalent constraint, $1>r_{ji}-\sigma_i>-1$. Since the kinematics must 
allow the decay $N_i\rightarrow N_jS$, the latter constraint requires that $1>\sqrt{r_{ji}}+\sqrt{\sigma_i}$. Importantly, we observe that 
the diagram only has an imaginary part for decays of the next-to-lightest neutrinos. Similarly, the imaginary part is non-vanishing provided the quadratic
equation for $|\vec l|$ in \eqref{eq: conditions for imaginary part of vertex correction SN cut} has real solutions, thus imposing
the condition $\delta>0$, here again, satisfied if $1>\sqrt{r_{ji}}+\sqrt{\sigma_i}$. The $l^0$ integration is trivial since its value is uniquely 
fixed. As usual, the remaining
integration over $\vec l$ is split into the radial and angular part. In spherical coordinates with $\vec k$ along the z-axis, the
azimuthal angle $\phi$ trivially integrates to $2\pi$, and $\theta$ corresponds to the inclination angle of the spherical coordinate system. The leftover integrals 
over $y$ and $|\vec
l|$ are not independent because of the constraint \eqref{eq: conditions for imaginary part of vertex correction SN cut}. That constraint 
has been plotted in Fig.~\ref{fig: kinematic constraints NS cut}, where we see that the kinematics are constrained to the 
ranges $-1\Leq y\Leq-\sqrt{1-\delta}$ and $M_i/2\left(1-\sqrt{\delta}\right)\Leq|\vec
l|\Leq M_i/2\left(1+\sqrt{\delta}\right)$. The integration is nonvanishing within this range, leading to the result,
\beq\label{eq: final result for I_jLL NS cut}
 \text{Im}\{I_{LL}\}^{N/S}=\frac{1}{32\pi}\left[-\sqrt{\delta}+r_{ji}\log\left(\frac{\sqrt{\delta+4r_{ji}\sigma_i}-\sqrt{\delta}}{
\sqrt{\delta+4r_{ji}\sigma_i}+\sqrt{\delta}}
\right)\right],\quad1\Geq\sqrt{r_{ji}}+\sqrt{\sigma_i}.
\eeq
Similar steps lead to the other chirality chain function, $\text{Im}\{I_{jRL}\}^{N/S}$, for the $N/S$ cut,
\beq\label{eq: final result for I_jLR NS cut}
 \text{Im}\{I_{jRL}\}^{N/S}=\frac{\sqrt{r_{ji}}}{32\pi}\log\left(\frac{\sqrt{\delta+4r_{ji}\sigma_i}-\sqrt{\delta}}{\sqrt{
\delta+4r_{ji}\sigma_i}+\sqrt{\delta}}\right),\quad1\Geq\sqrt{r_{ji}}+\sqrt{\sigma_i},
\eeq
as well as for the $H/S$ cuts,
\beq\label{eq: final result for I_jLL}
\begin{split}
 &\text{Im}\{I_{jRL}\}^{H/S}=\frac{\sqrt {r_{ji}}}{32\pi}\log\bigg|\frac{1-r_{ji}}{r_{ji}}\bigg|\Bigg|_{ \sigma_i=0}\quad,\\
 &\text{Im}\{I_{jLL}\}^{H/S}=\frac{1}{32\pi}\left(1+r_{ji}\log\bigg|\frac{1-r_{ji}}{r_{ji}}\bigg|\right)\Bigg|_{\sigma_i=0},
\end{split}
\eeq
while the $H/N$ cut gives a vanishing imaginary part $\text{Im}\{I_{jLL}\}^{H/N}=\text{Im}\{I_{jRL}\}^{H/N}=0$. The notation $|_{\sigma_i=0}$ 
means that the imaginary part is only nonzero if $\sigma_i=0$. Note that the $N/S$ cut is divergent in the infrared limit $\sigma_i=0$, where it is 
effectively equivalent to the decay $N_i\rightarrow N_jS$ followed by the scattering $N_jS\rightarrow LH$ mediated by the Higgs in the t-channel. The divergence, 
due to radiating massless scalars in the infrared collinear limit, is canceled by including the appropriate three-body decays as discussed below.

\subsubsection*{Vertex Corrections: 3-body final states}\label{subsubsec: three-body final state}
\begin{figure}[t]
\centerline{\includegraphics[width=15cm]{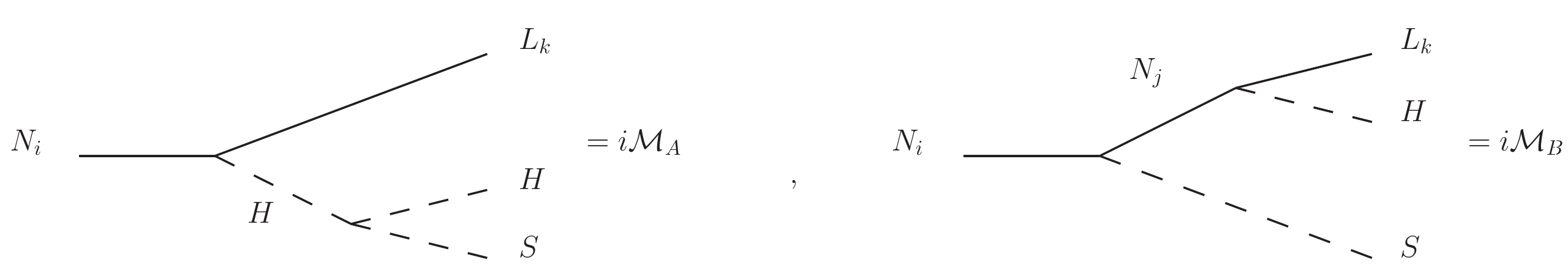}}
\caption{\footnotesize Two 3-body decay diagrams contributing to the $CP$-asymmetry at the same order in couplings as the loop-corrected 2-body decays. As 
discussed in the text, their inclusion is important in ensuring that the full $CP$ asymmetry is well-defined and free of infrared divergences.}
\label{fig: Three body decay channels}
\end{figure}

\begin{figure}[t]
\centering
\subfigure[]{
\includegraphics[width=15cm]{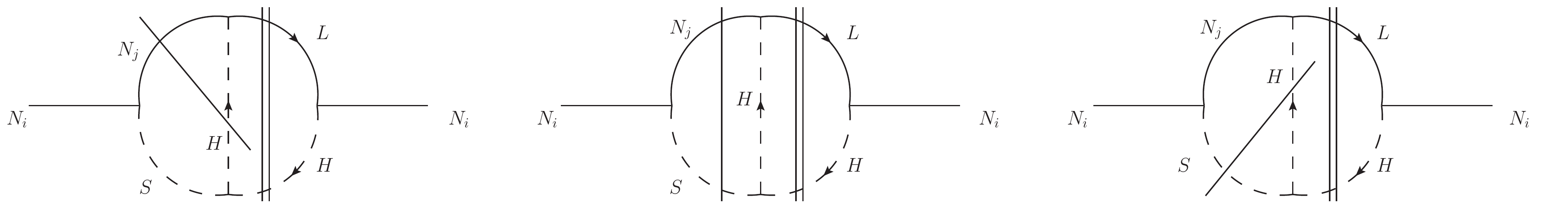}
\label{fig: Imaginary_part_bubble_diagram_vertex}}
\subfigure[]{\includegraphics[width=10cm]{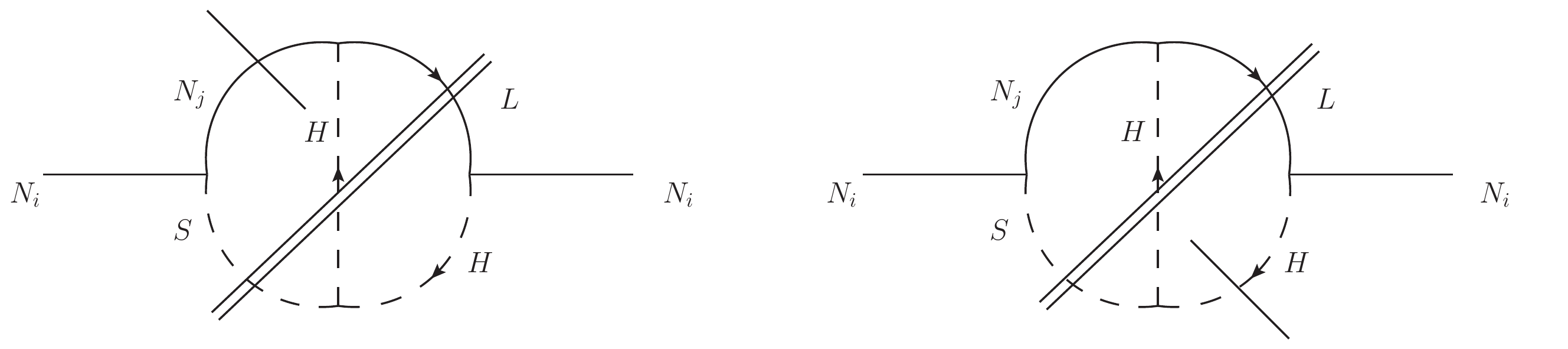}
\label{fig: Imaginary_bubble_diagram_tree}}
\caption[]{\footnotesize The bubble diagrams enumerating the cuts that contribute to the $CP$-asymmetry from both the vertex loop corrections (a) and 
3-body final states (b). The external state lines that are on-shell by definition are cut with a double line, while the Cutkosky cuts are shown with a single line.}
\label{fig: bubble diagrams}
\end{figure}

In Fig.~\ref{fig: Three body decay channels}, we show the two three-body final state amplitudes whose interference develops an imaginary part and 
contributes to the $CP$-asymmetry. The 
three-body final state $CP$-asymmetry
$\epsilon_{i}^{(3)}$ measures the difference $(\Gamma(N_i\rightarrow
LHS)-\Gamma(N_i\rightarrow\overline L\overline HS))/(\Gamma(N_i\rightarrow LH, LHS)+\Gamma(N_i\rightarrow \overline{LH},\overline{LH}S))$, with the total RHN 
decay rate  in the denominator. The three-body final state decay rate being subdominant due to
the reduced phase space, we can approximate $\epsilon_i^{(3)}$ as
\beq
 \epsilon_i^{(3)}\simeq \frac{\Gamma(N_i\rightarrow
LHS)-\Gamma(N_i\rightarrow\overline L\overline HS)}{\Gamma_i}.
\eeq
In general, the three-body $CP$-asymmetry arises from both $i\MM_A(i\MM_B)^*\sim\mathcal{O}(\lambda^2\beta\alpha)$ and 
$|i\MM_B|^2\sim\mathcal{O}(\lambda^2\alpha^2)$.
Only the former term, which enters at the same order as the vertex contribution, is of interest here. 
The two contributing cuts through the $H$ and the 
$N_j$ propagators are represented in Fig.~\ref{fig: Imaginary_bubble_diagram_tree} 
as bubble diagrams. The result is
\beq
 \epsilon_i^{(3)}=4\sum_{j}\left(\frac{\text{Im}\left\{(\lambda^\dagger\lambda)_{ji}\beta\alpha_{ij}\right\}}{
(\lambda^\dagger\lambda)_{ii}M_i}\text{Im}\{\mathcal I_{jLL}\}^{(3)}+\frac{\text{Im}\left\{(\lambda^\dagger\lambda)_{ji}\beta\alpha_{ij}
^*\right\}}{(\lambda^\dagger\lambda)_{ii} M_i}\text{Im}\{\mathcal I_{jRL}\}^{(3)}\right),
\eeq
with
\beq\label{eq: 3 body final state cp asymmetry}
 \begin{split}
 &\text{Im}\{\mathcal{I}_{jLL}\}^{(3)}=\frac{r_{ji}}{32\pi}\log\left(\frac{\sqrt{\delta+4r_{ji}\sigma_i+2\sigma_i}+\sqrt
{\delta}}{\sqrt{\delta+4r_{ji}\sigma_i+2\sigma_i}-\sqrt{\delta}}\right)-\frac{1}{32\pi}\left(1+r_{ji}\log\bigg|\frac{1-r_{ji}}{
r_ { ji}}\bigg|\right)\Bigg|_{\sigma_i=0}\quad,\\
&\text{Im}\{\mathcal{I}_{jRL}\}^{(3)}=\frac{\sqrt{r_{ji}}}{32\pi}\left[-\sqrt{\delta}
+\log\left(\frac{\sqrt{\delta+4r_{ji}\sigma_i+2\sigma_i}+\sqrt{\delta}}{\sqrt{\delta+4r_{ji}\sigma_i+2\sigma_i}-\sqrt{
\delta}}
\right)\right]-\frac{\sqrt{r_{ji}}}{32\pi}\log\bigg|\frac{1-r_{ji}}{r_{ji}}\bigg|\Bigg|_{\sigma_i=0}.
 \end{split}
\eeq
The first terms in $\text{Im}\{\mathcal{I}_{jLL}\}^{(3)}$ and $\text{Im}\{\mathcal{I}_{jRL}\}^{(3)}$ come from cutting the $N_j$ line, while the second 
terms come from cutting the $S$ line,  respectively combining with the $N/S$ and $H/S$ cuts of the tree-loop interference in (\ref{eq: final result 
for I_jLL NS cut}, \ref{eq: final result for I_jLR NS cut}, \ref{eq: final result for I_jLL}), leading to the \textit{corrected} vertex $CP$-asymmetry 
$\epsilon_i^v$,
\beq\label{eq: vertex correction correction to the hidden sector asymmetry}
 \epsilon_i^{v}=\sum_{j}\left(\frac{\text{Im}\left\{(\lambda^\dagger\lambda)_{ji}\beta\alpha_{ij}\right\}}{8\pi
(\lambda^\dagger\lambda)_{ii}M_i}\mathcal{F}^v_{jLL}(r_{ji},\sigma_i)+\frac{\text{Im}\left\{(\lambda^\dagger\lambda)_{ji}
\beta\alpha_ { ij }
^*\right\}}{8\pi(\lambda^\dagger\lambda)_{ii} M_i}\mathcal{F}^v_{jRL}(r_{ji},\sigma_i)\right),
\eeq
where
\beq\label{eq: new final result for I_{jLL}}
 \mathcal{F}^v_{jLL}(r_{ji},\sigma_i)\equiv\left(-\sqrt{\delta}+r_{ji}\log
G\right)\quad,\quad\mathcal{F}^v_{jRL}(r_{ji},\sigma_i)\equiv\left(-\sqrt{r_{ji}}\sqrt{\delta}+\sqrt{r_{ji}}\log
G\right),
\eeq
and
\beq
 G\equiv\frac{\sqrt{\delta+4r_{ji}\sigma_i+2\sigma_i}+\sqrt
{\delta}}{\sqrt{\delta+4r_{ji}\sigma_i+2\sigma_i}-\sqrt{\delta}}\frac{\sqrt{\delta+4r_{ji}\sigma_i}-\sqrt{\delta}}{\sqrt{
\delta+4r_{ji}\sigma_i}+\sqrt{\delta}}.
\eeq
Note that the infrared divergence at $\sigma_i=0$ has disappeared, resulting from a cancellation between \eqref{eq: 
final result for I_jLL NS cut}, \eqref{eq: final result for I_jLR NS cut} and \eqref{eq: 3 body final state cp asymmetry}. A simple graphical 
understanding of this cancellation emerges by comparing the $N/S$ cut diagram of Fig.~\ref{fig: Imaginary_part_bubble_diagram_vertex} with the $N_j$-line cut 
diagram of Fig.~\ref{fig: 
Imaginary_bubble_diagram_tree}. In general they have different kinematics, but they coincide in the infrared limit 
where all the internal lines of the respective diagrams are allowed to be on-shell, permitting the emission of soft particles. The inclusion 
of the two-body and three-body final state contributions renders the $CP$-asymmetry well-defined.

Note also that the contributions that are non-vanishing only in the $\sigma_i=0$ limit, i.e. those coming from the $H/S$ cut in \eqref{eq: final result for I_jLL} 
and 
from the $H$-line cut 
in \eqref{eq: 3 body final state cp asymmetry} also cancel out. This can again be understood by comparing the $H/S$-cut diagram of 
Fig.~\ref{fig: Imaginary_part_bubble_diagram_vertex} and the $H$-cut of Fig.~\ref{fig: Imaginary_bubble_diagram_tree}, for which the kinematics 
are identical.

\subsubsection*{Wave-function Corrections}\label{subsec: wave function correction}

Once again, because of the Majorana nature of the Right-handed neutrinos, there can be chirality mixing, leading to the following 4 chirality 
chains in the right-hand diagram of Fig.~\ref{fig: Hidden sector asymmetry channels}:
\beq
\begin{split}
 &\nu_R(\text{RH})\rightarrow\nu_R(\text{RH})\rightarrow\nu_R^C(\text{LH})\rightarrow L(\text{LH}),\\
 &\nu_R(\text{RH})\rightarrow\nu_R^C(\text{LH})\rightarrow\nu_R^C(\text{LH})\rightarrow L(\text{LH}),\\
 &\nu_R^C(\text{LH})\rightarrow\nu_R(\text{RH})\rightarrow\nu_R^C(\text{LH})\rightarrow L(\text{LH}),\\
 &\nu_R^C(\text{LH})\rightarrow\nu_R^C(\text{LH})\rightarrow\nu_R^C(\text{LH})\rightarrow L(\text{LH}).
\end{split}
\eeq
The two first chains contain only 1 chirality flip, the third contains 2 flips, and the last contains none. Each chain will
be labeled by the chiralities of the two first lines in the loop, i.e. the RR, RL, LR, LL chains respectively. The asymmetry then takes the following form,
\beq\label{eq: wave function correction to hidden sector asymmetry}
 \begin{split}
\epsilon_i^{w}=\sum_{l,j}\Bigg(&\frac{\text{Im}\{(\lambda^\dagger\lambda)_{li}\alpha_{lj}\alpha_{ij}^*\}}{8\pi
(\lambda^\dagger\lambda)_{ii}}\mathcal{F}^w_{jlLL}+\frac{\text{Im}\{(\lambda^\dagger\lambda)_{li}\alpha_{lj}^*\alpha_{ij}^*\}}{
8\pi(\lambda^\dagger\lambda)_{ii}}\mathcal{F}^w_{jlLR}\\&+\frac{\text{Im}\{(\lambda^\dagger\lambda)_{li}\alpha_{lj}\alpha_{ij}\}}{
8\pi(\lambda^\dagger\lambda)_{ii}}\mathcal{F}^w_{jlRL}+\frac{\text{Im}\{(\lambda^\dagger\lambda)_{li}\alpha_{lj}^*\alpha_{ij}\}}{
8\pi(\lambda^\dagger\lambda)_{ii}}\mathcal{F}^w_{jlRR}\Bigg).
 \end{split}
\eeq
Calculating the $\mathcal{F}^w$ loop functions is relatively simple as the imaginary part comes solely from the diagrams in which both lines in the
loop are cut, which uniquely defines
all the kinematics, trivializing the integrals. Thus we will simply state the final results,
\beq
 \begin{split} 
&\mathcal{F}^w_{jlLL}(r_{ji},\sigma_i)\equiv\frac{\sqrt{\delta}}{2}\frac{\sqrt{\delta+4r_{ji}}}{1-r_{li}},\quad\mathcal{F}^w_
{ jlLR }(r_{ji},\sigma_i)
\equiv\sqrt { \delta }\frac{\sqrt{r_{ji}}\sqrt{r_{li}}}{1-r_{li}},\\
&\mathcal{F}^w_{jlRL}(r_{ji},\sigma_i)\equiv\sqrt{\delta}\frac{\sqrt{r_{ji}}}{1-r_{li}},\quad\mathcal{F}^w_{jlRR}(r_{ji},
\sigma_i)\equiv\frac{\sqrt{
\delta } } {2}\frac{\sqrt{r_{li}}\sqrt{\delta+4r_{ji}}}{1-r_{li}}.
 \end{split}
\eeq
As noted earlier, we have used the shorthand notation $r_{ji}$, and $\delta=(1-r_{ji}-\sigma_i)^2-4r_{ji}\sigma_i$. The
kinematic constraint remains the same as for the vertex correction, $1>\sqrt{r_{ji}}+\sqrt{\sigma_i}$.

\subsubsection*{Summary}

Most significantly, the kinematic constraint $1>\sqrt{r_{ji}}+\sqrt{\sigma_i}$, prevents the lighter Neutrino flavor, $N_1$, from having any $CP$-odd decays 
through the Higgs portal, since by definition, $r_{21}>1$. Only the heavier flavors, $N_{2,3}$ can contribute to the $CP$-asymmetry through the hidden sector 
decays. For the remainder of this paper, we will generally focus on the minimal case with two heavy neutrinos $N_1$ and $N_2$, so that the hidden sector will 
play an important role through the decays of $N_2$. This presents us with the interesting possibility of taking $N_1$ parametrically light, where it 
could have 
other phenomenological consequences. At the same time, there is also the danger of significant washout of the asymmetry by scattering processes mediated by 
$N_1$. We will discuss the latter issue in some detail in subsequent sections.

The full $CP$ asymmetry is obtained by combining the above results for $\ep_i^v$ (\ref{eq: vertex correction correction to the hidden sector asymmetry}) and 
$\ep_i^w$ (\ref{eq: wave function correction to hidden sector asymmetry}). For the hierarchical $N_1-N_2$ regime, with $M_2/M_1\geq10$  that will be of interest 
later, 
 the $CP$-asymmetry can be approximated by the following simple expressions (see Appendix \ref{sec: app paramet of CP-asymmetry} for details),
\beq
\begin{split}
 &\epsilon_1\sim\frac{3M_1\sum_\alpha m_\alpha}{8\pi v^2},\\
 &\epsilon_2\sim\frac{3M_2\sum_\alpha m_\alpha}{16\pi v^2}+\left(\frac{\beta}{M_2}+\frac{|\alpha_{11}|}
{2}(1-\sigma_2)\right)\frac{|\alpha_{21}|}{8\pi
}\sqrt{\frac{M_1}{M_2}}(1-\sigma_2).
\label{cp_sum}
\end{split}
\eeq
The index $\alpha=1,2,3$, and $m_\alpha$ is the mass of the 
$\alpha$-th active neutrino. Assuming a normal hierarchy among the light neutrino masses, we set  
$\sum_\alpha m_\alpha\simeq m_3\simeq\sqrt{\Delta m_{31}^2}\sim0.05$eV.

\section{Two-stage Boltzmann evolution}\label{sec: The two-level Boltzmann evolution}

In minimal leptogenesis, the RHN sector $\lambda_{ij}N_jL_iH+M_iN_iN_i$ provides the ingredients for two of Sakharov's conditions to be satisfied; $L$ is 
violated due to the presence of both $M_i$ and $\lambda_{ij}$, while there are physical $CP$-odd phases in $\lambda_{ij}$. The third and final condition is 
satisfied dynamically as the expansion of the universe provides a mechanism for $L$-violating processes to go out of equilibrium. For this to happen, the rate 
$\Gamma_N$ of $L$-violating RHN decays must fall below the Hubble expansion rate $H$. This transition is controlled by the Gamow \textit{equilibrium parameter}, 
$K=\Gamma_N/H(T=M)$ \cite{Kolb:1990vq},\footnote{We will follow the literature and denote the Hubble-normalized $N$ decay rate as $K$, while the modified Bessel function that generically appears in the thermal rates will consistently be written as $K_i(z)$, distinguished by the extra argument $z=M/T$.} where the Hubble rate $H(T=M)$ sets the time scale, $t_H$, at which the equilibrium density becomes Boltzmann 
suppressed. Setting $K<1$ 
ensures the particle lifetime is longer than the Hubble time, $\tau_N>t_H$, and an excess abundance develops. In that case, the rate of decays will be large 
compared to that of inverse decays in order for the neutrino abundance to be able to reach equilibrium, effectively putting the system 
out of equilibrium.

In Higgs Portal Leptogenesis, we require at least two Majorana neutrinos and there are two major implications. On one hand, the $CP$-asymmetry from $N_2$ decays 
(\ref{cp_sum}) is enhanced for low masses, and can in fact become the dominant contribution. This suggests the possibility of establishing a `lower energy' theory 
of leptogenesis, 
mainly controlled by $N_2$ physics. On the other hand, the two RHN  flavors leads to a novel evolution in the total lepton asymmetry. In minimal leptogenesis, 
the lepton asymmetry is primarily generated in a temperature range near the lightest RHN mass, $T\sim M_1$, since the decays and scattering are 
out-of-equilibrium for lower 
temperatures. The difference here is that, even though most of the lepton asymmetry can be generated through $N_2$ decays and inverse decays at temperatures 
around $T\sim M_2$, the lighter neutrino flavor $N_1$ potentially remains in equilibrium and can mediate rapid washout of the $N_2$-generated asymmetry. These 
interactions will be studied carefully below, to identify regimes in which $N_1$ is sufficiently weakly coupled that these new washout processes are suppressed.

\subsection{Boltzmann equations}\label{subsec: Boltzmann equations scattering}

 In the minimal leptogenesis scenario, typically once the neutrino decays go out-of-equilibrium, all the scattering processes also go out-of-equilibrium. 
 The new feature in HPL is the possibility of having scattering processes in equilibrium during the period that a $CP$-asymmetry would be generated through out of 
equilibrium decays. The most 
significant are those $L$-violating scattering processes with an external $N_1$, whose abundance is not Boltzmann suppressed. The scattering processes that have 
an external $N_2$ are of course suppressed by the $N_2$ abundance which rapidly falls off exponentially. Among the scattering processes that violate the lepton 
number by $\Delta L=1$ units, we 
include the scattering $N_iL\leftrightarrow Q\overline t$ in the $s$-channel, and $N_iQ\leftrightarrow Lt$, $Nt\leftrightarrow LQ$ in the $t$-channel. From the hidden sector, one 
includes the $s$-channel processes $N_iL\leftrightarrow HS$ mediated by a Higgs, and $N_iS\leftrightarrow 
LH$ mediated by a neutrino. In the $t$-channel one has $N_iS\leftrightarrow LH$ and $N_iH\leftrightarrow LS$ both mediated by a Higgs. A full treatment 
of neutrino-mediated scattering is complicated because of the $\lambda_{ij}$ and $\alpha_{ij}$ flavor structures. For simplicity, we will ignore the 
flavor-mixing in these processes with intermediate neutrinos, e.g. $\sigma_{N_iS\leftrightarrow LH}=\sum_j\sigma_{N_iS\xleftrightarrow[j]{} LH}$, and assume the 
processes are dominated by one flavor. This is sufficient for order of magnitude estimates. Note that because of the $\alpha_{ij}$ coupling, we need to include 
interactions such as $N_iN_j\leftrightarrow HH$ mediated by $S$ in the $s$-channel, which can efficiently deplete the neutrino abundance, and in turn affect 
the lepton asymmetry washout \cite{Sierra:2014sta}.\footnote{In the context of standard leptogenesis, $\Delta N=2$ interactions, e.g. 
$N_1N_1\rightarrow HH$ mediated by a lepton in the $t$-channel, are negligible since they scale as $\lambda^4$ which is suppressed for 
$M_1\lesssim10^{15}$GeV. Interactions in the $\Delta N=2$ class have for instance been taken into account in the context of GUT theories in 
\cite{Plumacher:1996kc}.}

 As for the $\Delta L=2$ interactions, one has $LH\leftrightarrow \overline{LH}$ mediated by a 
neutrino in the $s$-channel, and $LL\leftrightarrow HH$, $\overline{LL}\leftrightarrow HH$ mediated by a neutrino in the $t$-channel. We start by 
describing the Boltzmann equations for the lepton asymmetry, which are the most complex, and then review the neutrino abundance and the general features. 
Further technical details are contained in Appendices B and C. Note that this work is concerned with the main dynamical features of the model presented above, 
focussing on the 
impact of the Higgs portal couplings. Thus, in deriving the Boltzmann equations, we study only the 
\textit{total} lepton asymmetry, ignoring the often significant effects on individual lepton 
flavors \cite{Barbieri:1999ma,*Abada:2006fw,*Nardi:2006fx,*Endoh:2003mz,*Pilaftsis:2004xx,*Dev:2014laa}. For our 
purposes, it will also be sufficient to utilize the $CP$-asymmetries calculated within zero-temperature field theory, although real-time thermal field theory 
provides a more complete formalism, see 
e.g. \cite{Kiessig:2011fw,*Garbrecht:2013iga,*Pilaftsis:2013xna,*Beneke:2010dz,*Kiessig:2011ga,*Frossard:2012pc,*Kiessig:2010pr}.

\subsubsection*{Lepton asymmetry}

Starting with the lepton asymmetry equation, we have
\beq
 \begin{split}
  n_\gamma^{eq}z_1H\frac{\ptl Y_{L-\overline{L}}}{\ptl z_1}=\sum_i\Bigg[&\frac{Y_i}{Y^{eq}_i}\left(\gamma^{eq}_{N_i\rightarrow 
LH}-\gamma^{eq}_{N_i\rightarrow \overline{LH}}\right)-\frac{Y_L}{Y^{eq}_L}\gamma^{eq}_{LH\rightarrow 
N_i}+\frac{Y_{\overline{L}}}{Y^{eq}_L}\gamma^{eq}_{\overline{LH}\rightarrow 
N_i}\\&
-\frac{Y_{L-\overline L}}{Y^{eq}_L}\left(\frac{Y_i}{Y^{eq}_i}\gamma^{eq}_{N_iL\rightarrow 
Qt}+\frac{Y_i}{Y^{eq}_i}\gamma^{eq}_{N_iL\rightarrow HS}+2\gamma^{eq}_{N_iQ\rightarrow Lt}+2\gamma^{eq}_{N_iH\rightarrow 
LS}\right)\Bigg]\\&
-\frac{Y_{L-\overline 
L}}{Y^{eq}_L}\left(\gamma^{eq}_{N_1S\xrightarrow[1]{}LH}+\gamma^{eq}_{N_2S\xrightarrow[1]{}LH}+\gamma^{eq}_{N_2S\xrightarrow[2]{}LH}\right)\\&
+\frac{Y_1}{Y^{eq}_1}\left(\gamma^{eq,sub}_{N_1S\xrightarrow[2]{}LH}-\gamma^{eq,sub}_{N_1S\xrightarrow[2]{}\overline{LH}}\right)-\frac{Y_L}{Y^{eq}_L}
\gamma^{eq,sub}_{LH\xrightarrow[2]{}N_1S}+\frac{Y_{\overline L}}{Y^{eq}_L}\gamma^{eq,sub}_{\overline{LH}\xrightarrow[2]{}N_1S} \\&
-2\frac{Y_L}{Y^{eq}_L}\sum_j\gamma^{eq,sub}_{LH\xrightarrow[j]{}\overline{LH}}+2\frac{Y_{\overline{L}}}{Y^{eq}_{\overline{L}}}\sum_j\gamma^{eq,sub}_{
\overline { LH } \xrightarrow [ j ] { } LH }\\&-2\frac{Y_{L-\overline L}}{Y^{eq}_L}\sum_j\gamma^{eq}_{LL\xrightarrow[j]{} 
HH}.
 \end{split} \label{DeltaL}
\eeq
In this expression, we use the following notation,
\beq
 z_i=\frac{M_i}{T},\qquad Y_i=\frac{n_i}{n^{eq}_\gamma},\qquad Y^{eq}_i=\frac{3}{8}z_i^2K_2(z_i),
\eeq
where $K_i(z)$ is a modified Bessel function of the second kind, along with the thermal cross sections, 
\beq\begin{split}
 &\gamma^{eq}_{i\rightarrow mn}(T)=n^{eq}_i\Gamma_{i\rightarrow 
mn}\left\langle\frac{M_i}{E}\right\rangle=n_i^{eq}\frac{K_1(z_i)}{K_2(z_i)}\Gamma_{i\rightarrow mn},\\
 &\gamma^{eq}_{ij\rightarrow 
mn}(T)=n^{eq}_in^{eq}_j\langle v\sigma_{ij\rightarrow 
mn\cdots}\rangle=g_ig_j\frac{T^4}{32\pi^4}\int_{w_{min}}^{\infty}dw\sqrt{w}K_{1}\left(\sqrt{w}
\right)\hat\sigma_{ ij\rightarrow
mn}\left(w\frac{m_i^2}{z_i^2}\right),
\end{split}
\eeq
where $w=s/T^2$, and the reduced cross section $\hat\sigma$ is given by
\beq
 \hat\sigma_{ij\rightarrow mn}(s)=\frac{1}{s}\delta\left(s,m_i^2,m_j^2\right)\sigma_{ij\rightarrow
mn}(s),\qquad\delta(a,b,c)=(a-b-c)^2-4bc.
\eeq

In Appendix \ref{sec: app Boltzmann Equations and Equilibrium}, we include further details about the Boltzmann equations and thermal cross sections. In writing 
the above equation, we have assumed $CP$-invariance, $\gamma^{eq}_{\overline{ij}\rightarrow \overline{mn}}=\gamma^{eq}_{ij\rightarrow mn}$ in the 
scattering 
processes, along with $CPT$-symmetry, $\gamma^{eq}_{\overline{ij}\rightarrow \overline{mn}}=\gamma^{eq}_{mn\rightarrow ij}$. $CP$-violating corrections appear 
in 
the scattering amplitudes at fourth order in the coupling constants, which is of higher order than we will consider here. That being said, it is necessary to 
make 
an exception when dealing with the subtracted rates as discussed below.

The superscript `$sub$' signifies that the process contains a real-intermediate-state (RIS) mediator that should be subtracted. Most famously, the process 
$LH\rightarrow \overline{LH}$ is mediated by a neutrino $N_j$ in the $s$-channel which can be on-shell for a sufficiently high center of mass 
energy. The real intermediate state represents the physical process $LH\rightarrow N_j\rightarrow \overline{LH}$. However, these processes are 
already accounted for by decays and inverse decays, and therefore need to be removed. Similarly, $N_1S\xrightarrow[2]{}LH$ has the real intermediate 
state $N_1S\rightarrow N_2\rightarrow LH$, which is also accounted for by decays and inverse decays. As it turns out, the $t$-channel process $N_iH\rightarrow 
LS$ also 
contains a real intermediate state, that needs to be removed. This point is explained in Appendix \ref{sec: app scattering cross sections}, where 
explicit formulas for the cross sections are displayed. For simplicity, all quarks and leptons, as well as the Higgs and scalar $S$ are considered 
massless. This can be justified because leptogenesis necessarily occurs at temperatures above sphaleron decoupling, $T>T_{sphaleron}\sim160\,\text{GeV}>m_H$ 
\cite{D'Onofrio:2014kta}, though typically we shall take $M_2>1~$TeV. The singlet $S$ mass is 
not yet stringently constrained, provided the Higgs portal coupling is not too large \cite{Aad:2014iia}, but we will typically 
take it to be of the same order as the Higgs mass. With these simplifications, the $t$-channel processes 
$N_iQ\leftrightarrow Lt$ and $N_it\rightarrow LQ$ have equal rates, and similarly for $N_iH\rightarrow LS$ and $N_iS\rightarrow LH$, which 
explains the factor of `2' sitting in front of these processes in Eq.~\eqref{DeltaL} above.\newline

\subsubsection*{Subtracted Rates and Real Intermediate States}

\begin{figure}[t]
\centerline{\includegraphics[width=18cm]{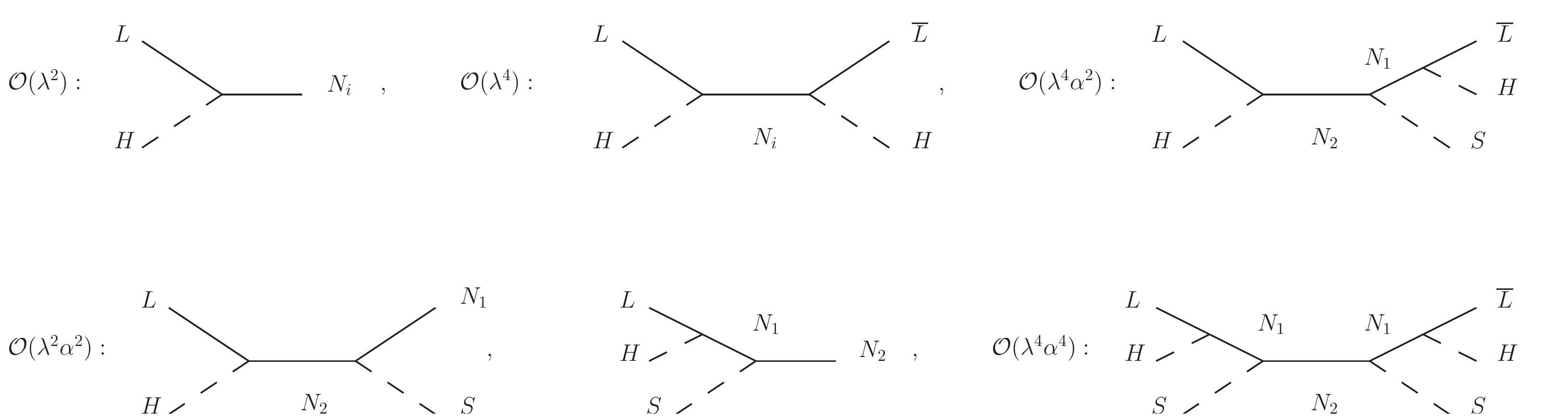}}
\caption{\footnotesize Classes of decay and scattering diagrams and sub-diagrams. We subtract the $N_i$ real intermediate states coupling through $\alpha$ and 
$\lambda$ order by order to avoid double counting, as discussed in the text.}
\label{fig: set subtracted rates}
\end{figure}

In this subsection, we summarize the procedure used to account for real intermediate states. The source terms in the Boltzmann equations are systematically 
expanded in each of the couplings and one needs to avoid double counting the RIS contributions that appear in (naively) higher order scattering processes.  
Doing this consistently in standard leptogenesis requires the inclusion of all processes up to and including two-to-three scattering and the associated 
$CP$-asymmetries 
\cite{Abada:2006ea, *Nardi:2007jp}. For HPL, we will do the same, extending the analysis to account for real intermediate states coupling via both the Yukawa 
$\lambda_{ij}$ and the singlet $\alpha_{ij}$ interactions. The relevant tree-level diagrams are displayed in Fig.~\ref{fig: set subtracted rates}, although it's 
important to account also for loop corrections that contribute to the $CP$-asymmetries in scattering. As already mentioned, we will focus on the impact of the 
additional singlet decay channel and ignore the issue of neutrino flavor mixing in scattering amplitudes, e.g. 
$\gamma^{eq}_{LH\rightarrow \overline{LH}}=\sum_i\gamma^{eq}_{LH\xrightarrow[i]{} \overline{LH}}$, which has been discussed in detail elsewhere. The RIS calculation 
of the $s$-channel cross sections is generally a nontrivial task once the flavor structure is taken into account. However, given this simplifying assumption, we can 
use the result $\gamma^{sub}=\gamma-\gamma^{on-shell}$ 
\cite{Kolb:1979qa,*Strumia:2006qk}.

To proceed to discuss the subtracted rates, we first make the following definitions associated with $N_2$ decays,
\beq\label{eq: N2 branching ratios}
\begin{split}
 &\Gamma_{2T}=\Gamma_2+\Gamma_{N_2\rightarrow N_1S},\\
 &\Gamma_2=\Gamma_{N_2\rightarrow LH}+\Gamma_{N_2\rightarrow\overline{LH}},\\
 &\epsilon_2\Gamma_2=\Gamma_{N_2\rightarrow LH}-\Gamma_{N_2\rightarrow\overline{LH}},
\end{split}
\quad\Longrightarrow\quad\begin{split}
		&\text{Br}(N_2\rightarrow LH)=\frac{1+\epsilon_2}{2}\frac{\Gamma_2}{\Gamma_{2T}},\\
		&\text{Br}(N_2\rightarrow \overline{LH})=\frac{1-\epsilon_2}{2}\frac{\Gamma_2}{\Gamma_{2T}},\\
		&\text{Br}(N_2\rightarrow N_1S)=\frac{\Gamma_{21}}{\Gamma_{2T}}=1-\frac{\Gamma_2}{\Gamma_{2T}},
	       \end{split}
\eeq
where $\Gamma_{21}$ is shorthand notation for $\Gamma_{N_2\rightarrow N_1S}$, while $N_1$ only decays to leptons, so that
\beq\label{eq: N1 branching ratios}
\begin{split}
 &\Gamma_{1T}=\Gamma_1,\\
 &\Gamma_1=\Gamma_{N_1\rightarrow LH}+\Gamma_{N_1\rightarrow\overline{LH}},\\
 &\epsilon_1\Gamma_1=\Gamma_{N_1\rightarrow LH}-\Gamma_{N_1\rightarrow\overline{LH}},
\end{split}
\quad\Longrightarrow\quad\begin{split}
		&\text{Br}(N_1\rightarrow LH)=\frac{1+\epsilon_1}{2},\\
		&\text{Br}(N_1\rightarrow \overline{LH})=\frac{1-\epsilon_1}{2}.
	       \end{split}
\eeq

\begin{figure}[t!]
\centering
\subfigure[]{
\includegraphics[width=7cm]{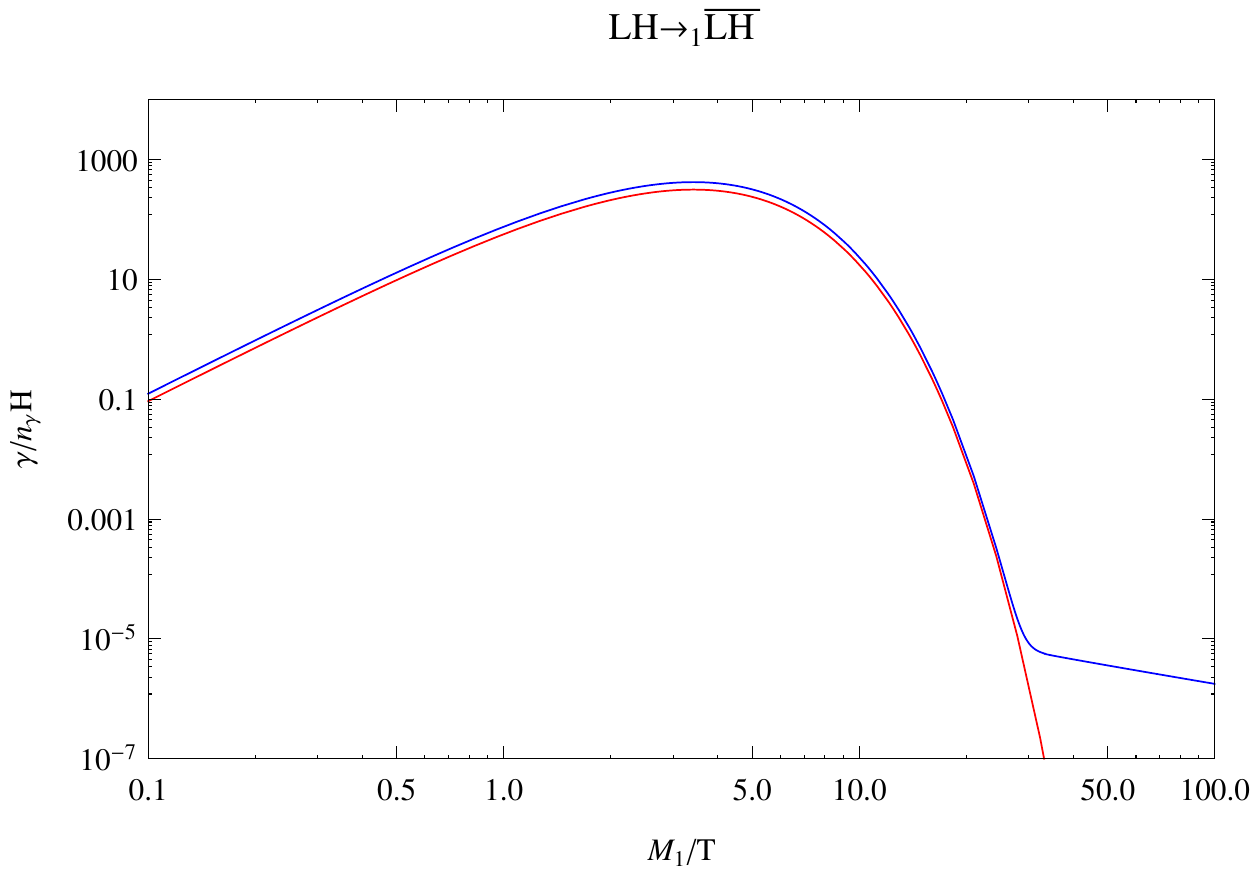}
\label{fig: LH to N1 to LbarHbar}}
\subfigure[]{\includegraphics[width=7cm]{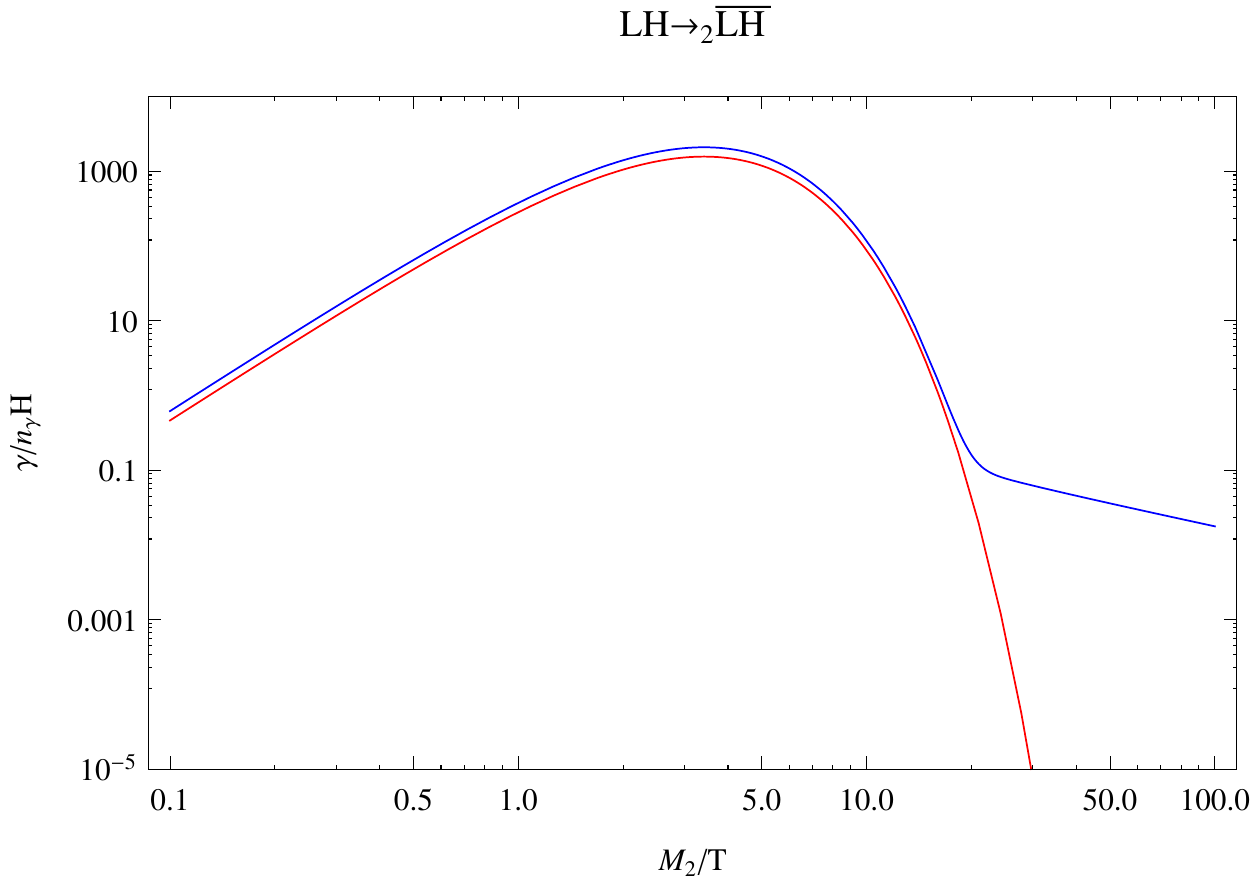}
\label{fig: LH to N2 to LbarHbar}}
\subfigure[]{
\includegraphics[width=7cm]{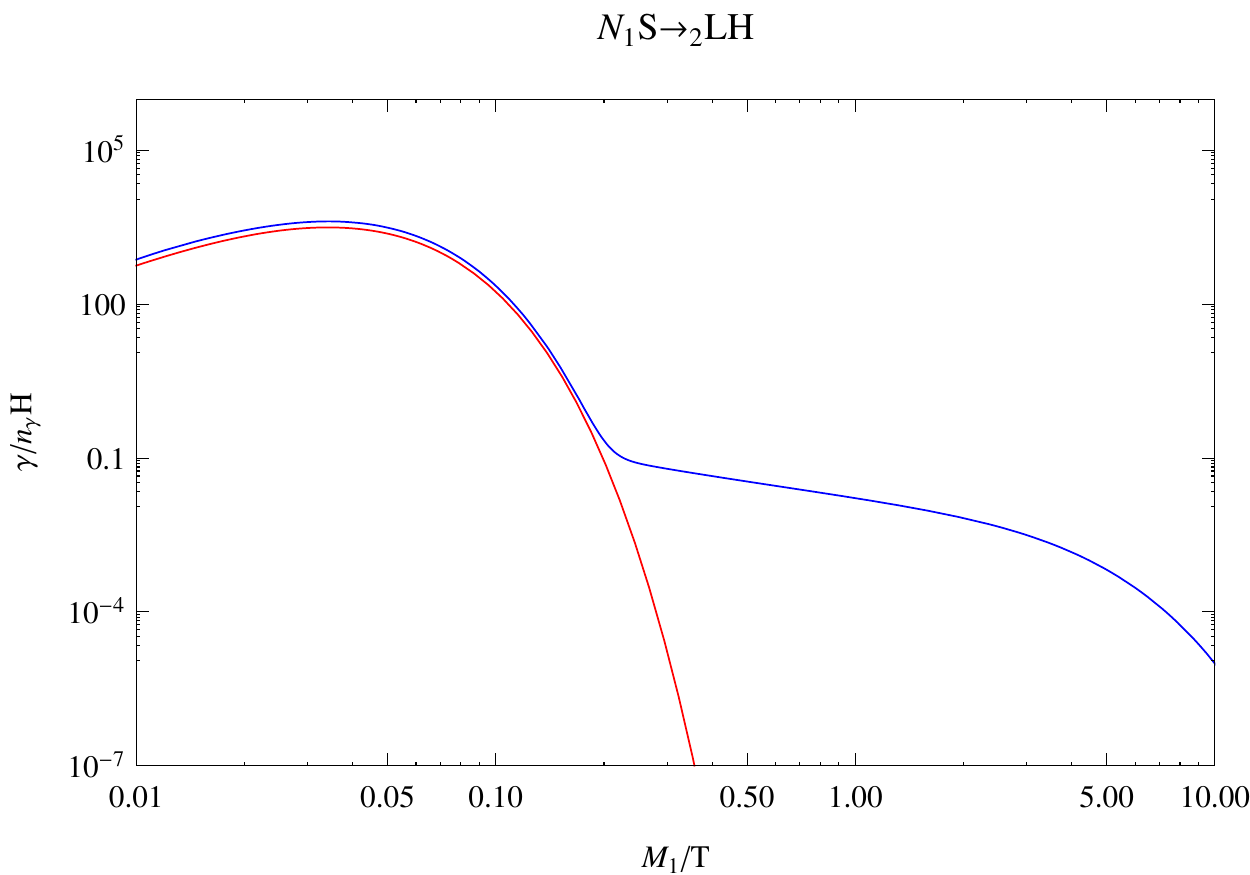}
\label{fig: N1S to N2 to LH}}\caption[]{\footnotesize Plots of the un-subtracted thermal rates (blue), along with 
their on-shell parts (red). The plots in (a), (b) and (c) are for $LH\xrightarrow[1]{}\overline{LH}$, $LH\xrightarrow[2]{}\overline{LH}$, and 
$N_1S\xrightarrow[2]{}LH$ respectively,  generated using $\lambda_1\sim0.002$, $\lambda_2\sim0.07$, 
$\alpha_{12}\sim0.07$, $M_2=10^{10}$GeV, $M_1=10^8$GeV.}
\label{fig: subtracted and unsubtracted rates}
\end{figure}

Building up the equations order by order, we have the following contributions:

\begin{itemize}
 \item $\mathcal{O}(\lambda^2)$: 1-to-2 decays,
 \beq\label{eq: lambda2 equations}
 \begin{split}
  n_\gamma^{eq}z_1H\frac{\ptl Y_{L-\overline{L}}}{\ptl 
z_1}\Bigg|_{\lambda^2}=&\sum_{i=1,2}\Bigg[\frac{Y_i}{Y^{eq}_i}\left(\gamma^{eq}_{N_i\rightarrow 
LH}-\gamma^{eq}_{N_i\rightarrow \overline{LH}}\right)-\frac{Y_L}{Y^{eq}_L}\gamma^{eq}_{LH\rightarrow 
N_i}+\frac{Y_{\overline{L}}}{Y^{eq}_L}\gamma^{eq}_{\overline{LH}\rightarrow 
N_i}\Bigg]
  \\=&\left(\frac{Y_1}{Y^{eq}_1}+1\right)\epsilon_1\gamma_{D_1}^{eq}+\left(\frac{Y_2}{Y^{eq}_2}+1\right)\epsilon_2\gamma_{D_2}^{eq}-\frac{Y_
{L-\overline L}}{2Y^{eq}_L}\left(\gamma^{eq}_{D_1}+\gamma^{eq}_{D_2}\right).
 \end{split}
 \eeq
This Boltzmann equation suffers from the well-known flaw that it does not respect Sakharov conditions. Indeed, even in equilibrium $Y_i=Y^{eq}_i$, a lepton 
asymmetry may be generated because of the non-vanishing source term. This is related to the fact that at $\mathcal{O}(\lambda^2)$, the set of interactions is 
incomplete because $\Delta L=2$ rates at (naive) $\mathcal{O}(\lambda^4)$ contain real intermediate state contributions that are in fact of order 
$\mathcal{O}(\lambda^2)$ and need to be included; see below.

  \item $\mathcal{O}(\lambda^4)$: $\Delta L=2$ scattering (2-to-2), 
 \beq\label{eq: lambda4 equations}
\begin{split}
 n_\gamma^{eq}z_1H\frac{\ptl Y_{L-\overline{L}}}{\ptl 
z_1}\Bigg|_{\lambda^4}=&\sum_{j=1,2}\Bigg[-2\frac{Y_L}{Y^{eq}_L}\gamma^{eq,sub}_{LH\xrightarrow[j]{}\overline{LH}}+2\frac{Y_{\overline{L}}}{Y^{eq}_{
\overline { L } } }\gamma^{eq,sub}_{
\overline { LH } \xrightarrow [ j ] { } LH 
}\Bigg]\\&=-2\epsilon_1\gamma_{D_1}^{eq}-2\epsilon_2\frac{\Gamma_2}{\Gamma_{2T}}\gamma_{D_2}^{eq}-\frac{Y_{L-\overline 
L}}{Y^{eq}_L}\left(2\gamma^{eq,sub,2}_{\Delta L=2}+2\gamma^{eq,sub,1}_{\Delta L=2}\right),
\end{split}
\eeq
where we used the definitions for the subtracted $\Delta L=2$ rates,
\beq\label{eq: delta l 2 on shell 2}
\begin{split}
 &\gamma^{eq,on-shell}_{LH\xrightarrow[2]{}\overline{LH}}=\gamma^{eq}_{LH\rightarrow N_2}\text{Br}(N_2\rightarrow 
\overline{LH})=\left(\frac{1-\epsilon_2}{2}\right)^2\frac{\Gamma_{2}}{\Gamma_{2T}}\gamma^{eq}_{D_2}\approx\frac{1-2\epsilon_2}{4}\frac{
\Gamma_{2}}{\Gamma_{2T}}\gamma^{eq}_{D_2},\\ 
&\gamma^{eq,sub}_{LH\xrightarrow[2]{}\overline{LH}}=\gamma^{eq,sub,(2)}_{\Delta L=2}-\frac{\epsilon_2}{2}\frac{\Gamma_{2}}{\Gamma_{2T}}\gamma^{eq}_{D_2},\\
 &\gamma^{eq,sub,(2)}_{\Delta L=2}\equiv\gamma^{eq}_{LH\xrightarrow[2]{}\overline{LH}}-\frac{1}{4}\frac{\Gamma_{2}}{\Gamma_{2T}}\gamma^{eq}_{D_2},\\ 
 &\Delta\gamma^{eq,sub,(2)}_{\Delta 
L=2}\equiv\gamma^{eq,sub}_{LH\xrightarrow[2]{}\overline{LH}}-\gamma^{eq,sub}_{\overline{LH}\xrightarrow[2]{}LH}=-\epsilon_2\frac{\Gamma_2}{\Gamma_{2T}}\gamma^{
eq } _ { D_2 },
\end{split}
\eeq
and
\beq\label{eq: delta l 2 on shell 1}
\begin{split}
 &\gamma^{eq,on-shell}_{LH\xrightarrow[1]{}\overline{LH}}=\gamma^{eq}_{LH\rightarrow N_1}\text{Br}(N_1\rightarrow 
\overline{LH})=\left(\frac{1-\epsilon_1}{2}\right)^2\gamma^{eq}_{D_1}\approx\frac{1-2\epsilon_1}{4}\gamma^{eq}_{D_2},\\ 
&\gamma^{eq,sub}_{LH\xrightarrow[1]{}\overline{LH}}=\gamma^{eq,sub,(1)}_{\Delta L=2}-\frac{\epsilon_1}{2}\gamma^{eq}_{D_1},\\
 &\gamma^{eq,sub,(1)}_{\Delta L=2}\equiv\gamma^{eq}_{LH\xrightarrow[1]{}\overline{LH}}-\frac{1}{4}\gamma^{eq}_{D_1},\\ 
 &\Delta\gamma^{eq,sub,(1)}_{\Delta 
L=2}\equiv\gamma^{eq,sub}_{LH\xrightarrow[1]{}\overline{LH}}-\gamma^{eq,sub}_{\overline{LH}\xrightarrow[1]{}LH}=-\epsilon_1\gamma^{eq}_{D_1}.
\end{split}
\eeq
The unsubtracted $\Delta L=2$ rates are $CP$-symmetric at $\mathcal{O}(\lambda^2)$, which implies that the subtracted rates are in 
fact $CP$-asymmetric. The functions $\gamma^{eq,sub,(i)}_{\Delta L=2}$ are the $CP$-conserving parts of the subtracted $\Delta L=2$ rates, and have been 
plotted in Fig.~\ref{fig: LH to N1 to LbarHbar} and \ref{fig: LH to N2 to LbarHbar}, showing that the $CP$-conserving subtracted rates are negligible.

Note that the RIS $CP$-asymmetry that comes from the $N_1$-mediated $\Delta L=2$ rate corrects the flaw above in the Boltzmann equation~\eqref{eq: lambda2 
equations}. 
The $\Delta L=2$ rates mediated by $N_2$ however do not fully correct the above Boltzmann equation. This is in contrast to the standard case, the reason being that 
$\text{Br}(N_2\rightarrow N_1S)>0$, so there is a second decay channel. Unlike in the standard case, for HPL, obtaining a consistent set of Boltzmann equations 
requires 
including the RIS contributions from higher order $\Delta L=2$ scattering processes; see Eq.~\eqref{eq: lambda4alpha2 equations} below.

 \item $\mathcal{O}(\lambda^4\alpha^2)$: $\Delta L=2$ scattering (2-to-3),
  \beq\label{eq: lambda4alpha2 equations}
  \begin{split}
   n_\gamma^{eq}z_1H\frac{\ptl Y_{L-\overline{L}}}{\ptl
z_1}\Bigg|_{\lambda^4\alpha^2}=&-4\frac{Y_L}{Y^{eq}_L}\gamma^{eq,sub}_{LH\xrightarrow[2]{}\overline{LH}S}+4\frac{Y_{\overline{L}}}{Y^{eq}_L}\gamma^{eq,sub}_{ 
  \overline{LH}\xrightarrow [ 2 ]{}LHS}\\
   =&-2\epsilon_2\frac{\Gamma_{21}}{\Gamma_{2T}}\gamma^{eq}_{D_2}-4\frac{Y_{L-\overline{L}}}{Y^{eq}_L}\gamma^{eq,(2-3)}_{\Delta L=2}\\
   &-2\Delta\gamma^{eq,sub}_{\Delta 
L=1}-2\epsilon_1\frac{\Gamma_{21}}{\Gamma_{2T}}\gamma^{eq}_{D_2},
  \end{split}
  \eeq
  where we have defined
  \beq
\begin{split}
 &\gamma^{eq,on-shell}_{LH\xrightarrow[2]{}\overline{LH}S}=\gamma^{eq,sub}_{LH\xrightarrow[2]{}N_1S}\frac{1-\epsilon_1}
 {2 } +\frac{(1-\epsilon_1)(1-\epsilon_2)}{4}\frac{\Gamma_{21}}{\Gamma_{2T}}\gamma^ { eq } _ { D_2},\\
 &\gamma^{eq,sub}_{LH\xrightarrow[2]{}\overline{LH}S}=\gamma^{eq}_{LH\xrightarrow[2]{}\overline{LH}S}-\gamma^{eq,(2-3)}_{\Delta 
L=2}+\frac{1}{4}\Delta\gamma^{eq,sub}_{\Delta L=1}+\frac{\epsilon_1+\epsilon_2}{4}\frac{\Gamma_{21}}{\Gamma_{2T}}\gamma^ { eq } _ { D_2},\\ 
 &\gamma^{eq,(2-3)}_{\Delta L=2}=\gamma^{eq}_{LH\xrightarrow[2]{}\overline{LH}S}-\frac{1}{4}\gamma^{eq}_{\Delta L=1},
\end{split}
\eeq
ignoring terms of order $\mathcal{O}(\lambda^6\alpha^2)$ and above. The functions $\gamma^{eq}_{\Delta L=1}$ and $\Delta\gamma^{eq,sub}_{\Delta L=1}$ 
are the $CP$-conserving and $CP$-violating parts of the $N_1S\xrightarrow[2]{}LH$ scattering terms, as is defined in equations \eqref{eq: delta l 1 on shell} 
below. 

  In Eq.~\eqref{eq: lambda4alpha2 equations}, the factors of 4 arise on accounting for both $LH\xrightarrow[2]{}\overline{LH}S$ and 
$LHS\xrightarrow[2]{}\overline{LH}$, which are $CPT$ conjugates of each 
other.  The first term on the first 
line is exactly what is needed to combine with the RIS term from Eq.~\eqref{eq: lambda4 equations} and correct the flaw in \eqref{eq: lambda2 
equations}. The second line of Eq.~\eqref{eq: lambda4alpha2 equations} in fact contains terms that correspond to RIS contributions  at  
$\mathcal{O}(\lambda^2\alpha^2)$.

 \item $\mathcal{O}(\lambda^2\alpha^2)$: $\Delta L=1$ scattering (2-to-2) and 1-to-3 decays,
 
 As noted above, at $\mathcal{O}(\lambda^4\alpha^2)$ there are additional uncompensated terms. 
 This is because the scattering processes at that order 
contain real intermediate states of $\mathcal{O}(\lambda^2\alpha^2)$, the same order as the $N_1S\xrightarrow[2]{}LH$ scattering process and 
the $N_2\xrightarrow[1]{}LHS$ three-body final state decay rate. In order to include the scattering properly, it is necessary to use the subtracted rate since the 
on-shell piece is equivalent to $N_1S\rightarrow N_2\rightarrow LH$ which has already been accounted for. We have 
 \beq\label{eq: order lambda2alpha2 equations}
\begin{split}
 n_\gamma^{eq}z_1H\frac{\ptl Y_{L-\overline{L}}}{\ptl 
z_1}\Bigg|_{\lambda^2\alpha^2}=&\frac{Y_1}{Y^{eq}_1}\left(\gamma^{eq,sub}_{N_1S\xrightarrow[2]{}LH}-\gamma^{eq,sub}_{N_1S\xrightarrow[2]{}\overline{LH}}
\right)-\frac{Y_L}{Y^{eq}_L}
\gamma^{eq,sub}_{LH\xrightarrow[2]{}N_1S}+\frac{Y_{\overline 
L}}{Y^{eq}_L}\gamma^{eq,sub}_{\overline{LH}\xrightarrow[2]{}N_1S}\\
 &+\frac{Y_2}{Y^{eq}_2}\left(\gamma^{eq}_{N_2\rightarrow 
LHS}-\gamma^{eq}_{N_2\rightarrow 
\overline{LH}S}\right)-\frac{Y_L}{Y^{eq}_L}\gamma^{eq}_{N_2\rightarrow\overline{LHS}}+\frac{Y_{\overline{L}}}{Y^{eq}_L}\gamma^{eq}_{N_2\rightarrow LHS} 
\\=&\left(\frac{Y_1}{Y^{eq}_1}+1\right)\Delta\gamma^{eq,sub}_{\Delta 
L=1}-\frac{Y_{L-\overline L}}{2Y^{eq}_L}\gamma^{eq,sub}_{\Delta L=1}\\
 &+\left(\frac{Y_2}{Y^{eq}_2}+1\right)\epsilon_1\gamma_{N_2\rightarrow N_1S}-\frac{Y_{L-\overline{L}}}{2Y^{eq}_L}\left(\gamma^{eq}_{N_2\rightarrow 
LHS}+\gamma^{eq}_{N_2\rightarrow \overline{LH}S}\right),
\end{split}
\eeq
where
\beq\label{eq: delta l 1 on shell}
\begin{split}
 &\gamma^{eq,on-shell}_{N_1S\xrightarrow[2]{}LH}=\gamma^{eq}_{N_1S\rightarrow N_2}\text{Br}(N_2\rightarrow 
LH)=\frac{1+\epsilon_2}{2}\frac{\Gamma_{21}}{\Gamma_{2T}}\gamma^{eq}_{D_2},\\ 
&\gamma^{eq,sub}_{N_1S\xrightarrow[2]{}LH}=\gamma^{eq}_{N_1S\xrightarrow[2]{}LH}-\frac{1}{2}\frac{\Gamma_{21}}{\Gamma_{2T}}\gamma^{eq}_{
D_2}-\frac{\epsilon_2}{2}\frac{\Gamma_{21}}{\Gamma_{2T}}\gamma^{eq}_{D_2},\\
 &\gamma^{eq,sub}_{\Delta L=1}\equiv\gamma^{eq,sub}_{N_1S\xrightarrow[2]{}LH}+\gamma^{eq,sub}_{N_1S\xrightarrow[2]{}\overline{LH}}=\gamma^{eq}_{\Delta 
L=1}-\frac{\Gamma_{21}}{\Gamma_{2T}}\gamma^{eq}_{D_2},\\
 &\Delta\gamma^{eq,sub}_{\Delta 
L=1}\equiv\gamma^{eq,sub}_{N_1S\xrightarrow[2]{}LH}-\gamma^{eq,sub}_{N_1S\xrightarrow[2]{}\overline{LH}}=\Delta\gamma^{eq}_{\Delta 
L=1}-\epsilon_2\frac{\Gamma_{21}}{\Gamma_{2T}}\gamma^{eq}_{D_2}.
\end{split}
\eeq
The function $\gamma^{eq}_{\Delta L=1}/2$ determines the 
$CP$-conserving part of the scattering rate, and has been plotted in Fig.~\ref{fig: N1S to N2 to LH}, along with the RIS rate 
$\gamma^{eq}_{D_2}\Gamma_{21}/(2\Gamma_{2T})$. This shows that $\gamma^{eq,sub}_{\Delta L=1}$ is negligible. The $CP$-asymmetry in the scattering, 
$\Delta\gamma^{eq}_{\Delta 
L=1}$, is largely inherited from the on-shell part in such way that $\Delta\gamma^{eq,sub}_{\Delta L=1}$ is again negligible. We can convince ourselves of this 
by taking the ratio $\Delta\gamma^{eq,sub}_{\Delta L=1}/\gamma^{eq}_{\Delta L=1}\approx\Delta\gamma^{eq}_{\Delta L=1}/\gamma^{eq}_{\Delta 
L=1}-\epsilon_2\approx0$, as the $CP$-asymmetry in $\Delta L=1$ scatterings is equal to that of the neutrino leptonic decays. The 
$CP$-asymmetry in three-body final state decay rates can only come from the kinematic point where the intermediate line is on-shell, which means
\beq
\begin{split}
 &\gamma^{eq,on-shell}_{N_2\rightarrow LHS}=\gamma^{eq}_{N_2\rightarrow N_1S}\frac{1+\epsilon_1}{2},\\
 &\Delta\gamma^{eq}_{N_2\rightarrow LHS}=\gamma^{eq,on-shell}_{N_2\rightarrow LHS}-\gamma^{eq,on-shell}_{N_2\rightarrow 
\overline{LH}S}=\epsilon_1\gamma_{N_2\rightarrow N_1S}.
\end{split}
\eeq
Eq.~\eqref{eq: order lambda2alpha2 equations} displays the same flaw as Eq.~\eqref{eq: lambda2 equations} in failing to follow Sakharov's criteria. 
Following the same logic as at $\mathcal{O}(\lambda^2)$, it is necessary to include (naively) higher order contributions, namely
$\Delta L=2$ scattering processes that contain RIS at $\mathcal{O}(\lambda^2\alpha^2)$. Indeed, the second to last line of 
equation \eqref{eq: lambda4alpha2 equations} fully corrects this problem at the scattering level. The last line of the same equation \textit{partially} corrects 
the 
corresponding flaw at the level of three-body final state decays. In order to correct this rate completely, we need to consider yet higher order interactions, 
$\mathcal{O}(\lambda^4\alpha^4)$; see below.

 \item $\mathcal{O}(\lambda^4\alpha^4)$: $\Delta L=2$ scattering (3-to-3),
 
 The 3-to-3 scattering process $LHS\xrightarrow[2]{}\overline{LH}S$ contains a real intermediate state, 
\beq
 \gamma^{eq,son-shell}_{LHS\xrightarrow[2]{}\overline{LH}S}\supset\text{Br}(LH\rightarrow N_1)\gamma^{eq}_{N_1S\rightarrow N_2}\text{Br}(N_2\rightarrow 
 N_1S)\text{Br}(N_1\rightarrow \overline{LH})\supset\frac{\epsilon_1}{2}\frac{\Gamma_{21}}{\Gamma_{2T}}\gamma^{eq}_{N_2\rightarrow N_1S},
\eeq
so that 
\beq\label{eq: lambda4alpha4 equations}
\begin{split}
 n^{eq}_\gamma z_1H\frac{\ptl Y_{L-\overline L}}{\ptl z_1}=&-2\frac{Y_{L}}{Y^{eq}_L}\gamma^{eq,sub}_{LHS\xrightarrow[2]{}\overline{LH}S}+2\frac{Y_{\overline 
L}}{Y^{eq}_L}\gamma^{eq,sub}_{\overline{LH}S\xrightarrow[2]{}LHS}\\
 &\supset-2\epsilon_1\frac{\Gamma_{21}}{\Gamma_{2T}}\gamma^{eq}_{N_2\rightarrow N_1S},
\end{split}
\eeq
which combines with the $-2\epsilon_1\gamma^{eq}_{D_2}\Gamma_{21}/\Gamma_{2T}$ term of equation \eqref{eq: lambda4alpha2 equations}, leading to the combination 
$-2\epsilon_1\gamma^{eq}_{N_2\rightarrow N_1S}$ that ultimately corrects the above flaw at the three-body final state decay level.

\end{itemize}

In conclusion, one obtains the correct Boltzmann equations at order $\mathcal{O}(\lambda^2)$, by combining the equations \eqref{eq: lambda2 equations}, \eqref{eq: 
lambda4 equations} and \eqref{eq: lambda4alpha2 equations} at order 
$\mathcal{O}(\lambda^2)$, $\mathcal{O}(\lambda^4)$, and $\mathcal{O}(\lambda^4\alpha^2)$ respectively. If one wishes to include the 
$\Delta L=1$ scattering and decays at order $\mathcal{O}(\lambda^2\alpha^2)$, it is necessary to combine the contributions of
$\mathcal{O}(\lambda^2\alpha^2)$, $\mathcal{O}(\lambda^4\alpha^2)$ and $\mathcal{O}(\lambda^4\alpha^4)$ in equations \eqref{eq: order lambda2alpha2 equations},
\eqref{eq: lambda4alpha2 equations} and \eqref{eq: lambda4alpha4 equations}.

The need to include all these varied contributions to obtain the correct Boltzmann equations should not come as a surprise. Since there are two $N_2$ 
decay channels, whenever the decay  $N_2\rightarrow LH$ is part of a scattering process, we can write down an additional scattering 
diagram which has the $N_2\rightarrow N_1S\rightarrow LH$ decay chain as a sub-diagram. Because both $N_2\rightarrow LH$ and $N_2\rightarrow N_1S\rightarrow LH$ 
can happen on-shell, they both contribute at the same order and therefore combine to provide a complete set of scattering contributions; complete in the sense 
that $\text{Br}(N_2\rightarrow LH)+\text{Br}(N_2\rightarrow N_1S)=1$. This explains the necessity to  
 include all terms of both $\mathcal{O}(\lambda^4)$ and $\mathcal{O}(\lambda^4\alpha^2)$ to obtain the correct Boltzmann equations at $\mathcal{O}(\lambda^2)$.

We can also understand this conclusion at the level of unitarity and $CPT$ invariance, which requires that $\sum_j|\MM(i\rightarrow j)|^2=\sum_j|\MM(j\rightarrow 
i)|^2=\sum_j|\MM(\overline{i}\rightarrow \overline{j})|^2$. At $\mathcal{O}(\lambda_1^4)$ 
one has,
\beq\label{eq: unitarity and cpt N1}
\begin{split}
 \left.|\MM(HL\xrightarrow[1]{} X)|^2\right|_{{\cal O}(\lambda_1^4)}=&|\MM(HL\rightarrow N_1)|^2+|\MM(HL\xrightarrow[1]{} LH)^{sub}|^2+|\MM(HL\xrightarrow[1]{} 
\overline{HL})^{sub}|^2\\
 =&|\MM(HL\rightarrow N_1)|^2+|\MM(HL\xrightarrow[1]{} LH)|^2+|\MM(HL\xrightarrow[1]{} \overline{HL})|^2\\
 &-|\MM(HL\rightarrow N_1)|^2\text{Br}(N_1\rightarrow LH)-|\MM(HL\rightarrow N_1)|^2\text{Br}(N_1\rightarrow \overline{LH})\\
 =&|\MM(HL\xrightarrow[1]{} LH)|^2+|\MM(HL\xrightarrow[1]{} \overline{HL})|^2.
\end{split}
\eeq
At this order both $|\MM(HL\xrightarrow[1]{} LH)|^2$ and $|\MM(HL\xrightarrow[1]{} \overline{HL})|^2$ are $CP$-symmetric in which case unitarity and $CPT$ 
invariance 
are straightforwardly satisfied. Now, at $\mathcal{O}(\lambda_2^4\alpha_{21}^2)$, it is necessary to include higher order scattering processes in order to 
obtain the same conclusion,
\beq\label{eq: unitarity and cpt N2}
 \begin{split}
\left. |\MM(HL\xrightarrow[2]{} X)|^2\right|_{{\cal O}(\lambda_2^4\alpha_{21}^2)}=&|\MM(HL\rightarrow N_2)|^2+|\MM(HL\xrightarrow[2]{} N_1S)^{sub}|^2\\
      &+|\MM(HL\xrightarrow[2]{} 
LH)^{sub}|^2+|\MM(HL\xrightarrow[2]{} \overline{HL})^{sub}|^2\\&+|\MM(HL\xrightarrow[2]{} LHS)^{sub}|^2+|\MM(HL\xrightarrow[2]{} \overline{HL}S)^{sub}|^2\\
 =&|\MM(HL\rightarrow N_2)|^2+|\MM(HL\xrightarrow[2]{} N_1S)^{sub}|^2+|\MM(HL\xrightarrow[2]{} 
LH)|^2+|\MM(HL\xrightarrow[2]{} \overline{HL})|^2\\&+|\MM(HL\xrightarrow[2]{} LHS)|^2+|\MM(HL\xrightarrow[2]{} \overline{HL}S)|^2\\
 &-|\MM(HL\rightarrow N_2)|^2\left(\text{Br}(N_2\rightarrow LH)+\text{Br}(N_2\rightarrow \overline{LH})\right)\\
 &-\left(|\MM(HL\xrightarrow[2]{} N_1S)^{sub}|^2+|\MM(HL\rightarrow N_2)|^2\text{Br}(N_2\rightarrow N_1S)\right)\\
 =&|\MM(HL\xrightarrow[2]{} LH)|^2+|\MM(HL\xrightarrow[2]{} \overline{HL})|^2+|\MM(HL\xrightarrow[2]{} 
LHS)|^2+|\MM(HL\xrightarrow[2]{} \overline{HL}S)|^2,
 \end{split}
\eeq
where we have used the relation
\beq
 \begin{split}
  |\MM(HL\xrightarrow[2]{} LHS)^{sub}|^2=|\MM(HL\xrightarrow[2]{} N_1S)|^2\text{Br}(N_1\rightarrow LH),
 \end{split}
\eeq
and further split the rate $|\MM(HL\xrightarrow[2]{} N_1S)|^2$ into the subtracted and RIS parts.  Recall that $\text{Br}(N_1\rightarrow 
LH)+\text{Br}(N_1\rightarrow \overline{LH})=1$. Thus the unitarity+$CPT$ constraint is again consistently satisfied at this order.

We are now in position to summarize the final Boltzmann equations. In practice, we can ignore the 1-to-3 decay rates, which are numerically subdominant compared to the 1-to-2 
decays, and similarly we can ignore the 2-to-3 scatterings, which are subdominant compared to the 2-to-2 rates. The subtracted rates 
$\gamma^{eq,sub}_{N_1S\rightarrow LH}$ and 
$\gamma^{eq,sub}_{\overline{LH}\xrightarrow[i]{} 
LH}$ can also be ignored as suggested by the plots in Fig.~\ref{fig: subtracted and unsubtracted rates}.

To simplify the above discussion, we have considered the subset of 
interactions that contain the $\alpha$ and $\lambda$ coupling constants. In this paper, we also consider the set of 
interactions involving the coupling $\beta$. However, among the set of scatterings one considers, there is no additional real intermediate state 
from this source, and we can directly re-write 
the Boltzmann 
equation for the final lepton asymmetry as,
\beq\label{eq: HPL Boltzmann equations for asymmetry}
\begin{split}
 z_1\frac{\ptl Y_{L-\overline L}}{\ptl
z_1}=&\epsilon_1D_1\left(\frac{Y_1}{Y^{eq}_1}-1\right)+\epsilon_2D_2\left(\frac{Y_2}{Y^{eq}_2}-1\right)\\&-Y_ { L-\overline 
L}\left(W_{ID_1}+W_{S_1}+W_{ID_2}+W_{S_2}\right),
\end{split}
\eeq
where we have used the notation of the decay, scattering and washout functions $D$, $W$ and $S$, defined in 
Appendix \ref{sec: app Boltzmann Equations and Equilibrium}, 
\beq
 D_{i}=\frac{\gamma^{eq}_{D_i}}{n_\gamma^{eq}H}=K_{i}z_i^2Y^{eq}_i\frac{K_1(z_i)}{K_2(z_i)},\qquad W_{ID_i}=\frac{1}{2Y^{eq}_L}D_i,\qquad 
Y^{eq}_i=\frac{3}{8}z_i^2K_2(z_i).
\eeq
The equilibrium parameter for leptonic decays $K_i$ is defined as,
\beq\label{Ki}
 K_i=\frac{\Gamma_i}{H(T=M_i)}=\frac{\tilde m_i}{m_*},
 \eeq
where the effective light neutrino mass scales $\tilde m_i$ and $m_*$ are
 \beq\label{m_eff}
 \tilde m_i=\frac{(\lambda^\dagger\lambda)_{ii}v^2}{M_i},\qquad 
m_*=8\pi v^2\sqrt{\frac{8\pi^3g_*}{90M_p^2}}\sim1.05\cdot10^{-3}\text{eV},
\eeq
emerging from the see-saw mechanism, with $g_*\sim 100$ the total number of degrees of freedom. We assume here the normal hierarchy among light neutrino 
masses. The 
washout functions are written in terms of the scattering function, $S_{ia\rightarrow mn}=\gamma^{eq}_{ia\rightarrow mn}/(n_\gamma^{eq}H)$,
\beq
 \begin{split}
  &W_{S_1}=\frac{1}{Y^{eq}_L}\left(2S_{N_1t\rightarrow LQ}+S_{N_1H\rightarrow 
LS}+S_{N_1S\xrightarrow[1]{}LH}\right)+\frac{Y_1}{Y^{eq}_LY^{eq}_1}\left(S_{N_1L\rightarrow Qt}+S_{N_1L\rightarrow 
HS}\right),\\
  &W_{S_2}=\frac{1}{Y^{eq}_L}\left(2S_{N_2t\rightarrow LQ}+S_{N_2H\rightarrow 
LS}+S_{N_2S\xrightarrow[1]{}LH}+S_{N_2S\xrightarrow[2]{}LH}\right)+\frac{Y_2}{Y^{eq}_LY^{eq}_2}\left(S_{N_2L\rightarrow 
Qt}+S_{N_2L\rightarrow 
HS}\right).
 \end{split}
\eeq
The functions $D,S,W$ have been defined to facilitate writing the 
Boltzmann equations in a manner that is independent of the choice of reference mass scale in the definition of the time variable. The results of integrating 
the 
Boltzmann equations can be qualitatively understood by considering the transition points where various rates go in and out of equilibrium. As discussed in 
Appendix B, this is conveniently tracked with the \textit{thermal equilibrium parameters} $\mathcal{K}$. For decays $N_i\rightarrow LH+\overline{LH}$, one has 
$\mathcal{K}_i=\langle\Gamma_i\rangle/H=D_i/Y^{eq}_i$, for inverse decays 
$LH+\overline{LH}\rightarrow N_i$ one has $\mathcal{K}_{ID}=\langle\Gamma_{ID}\rangle/H=D_i/2Y^{eq}_L$, whereas for scattering one 
has $\mathcal{K}_{ia\rightarrow mn}=n^{eq}_i\langle v\sigma_{ia\rightarrow mn}\rangle/H=S_{ia\rightarrow mn}/Y^{eq}_i$.

\subsubsection*{Neutrino abundance}

The Boltzmann equations for the RHN abundances can be determined in a similar manner to the lepton asymmetry discussed above,
\beq\label{eq: HPL Boltzmann equations for abundance}
\begin{split}
 z_1\frac{\ptl Y_1}{\ptl 
z_1}=&-\left(\frac{Y_1}{Y^{eq}_1}-1\right)\left(D_1+D_{21}+S_1
\right)+\left(\frac{Y_2}{Y^{eq}_2}-1\right)D_{21}\\&-\left(\frac{Y_1Y_2}{Y^{eq}_1Y^{eq}_2}-1\right)S_{N_1N_2\rightarrow 
HH}-\left(\frac{Y_1^2}{Y^{eq 
2}_1}-1\right)S_{N_1N_1\rightarrow HH},\\
 z_1\frac{\ptl Y_2}{\ptl 
z_1}=&-\left(\frac{Y_2}{Y^{eq}_2}-1\right)\left(D_2+D_{21}+S_2
\right)+\left(\frac{Y_1}{Y^{eq}_1}-1\right)D_{21}\\&-\left(\frac{Y_1Y_2}{Y^{eq}_1Y^{eq}_2}-1\right)S_{N_1N_2\rightarrow 
HH}-\left(\frac{Y_2^2}{Y^{eq 
2}_2}-1\right)S_{N_2N_2\rightarrow HH},
\end{split}
\eeq
with
\beq
\begin{split}
 &S_1=2S_{N_1L\rightarrow Qt}+4S_{N_1Q\rightarrow Lt}+2S_{N_1L\rightarrow 
HS}+4S_{N_1H\rightarrow LS}+2S_{N_1S\xrightarrow[1]{} LH},\\
 &S_2=2S_{N_2L\rightarrow Qt}+4S_{N_2Q\rightarrow Lt}+2S_{N_2L\rightarrow 
HS}+4S_{N_2H\rightarrow LS}+2S_{N_2S\xrightarrow[1]{} LH}+2S_{N_2S\xrightarrow[2]{} LH}.
\end{split}
\eeq
The subtracted rate for $N_1S\rightarrow LH$ is very small, and has been ignored. The decay function for $N_2\rightarrow N_1S$ is defined as
\beq
 D_{21}=K_{21}z_2^2\frac{K_1(z_2)}{K_2(z_2)}Y^{eq}_2,\qquad z_2=\frac{M_2}{T}=\frac{M_2}{M_1}z_1,
\eeq
with $K_{21}$ defined by analogy to $K_1$,
\beq
 K_{21}=\frac{\Gamma_{21}}{H(T=M_2)}=|\alpha_{21}|^2\frac{v^2}{2m_*M_2}\sim|\alpha_{21}|^2\left(1.44\cdot\frac{10^{16}\text{GeV}}{M_2}\right).
\eeq
We have implicitly assumed the hierarchy $\{M_1,m_S\}\ll M_2$ in writing down $K_{21}$ above; the exact decay rate is calculated in 
Appendix \ref{sec: app scattering cross sections}. Once again, the thermal equilibrium parameter for this process is $\mathcal{K}_{N_2\rightarrow 
N_1S}=D_{21}/Y^{eq}_2$, while for inverse decays 
$N_1S\rightarrow N_2$ one has $\mathcal{K}_{N_1S\rightarrow N_2}=D_{21}/Y^{eq}_1$.  

\begin{figure}[t]
\centerline{\includegraphics[width=18cm]{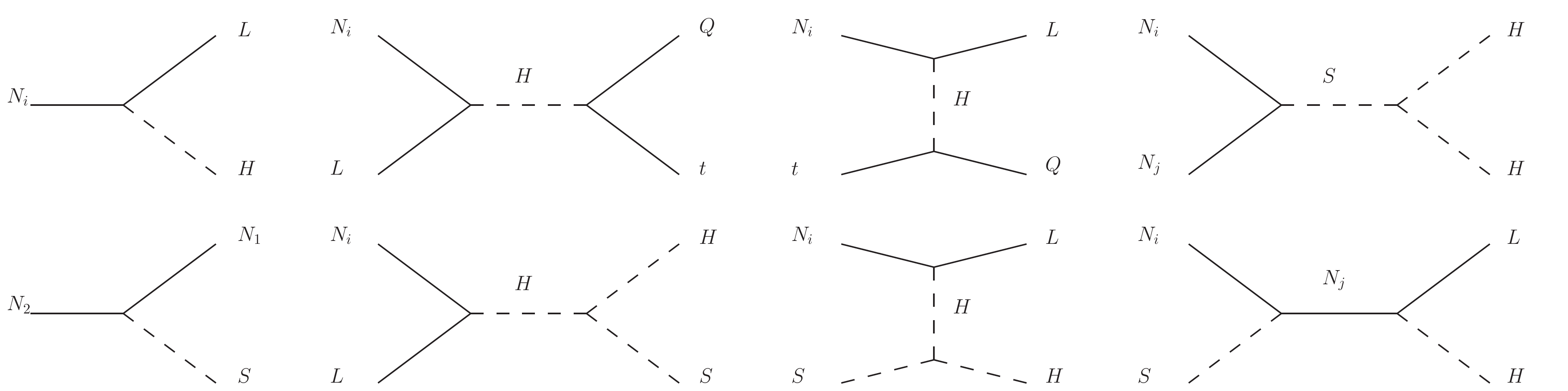}}
\caption{\footnotesize A list of the decay and scattering processes included in our Boltzmann equations. This list is not exhaustive, though it includes the dominant contributions. 
In the limit where the quarks, Higgs and singlet $S$ are massless, the $t$-channel diagrams are counted twice since it 
is possible to swap $Q-t$ and $H-S$.}
\label{fig: scattering cross sections summary}
\end{figure}

As a summary, Fig.~\ref{fig: scattering cross sections summary} lists
the scattering processes that are relevant for the equations.

\subsubsection*{Physical Regimes}

The underlying dynamics of this system is in the end quite similar to the simpler system 
which only accounts for decays and inverse decays (plus RIS contributions), which will be discussed in Section~\ref{sec: app Toy model of the 2-level Boltzmann 
equations}. 
At first sight it is surprising that the 2-to-2 scattering processes involving $N_1$, which can remain in equilibrium after $N_2$ decays, do not have a more 
significant role in washing out the asymmetry. Indeed, this intuition is realized if the couplings are sufficiently large as will be seen in the next section. 
However, there are natural parameter regimes in which the rates involving $N_1$ which change lepton number can be out of equilibrium while the rates that change 
the number density remain active. This feature is crucial for realizing viable HPL and will be discussed in more detail subsequently. For now, we briefly 
summarize the generation of a lepton asymmetry by breaking the evolution into three distinct phases:

\begin{enumerate}
 \item \textbf{The $N_2$ phase}:
 
 This phase takes place at temperatures $T\sim M_2$, and is marked by $N_2$ interactions going out of equilibrium, efficiently generating a 
primary lepton asymmetry. If there is a large mass hierarchy between $N_1$ and $N_2$, the rates involving $N_1$ may be sufficiently small (e.g. the decays which 
are proportional to the mass) that they are out of equilibrium, i.e. $D_1 \ll D_2, D_{21}$. In this case, the physics of this phase is almost that of a 
one-flavor system: the 
lepton asymmetry is generated through $N_2$ decays to leptons until the $N_2$ abundance becomes negligible. A second subdominant process can still be important, 
namely the mixing $N_2-N_1$ that allows decays and inverse decays $N_2\leftrightarrow N_1S$ if $\al_{12}$ is sufficiently large. Because of this channel, the 
branching ratio of $N_2$ into leptons is reduced as compared to 
the one-flavor system, making the production of the lepton asymmetry less efficient. At the same time, this very channel populates $N_1$, 
which is then kicked out-of-equilibrium momentarily. Its ability to return to equilibrium depends on the rate of the inverse decay 
$N_1S\rightarrow N_2$, as the decays $N_1\rightarrow LH$ are generally out-of-equilibrium in this phase. The overpopulation of $N_1$ is not very 
important during this phase, but will have an effect on the lepton asymmetry at the later stage when $N_1$ decays come into equilibrium. 
The resulting primary lepton asymmetry can be parametrized by an efficiency factor $\kappa_2$,
\beq
 Y^{(2)}_{L-\overline L}=\epsilon_2\kappa_2Y^{eq}_2(0).
\eeq
This phase typically ends when $T\sim M_2/10$, i.e. $z_1 \sim10M_1/M_2$.

 \item \textbf{The intermediate phase}:
 
 Given a sizeable mass hierarchy between $N_1$ and $N_2$, the second phase is marked by a large temperature gap once $N_2$ has effectively disappeared, and 
before $N_1$ interactions come into equilibrium.  Neither $N_2$- nor $N_1$-interactions are 
able to affect the lepton asymmetry, or the $N_1$ abundance, and the system effectively free streams leading to a plateau in $Y_{L-\overline{L}}$. This phase 
lasts for as long as the $N_1$ interactions remain out-of-equilibrium, and characteristically for a temperature range similar to the mass ratio. For example, if the 
decays 
and inverse decays dominate, the approximations discussed in Appendix \ref{sec: app Boltzmann Equations and Equilibrium} 
indicate that the phase ends 
when $z_1\sim\sqrt{2/K_1}$. If scattering effects are also significant, then the transition to the $N_1$ phase can occur somewhat earlier.

 \item \textbf{The $N_1$ phase}:
 
  This phase is marked by $N_1$ interactions being in-equilibrium
which efficiently deplete the 
neutrino abundance and lepton 
asymmetry. Assuming again a sizeable mass hierarchy, since the $N_2$ abundance is negligible and $N_2$ interactions are 
effectively 
turned off, $D_2,D_{21}\ll D_1$, the dynamics again approximates a purely one-flavor system. The distinction is that the 
initial lepton asymmetry is not zero, having been generated in the $N_2$ phase, and there is a possible $N_1$ overabundance $Y_1>Y^{eq}_1$ due to  
$N_2\rightarrow N_1S$ decays during the first phase. The final lepton asymmetry results from the competition between  $N_1$ mediated processes that wash out
the pre-existing lepton asymmetry from the $N_2$ phase, and those at the end of the $N_1$ phase that contribute to the asymmetry. The result can again be 
parametrized via an efficiency factor $\kappa_1$,
\beq
 Y_{L-\overline L}=Y^{(2)}_{L-\overline L}e^{-\int_{\mathcal{Z}}^{\infty}dz'(W_1/z')}+\epsilon_1\kappa_1Y_1^{eq}(0).
\eeq
where the washout function $W_1=W_{ID_1} + W_{S_1}$ is discussed above.
The variable $\mathcal{Z}$ marks the transition point after the $N_2$ phase, once the washout processes become active.
\end{enumerate}
Summarizing the full 2-level process, the final asymmetry resulting from the three phases can be parametrized by the two efficiency factors 
$\kappa_1$ and $\kappa_2$,
\beq\label{eq: general parametrization of the final asymmetry}
 Y^f_{L-\overline L}=\epsilon_2\left(\kappa_2e^{-\int_{\mathcal{Z}}^\infty dz'(W_1/z')}+\frac{\epsilon_1}{\epsilon_2}\kappa_1\right)Y^{eq}_2(0).
\eeq

We proceed in the next section to consider explicit examples which exhibit these features in detail. However, before considering the general case, we will first 
study a simplified toy model that allows some analytic understanding of the physics.

\subsection{Toy model of the 2-stage evolution}\label{sec: app Toy model of the 2-level Boltzmann equations}

In this subsection, in order to isolate some of the dominant physical effects, we study a toy model of the 2-level Boltzmann equations, accounting only for 
decays and inverse decays and ignoring the impact of 2-to-2 scattering.
For simplicity, we also take the $CP$-asymmetry to be constant, using $\epsilon_1=\epsilon_2=10^{-8}$, although this constraint will be relaxed towards 
the end of the section. The Boltzmann equations are as written in Eq.~\eqref{eq: HPL Boltzmann equations for 
asymmetry} and \eqref{eq: HPL Boltzmann equations for abundance}, of section \ref{subsec: Boltzmann equations scattering}, but without
scattering,
\beq\label{eq: 2stage Boltzmann equation without scattering}
\begin{split}
 z_1\frac{\ptl Y_1}{\ptl 
z_1}=&-\left(\frac{Y_1}{Y^{eq}_1}-1\right)\left(D_1+D_{21}\right)+\left(\frac{Y_2}{Y^{eq}_2}-1\right)D_{21},\\
 z_1\frac{\ptl Y_2}{\ptl 
z_1}=&-\left(\frac{Y_2}{Y^{eq}_2}-1\right)\left(D_2+D_{21}\right)+\left(\frac{Y_1}{Y^{eq}_1}-1\right)D_{21},\\
 z_1\frac{\ptl Y_{L-\overline L}}{\ptl
z_1}=&\epsilon_1D_1\left(\frac{Y_1}{Y^{eq}_1}-1\right)+\epsilon_2D_2\left(\frac{Y_2}{Y^{eq}_2}-1\right)-Y_ { L-\overline 
L}\frac{D_1+D_2}{2Y^{eq}_L}.
\end{split}
\eeq
The decay functions are as defined above, and we take the  initial conditions as a 
vanishing lepton asymmetry $Y_{L-\overline L}=0$, and equilibrium initial abundances $Y_{1,2}=Y^{eq}_{1,2}$. 
The $N_2$-phase ends at around $\mathcal{Z}\sim10M_1/M_2<1$, and we can numerically integrate $\int_{\mathcal{Z}}^\infty 
dz'W_{ID_1}/z'\sim1.2K_1$. Applying the general result of \eqref{eq: general parametrization of the final 
asymmetry} leads to an approximate final asymmetry,
\beq\label{eq: final asymmetry for 2 stage leptogenesis no scattering}
 Y^f_{L-\overline 
  L}\simeq \frac{3}{4}\left(\epsilon_2\kappa_2e^{-1.2K_1}+\epsilon_1\kappa_1\right).
\eeq

\subsubsection*{$N_2$ efficiency factor: $\kappa_2$}

\begin{figure}[t!]
\centerline{\includegraphics[width=14cm]{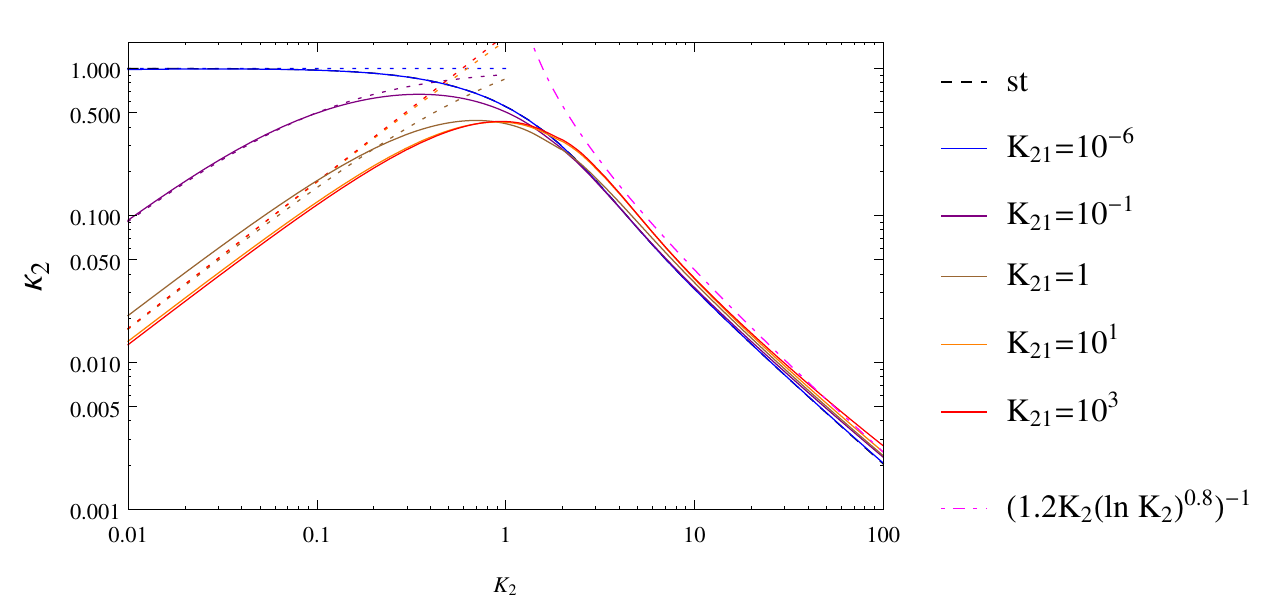}}
\caption{\footnotesize Plots of the efficiency factor $\kappa_2$, as a function of $K_2$, for various values of $K_{21}$. In the limit $K_{21}\ll 
K_2$, the efficiency factor approaches its value in  one-flavor standard leptogenesis  (shown here in black, dashed, buried under the blue 
line). Increasing $K_{21}$ has a significant effect on the efficiency factor at low $K_2$. The asymptotic behavior is also shown (dotted lines) 
for low $K_2$, from \eqref{eq: efficiency kappa2 approximation low K21} and \eqref{eq: efficiency kappa2 approximation high K21}. }
\label{fig: 2 stage leptogenesis scattering less efficiency factors}
\end{figure}

In the $N_2$ phase, the scenario of interest here is characterized by having $N_1$ out-of-equilibrium with $D_1\ll 
D_2,D_{21}$. Integrating the $Y_{L-\overline L}$ equation in \eqref{eq: 2stage Boltzmann equation without 
scattering} then leads to the efficiency factor $\kappa_2$,
\beq\label{eq: 2stage scatteringless efficiency factor first phase}
\begin{split} 
\kappa_2Y^{eq}_2(0)=&\int_0^{\mathcal{Z}}dz'\frac{D_2}{z'}\left(\frac{Y_2}{Y^{eq}_2}-1\right)e^{-\int_{z'}^{\mathcal{Z}}dz''(W_{ID_2}/z'')},\\=&-\frac{K_2}
{K_2+K_{21}}\int_0^{\mathcal{Z}}dz'\left(\frac{\ptl Y_2}{\ptl 
z'}-\frac{D_{21}}{z'}\left(\frac{Y_1}{Y^{eq}_1}-1\right)\right)e^{-\int_{z'}^{\mathcal{Z}}dz''(W_{ID_2}/z'')}.
\end{split}
\eeq
For concision, we have used the variable $z$, although strictly this is $z_1$. Evaluating $\kappa_2$ numerically, we obtain the contours shown in 
Fig.~\ref{fig: 2 stage leptogenesis scattering less 
efficiency factors}, as a function of $K_2$ for various values of $K_{21}$. To obtain an analytic approximation we 
focus on the regime $K_2\ll1$, where $K_{21}$ has the largest effect. In this limit, the washout from inverse decays is quite 
limited, $\exp(-\int_{z'}^{z}dz''W_{ID_2}/z'')\sim1$, so that the first part of the efficiency factor is trivial to integrate 
$\int_0^zdz'\ptl Y_2/\ptl z'=Y_2(z)-Y_2(0)\rightarrow -Y_2(0)$, since by definition we integrate to the point where the $N_2$ 
abundance drops to zero. Thus
\beq 
\kappa_2Y^{eq}_2(0)\simeq\frac{K_2}{K_2+K_{21}}Y^{eq}_2(0)+\frac{K_2}{K_2+K_{21}}\int_0^{\mathcal{Z}}dz'\frac{D_{21}}{z'}\left(\frac{Y_1}{Y^{eq}_1}-1\right),
\eeq
owing to the initial condition $Y_2(0)=Y^{eq}(0)$. Taking the limit $K_{21}\ll K_2$, the equations decouple in such a way that 
$Y_1\sim Y^{eq}_1$, and the second term above is subdominant leading to the efficiency factor,
\beq\label{eq: efficiency kappa2 approximation low K21}
 \kappa_2\sim \frac{K_2}{K_2+K_{21}}\quad {\rm when} \quad K_{21}\ll K_2\ll1.
\eeq
In the decoupled limit, the efficiency factor logically tends to the one-flavor value. If instead we take the limit $K_{21}\gg K_2$, 
the second term becomes significant, if not dominant, and we have to integrate the equations explicitly. In this limit, the branching ratio of 
$N_2$ into leptons is small, so that $D_{21}\gg D_2$. At the same time, $D_{21}\gg 
D_1$ in the $N_2$ phase. As a result, the $Y_{1,2}$ equations simplify to $\ptl Y_2/\ptl z_1=-\ptl Y_1/\ptl z_1$, so that the total number 
density $Y_1+Y_2$ is constant. This makes sense, since $N_2$'s decay dominantly into $N_1$'s. We thus find,
\beq
 Y_2(\mathcal{Z})+Y_1(\mathcal{Z})=Y_2(0)+Y_1(0)=2Y^{eq}_1(0).
\eeq
The first phase ends when $Y_2(\mathcal{Z})=0$, so that $Y_1(\mathcal{Z})=2Y^{eq}_1(0)$, and the 
term $Y_1(\mathcal{Z})/Y^{eq}_1(\mathcal{Z})-1\sim1$. 
The calculation of $\int_0^zdz'D_{21}/z'$ is most easily performed numerically, and for $z\geq10$ the integral converges to 
$1.7$. We conservatively take this result to obtain
\beq\label{eq: efficiency kappa2 approximation high K21}
 \kappa_2\sim1.7\frac{K_2 K_{21}}{K_2+K_{21}}\quad {\rm when} \quad K_{21}\gg K_2\ll1.
\eeq
Fig.~\ref{fig: 2 stage leptogenesis scattering less efficiency factors} exhibits both the numerical results for $\kappa_2$ along with the approximations 
\eqref{eq: efficiency kappa2 approximation low K21} and \eqref{eq: efficiency kappa2 approximation high K21}; the agreement is good in the region $K_2\geq 
K_{21}$, but less so for  $K_{21}\geq K_2$. The $K_2\gg1$ regime is similar to a one-flavor case, and the coupling $N_2-N_1$ 
does not have a large effect. Thus we can refer to the established literature \cite{Buchmuller:2004nz} for an approximate expression for $\ka_2$ in this region,
\beq
 \kappa_2\simeq\frac{1}{1.2K_2(\log K_2)^{0.8}} \quad {\rm when} \quad K_2\gg1.
\eeq

\begin{figure}[t!]
\centering
\subfigure[]{
\includegraphics[width=8.5cm]{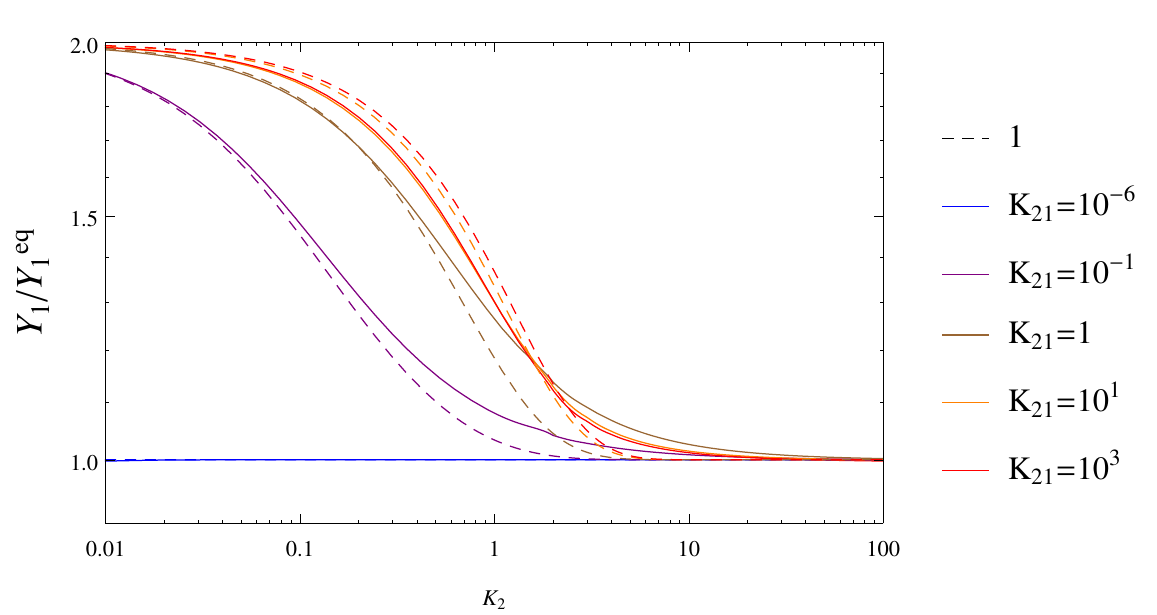}
\label{fig: N1Overabundance}}
\subfigure[]{\includegraphics[width=9.5cm]{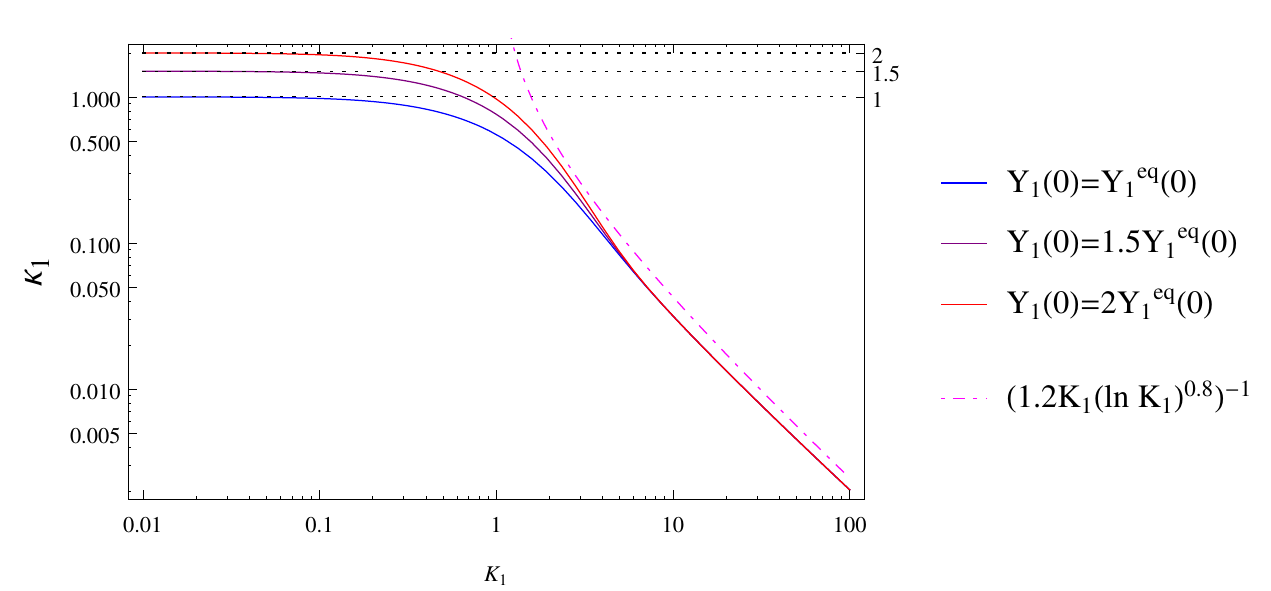}
\label{fig: N1EfficiencyFactor_OverabundanceEffect}
}\caption[]{\footnotesize The plot on the left (a) shows the $N_1$ abundance $Y_1(\mathcal{Z})$ offset compared to equilibrium at the end of the first phase 
due 
to the 
$N_2\rightarrow N_1S$ 
decay, referring to equation \eqref{eq: exact formula for abundance offset} along with the approximations \eqref{eq: efficiency kappa1 low K1 approximation}. 
These values are also the initial conditions for the abundance at the start of the very last phase. The plot on the right (b) represents the efficiency 
factor $\kappa_1$ for various initial values of $Y_1(\mathcal{Z})$ at the start of the $N_1$ phase, as given in the left plot.}
\label{fig: N1 Efficiency factor kappa1}
\end{figure}

\begin{figure}[t!]
\centering
\subfigure[]{
\includegraphics[width=16cm]{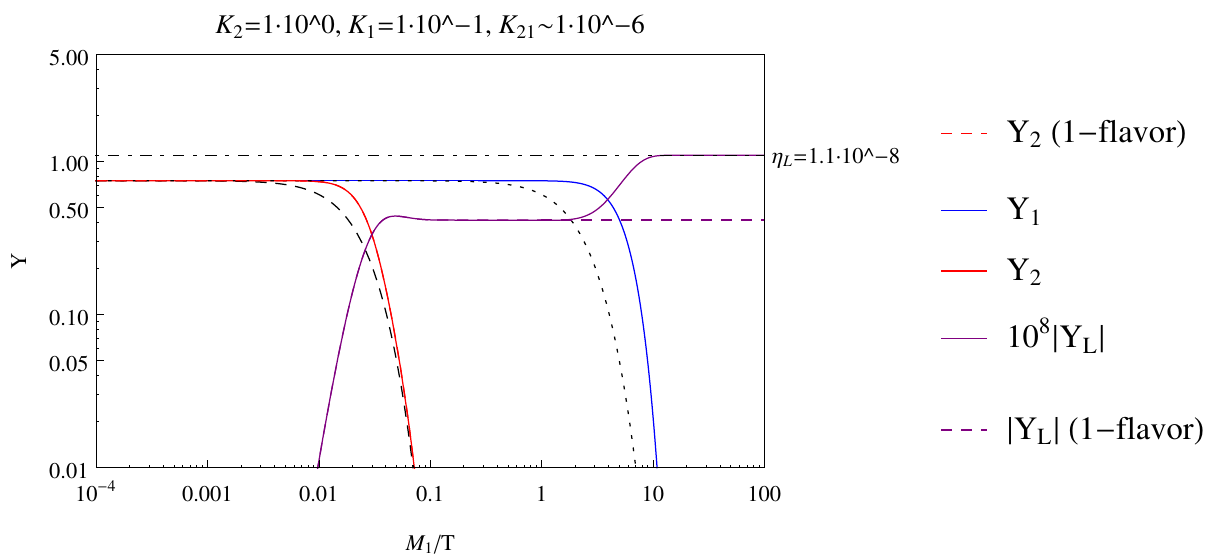}
\label{fig: 1 flavor vs 2 flavors K21=0}}

\subfigure[]{
\includegraphics[width=9cm]{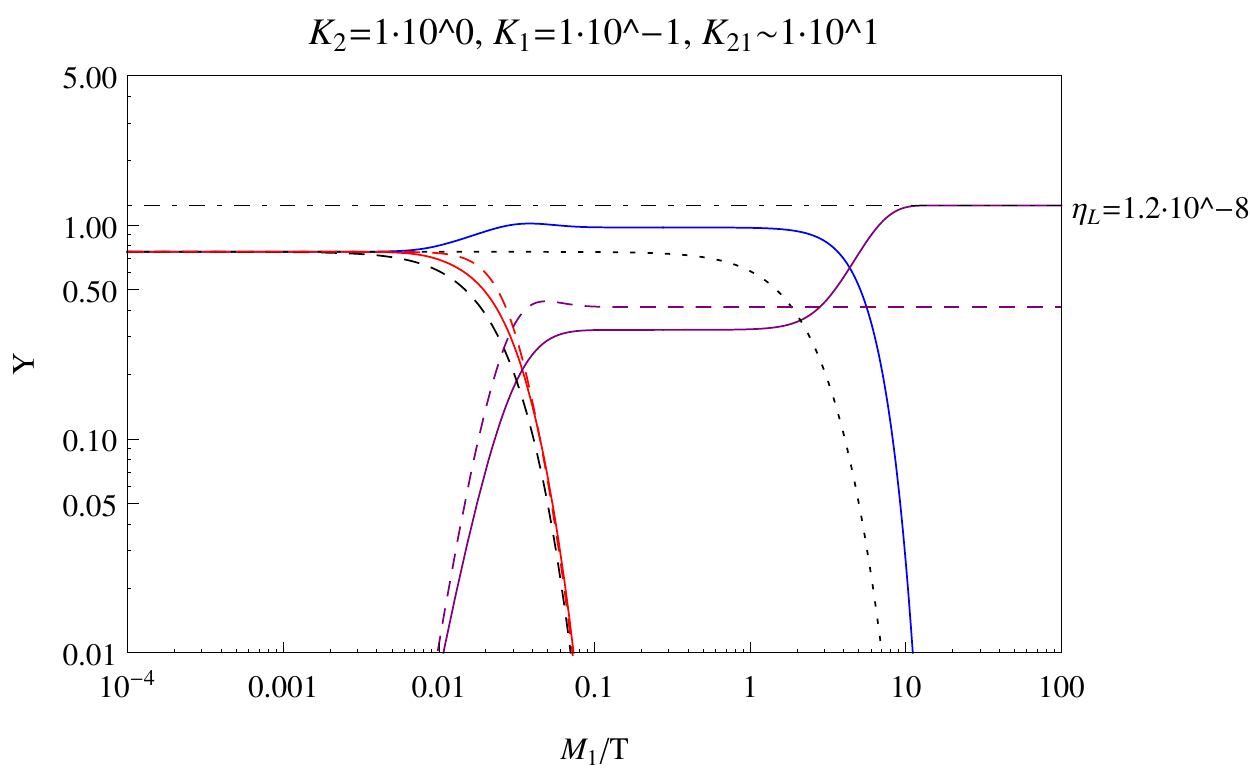}
\label{fig: 1 flavor vs 2 flavors K21 not 0 log log}}
\subfigure[]{\includegraphics[width=9cm]{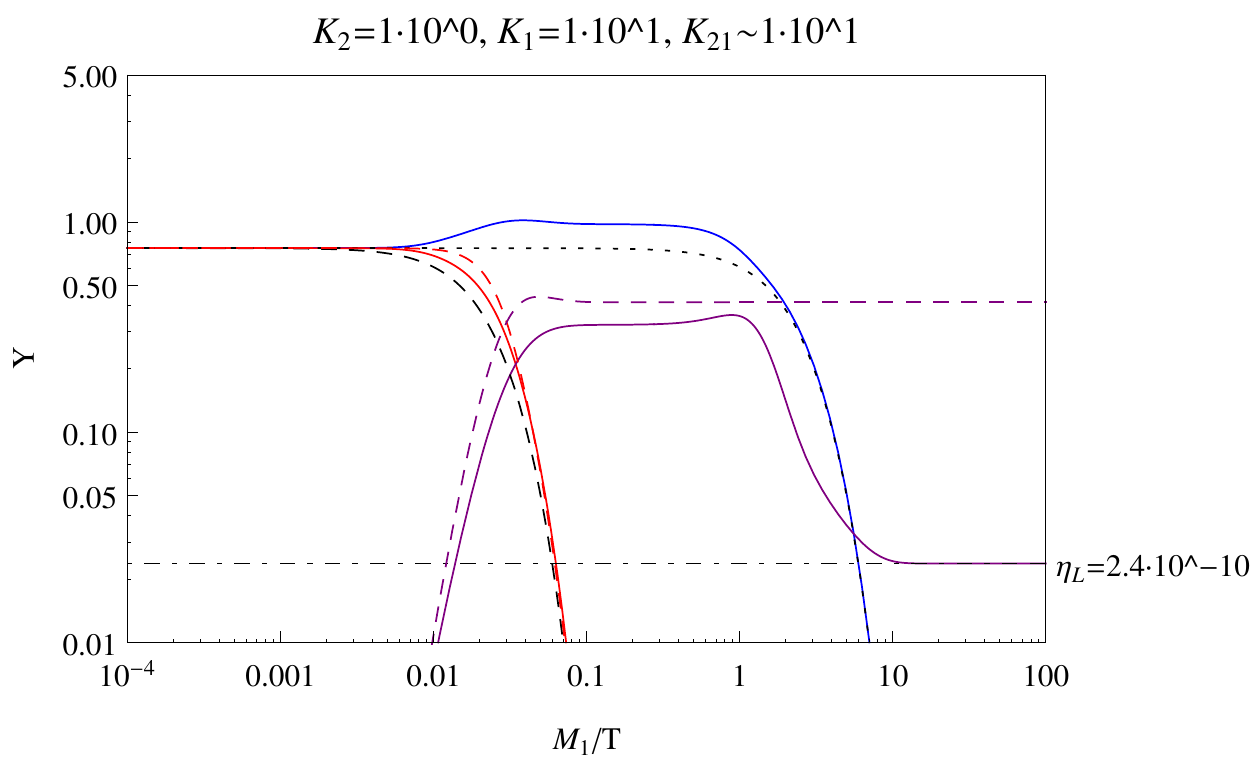}
\label{fig: 1 flavor vs 2 flavors K21 not 0 Large K1 loglog}
}\caption[]{\footnotesize Plot (a) shows the $N_{2}$ and $N_1$ abundances (red and blue)) and $Y_{L-\overline L}$ (purple) in the toy 2-flavor model, compared 
to the $N_2$ abundance (in dashed red) and the lepton asymmetry (in dashed purple) in the one-flavor case. Plots (b,c) show further examples of the 2-level toy 
model. The coupling $N_2\rightarrow N_1S$ allows $N_2$ to cope better with expansion, which 
causes a reduction of the generated lepton asymmetry. In turn, this results in the $N_1$ abundance being further out-of-equilibrium. The color 
coding is the 
same in all these and subsequent plots.}
\label{fig: 1 flavor vs 2 flavors K21 not 0}
\end{figure}

\subsubsection*{$N_1$ efficiency factor: $\kappa_1$}

The $N_1$ phase is characterized by negligible $N_2$ abundance, and with $N_2$ interactions being out-of-equilibrium $D_2,D_{21}\ll D_1$. Thus the physics is 
once again equivalent to the one-flavor case. The efficiency factor $\kappa_1$ is obtained by integrating the Boltzmann equation for the lepton asymmetry,
\beq
\begin{split}
 \kappa_1Y^{eq}_1(0)=&\int_{\mathcal{Z}}^\infty dz'\frac{D_1}{z'}\left(\frac{Y_1}{Y^{eq}_1}-1\right)e^{-\int_{z'}^\infty dz''(W_{ID_1}/z'')},\\
 =&-\int_{\mathcal{Z}}^\infty dz'\frac{\ptl Y_1}{\ptl z'}e^{-\int_{z'}^\infty dz''(W_{ID_1}/z'')}.
\end{split}
\eeq
In the limit $K_1\ll1$, the washout is minimal, leaving
\beq
 \kappa_1Y^{eq}_1(0)=Y_1(\mathcal{Z}).
\eeq
The efficiency factor depends on the value of the abundance once the third phase starts. Since $Y_1$ is constant in the intermediate phase, this is the same as 
the final abundance at the end of the first phase. In Fig.~\ref{fig: N1Overabundance}, we show numerical results for $Y_1(\mathcal{Z})$ as a function of $K_2$ 
for various values of $K_{21}$. In 
calculating $\kappa_2$, we estimated that $Y_1(\mathcal{Z})=2$ at the end of the first phase, which is true in the limit $K_2\sim0$, and $\ptl(Y_1+Y_2)/\ptl 
z_1=0$. When $K_2\ll1$ but not zero, $Y_1+Y_2$ is no longer constant, and instead we have
\beq\label{eq: exact formula for abundance offset}
 Y_1(\mathcal{Z})=2Y^{eq}_1(0)-\int_0^{\mathcal{Z}}dz'\frac{D_2}{z'}\left(\frac{Y_2}{Y^{eq}_2}-1\right).
\eeq
This function is plotted in Fig.~\ref{fig: N1Overabundance}, along with the analytical approximations. Thus, to a good approximation, we have
\beq\label{eq: efficiency kappa1 low K1 approximation}
 \kappa_1=\frac{Y_1(\mathcal{Z})}{Y^{eq}_1(0)}=1+\frac{K_{21}}{K_2+K_{21}}e^{-K_2}\quad {\rm when} \quad K_1 \ll 1.
\eeq
For large $K_2\gg1$, the behavior approaches the decoupled limit. Numerical results for the efficiency factor $\kappa_1$ are displayed in Fig.~\ref{fig: 
N1EfficiencyFactor_OverabundanceEffect} as a function of $K_1$ for varying initial conditions. At the other end of the spectrum, as in standard leptogenesis, 
the efficiency 
factor is \cite{Buchmuller:2004nz},
\beq\label{eq: efficiency kappa1 high K1 approximation}
 \kappa_1\sim\frac{1}{1.2K_1(\ln K_1)^{0.8}} \quad {\rm when} \quad K_1 \gg 1.
\eeq

\subsubsection*{Final lepton asymmetry in the toy model}

In Fig.~\ref{fig: 1 flavor vs 2 flavors K21=0}, we show an example of a decoupled system, with the parameters 
$\{K_2,K_1,K_{21}\}=\{1,0.1,10^{-6}\}$. From Fig.~\ref{fig: 2 stage leptogenesis scattering less efficiency factors}, we 
find the efficiency factor $\kappa_2\sim0.5$, while Fig.~\ref{fig: N1Overabundance} tells us that the over-abundance at the exit of the first phase will be 
zero 
(decoupling limit), and Fig.~\ref{fig: N1EfficiencyFactor_OverabundanceEffect} then implies that  $\kappa_1\sim1$. Inserting these values into Eq.~\eqref{eq: 
final asymmetry for 2 stage leptogenesis no scattering} leads to $Y^f_{L-\overline L}\sim1.1\cdot10^{-8}$, which is in excellent agreement 
with the numerically determined result. Two examples of coupled system are shown in Fig.~\ref{fig: 1 flavor vs 2 flavors K21 not 0 
log log} and \ref{fig: 1 flavor vs 2 flavors K21 not 0 Large K1 loglog} with the parameters $\{K_2,K_1,K_{21}\}=\{1,0.1,10\}$ and 
$\{K_2,K_1,K_{21}\}=\{1,10,10\}$ respectively. Using Figs.~\ref{fig: 2 stage leptogenesis scattering less efficiency factors}, and \ref{fig: 
N1 Efficiency factor kappa1}, we find the efficiency factors $\{\kappa_2,\kappa_1\}=\{0.4,1.3\}$ and $\{\kappa_2,\kappa_1\}=\{0.4,0.03\}$  and obtain 
$Y^f_{L-\overline L}\sim1.3\cdot10^{-8}$ and $Y^f_{L-\overline L}=2.25\cdot10^{-10}$ respectively.

With this understanding of the toy model, we turn in the next section to an analysis of the full system including scattering. It should already be apparent 
that 
viable models will be those in which $N_1$ is sufficiently weakly coupled that the most dangerous effect, rapid washout via $N_1$-mediated inverse decays and 
2-to-2 scattering, is suppressed. With this constraint, the residual effects of scattering are generally quite small.

\section{Results in the Hierarchical regime}\label{sec: Results in the Hierarchical regime}

The focus in this section will be on studying the solutions to the Boltzmann equations in the hierarchical regime, $M_2/M_1 \gg 1$. As shown in Fig.~\ref{fig: 
Asymmetry functions and 
approximation} in Appendix A, the exact and large $M_2/M_1$ expressions for the $CP$ asymmetry are within a factor of two for $M_2/M_1 > 10$, which will serve 
as a practical definition of this regime. Qualitatively, the physical behaviour should be similar for all mass ratios outside the resonant regime 
\cite{Pilaftsis:2003gt,Pilaftsis:2005rv}, $M_2-M_1\sim\Gamma_{1,2}/2$, which we will not consider here.

The two flavor HPL model is distinct from standard leptogenesis in at least two ways. The first difference concerns the mass dependence
of the $CP$-asymmetry. As discussed in Sec.~\ref{sec: HPL CP-asymmetry}, and again in Appendix~\ref{sec: app paramet of CP-asymmetry}, the $CP$-asymmetry 
presents distinct high and the low mass regimes. The high mass regime, $M_2\gtrsim 10^8$GeV, is determined by the Standard Yukawa contribution to the 
$CP$-asymmetry. 
The low mass regime is instead determined by the hidden sector contribution to the $CP$-asymmetry, proportional to the trilinear coupling $\beta/M_i$. 
This liberates the model from the Davidson-Ibarra bound on the $CP$-asymmetry, and allows for viable low scale scenarios. 

The other significant difference with standard leptogenesis  concerns the dynamics. In the minimal model, the main contribution to the $CP$-asymmetry
comes from the decays and inverse decays into leptons, and the scattering processes are largely subdominant. 
This is in part because as the temperature falls below $T\sim M_1$, all the $L$-violating scattering rates are suppressed due to Boltzmann suppression of the 
neutrino abundance. For HPL, the situation is different due to the emphasis on the $CP$-asymmetry generated by $N_2$ decays, and the importance of the evolution 
between $T\sim M_2$ and $T\sim M_1$. As a consequence, scattering processes involving $N_1$ have the potential to affect the lepton asymmetry quite 
significantly, and need to be considered carefully.

\subsection{Viable Scenarios}\label{subsec: high mass scale}

We will impose two requirements on realistic scenarios, namely the ability to reproduce the observed baryon asymmetry, and similarly that they admit a consistent light neutrino 
mass spectrum. The first requirement translates within leptogenesis to a specific lepton asymmetry at the temperature where $B+L$-violating sphaleron processes fall 
out of equilibrium. For the Standard Model field content, the equilibrated lepton and baryon asymmetries are related by $\eta_L=-(51/28)\eta_B$ 
\cite{Harvey:1990qw}. The additional singlet in the Higgs portal model only affects this by changing the critical temperature of the electroweak crossover 
\cite{Ahriche:2007jp}. Since we assume $\langle S\rangle=0$, then at least for relatively weak $H-S$ mixing the impact should be small \cite{Pilaftsis:2008qt}. We 
therefore require $|\eta_L|\simeq1\cdot10^{-9}$, given the 
Planck result for $\Omega_bh^2$ \cite{Ade:2013zuv}, which translates to the baryon-to-photon ratio $\eta_B\simeq(6.04\pm0.09)\cdot10^{-10}$ \cite{Coc:2014oia}.

For the second requirement, since the see-saw mechanism is a motivating factor for leptogenesis, we also require consistency with current data on the mass squared 
differences, e.g. $\Delta m_{21}^2=m_{2}^2-m_1^2\simeq(7.5\pm0.5)\cdot10^{-5}\text{eV}^2$ 
\cite{Capozzi:2013csa,*Fogli:2012ua,*GonzalezGarcia:2012sz,*Tortola:2012te}. The see-saw mechanism determines an \textit{effective} light neutrino mass $\tilde 
m_{i}\sim\lambda_i^2v^2/M_i$, which we can 
trade for the thermal equilibrium parameters $K_i=\tilde 
m_i/m_*$ from (\ref{Ki}), and write down an \textit{effective} mass squared difference, $K_2^2-K_1^2\simeq\mathcal{O}\left(\Delta 
m_{21}^2/m_*^2\right)$. However, this relation relies on the equality $\tilde m_i\simeq m_i$ between the effective and the physical light neutrino 
masses, which only holds when the neutrino flavor structure is nearly diagonal \cite{Plumacher:1996kc}. More generally, the precise relation can 
be relaxed, so we will consider models to be viable if the interactions are in the range $K_2\sim\mathcal{O}(1-10)$.

 In the rest of this subsection, we present example scenarios
  that satisfy the above constraints on the neutrino masses and the lepton asymmetry.
 For each case we present three figures: (i) the Boltzmann evolution of the neutrino abundances and the lepton asymmetry, (ii) the relevant thermal 
rates of decays, inverse decays and scattering, and finally (iii) the $CP$-asymmetry `landscape' in which the theory is situated.
In the following subsection, we provide further details showing the impact of varying the parameters of the 
theory, while relaxing the constraints imposed here on viable models.
In particular, we show that  the dynamics of the high mass regime is most sensitive to $\{K_1,K_2\}$, whereas the low mass dynamics 
responds to $\{K_2,K_1,\alpha,\beta\}$.

It is useful to distinguish `high' and `low' mass regimes, based primarily on the mass dependence of the $N_2$ $CP$-asymmetry.   We focus below on the relative impact of 
two-to-two scattering processes, compared to the toy model discussed above.

\begin{figure}[t]
\centering
\subfigure[$\{M_2,M_1,K_2,K_1,K_{21},\beta\}=\{10^9\text{GeV},10^5\text{GeV},8,0.2,40,50\text{GeV}\}$]{
\includegraphics[width=17cm]{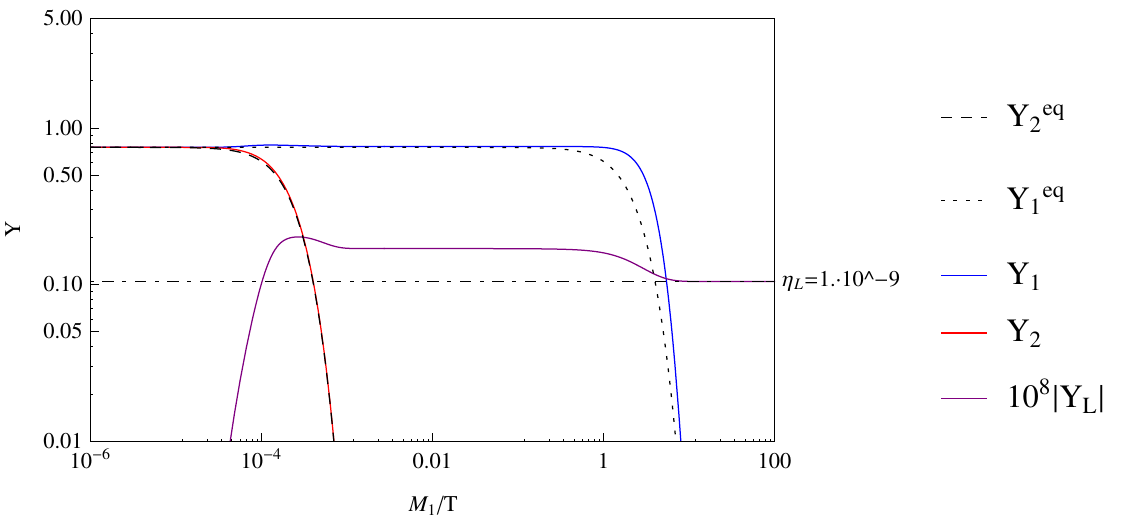}
\label{fig: HighMassExample3 Loglog}}

\subfigure[]{
\includegraphics[width=9cm]{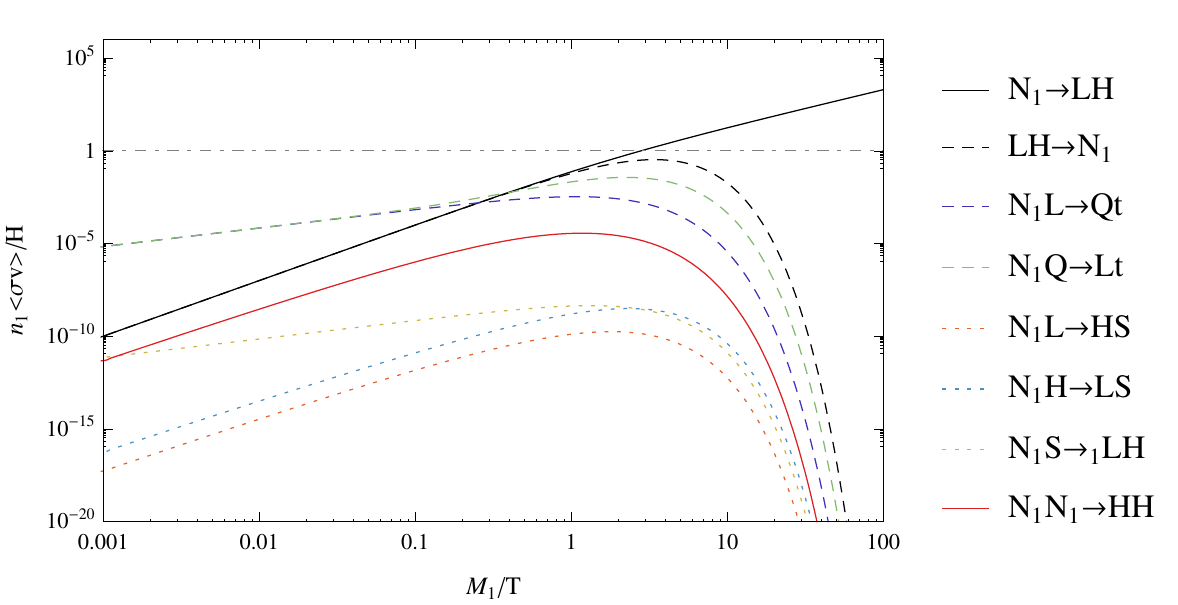}
\label{fig: HighMassExample3 scattering}}
\subfigure[]{\includegraphics[width=9cm]{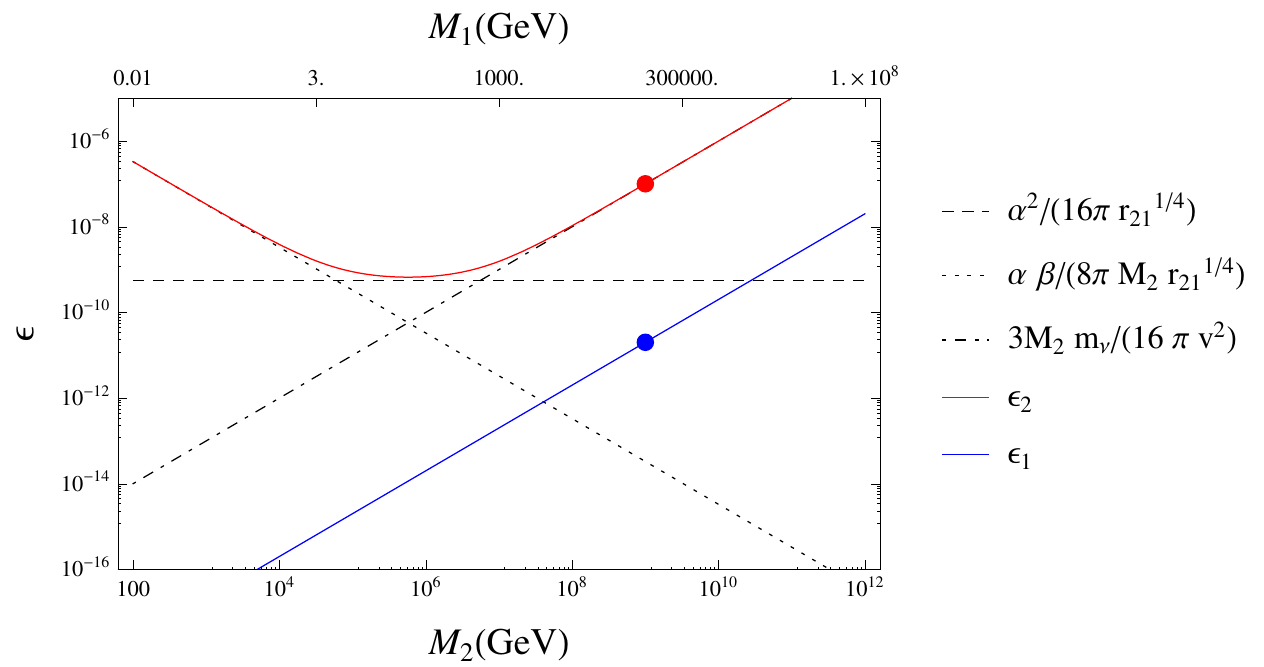}
\label{fig: HighMassExample3 asymmetry}
}\caption[]{\footnotesize These plots show a viable example of the high scale scenario. We show the evolution of the abundances and (rescaled) asymmetry (a), 
and the relative rates (b) as a function of temperature. In this regime, the $CP$-asymmetry is dominated by the standard Yukawa contribution, 
$\epsilon_i\propto m_\nu M_i/v^2$, and the final plot (c) displays the $CP$-asymmetries $\epsilon_{1,2}$ as functions of $M_2$, for a constant mass ratio, 
$M_2/M_1=10^4$. With a large mass hierarchy, there would be plenty of time for $N_1$-mediated processes to wash out all of the lepton asymmetry unless the 
$N_1$ interactions are sufficiently suppressed, hence the very small Yukawa $\lambda_1$.
}
\label{fig: High Mass Example 3}
\end{figure}

\begin{figure}[t!]
\centering
\subfigure[$\{M_2,M_1,K_2,K_1,K_{21},\beta\}=\{20\text{TeV},2\text{TeV},7,0.001,3\cdot10^5,50\text{GeV}\}$]{
\includegraphics[width=17cm]{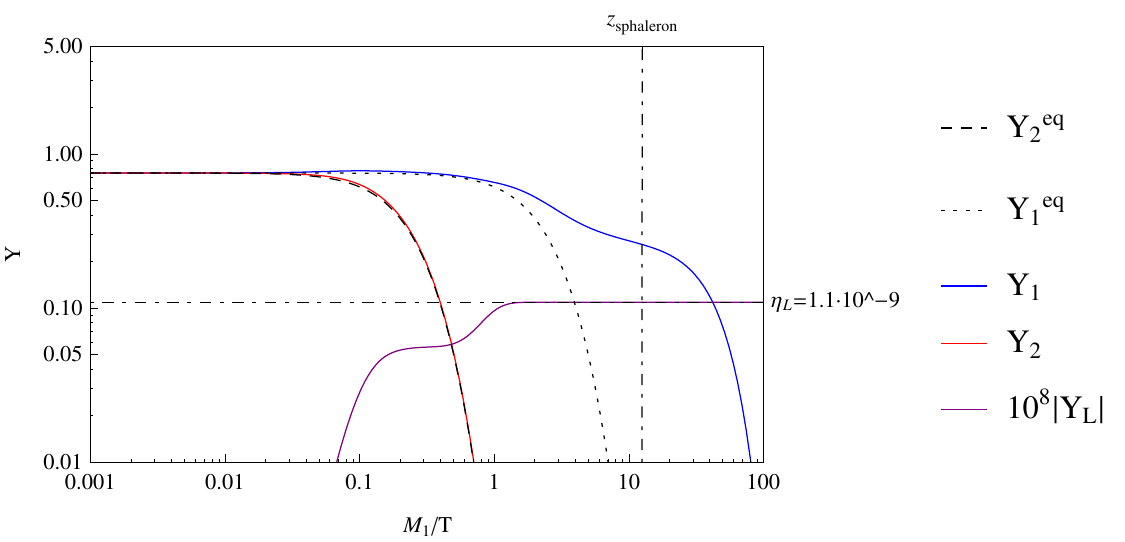}
\label{fig: LowMassExample2 Loglog}}

\subfigure[]{
\includegraphics[width=9cm]{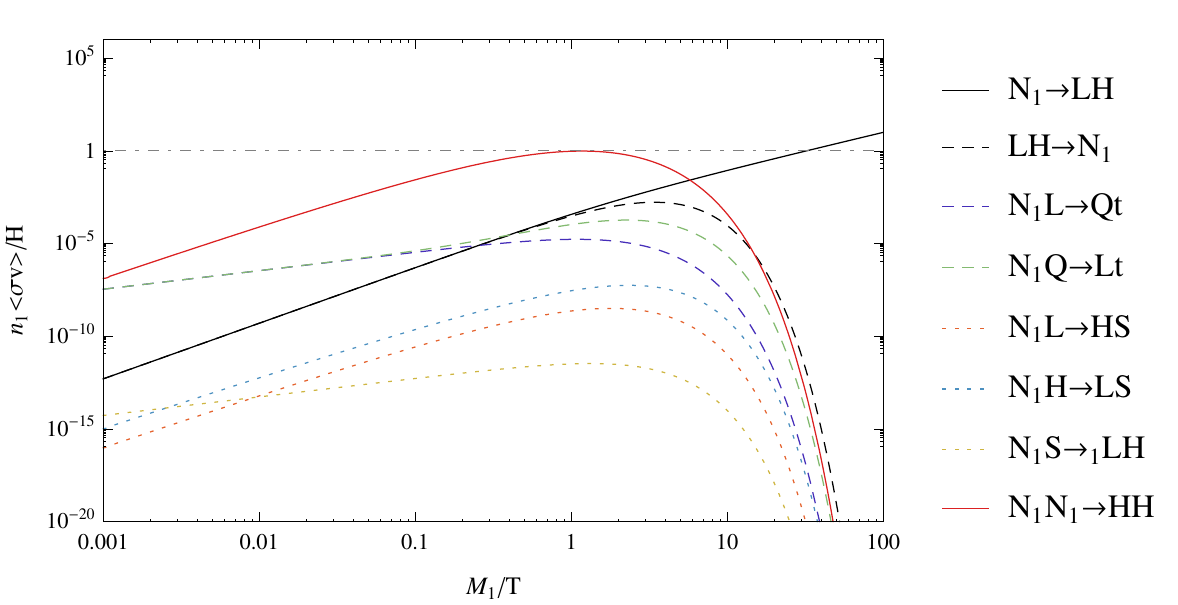}
\label{fig: LowMassExample2 scattering}}
\subfigure[]{\includegraphics[width=9cm]{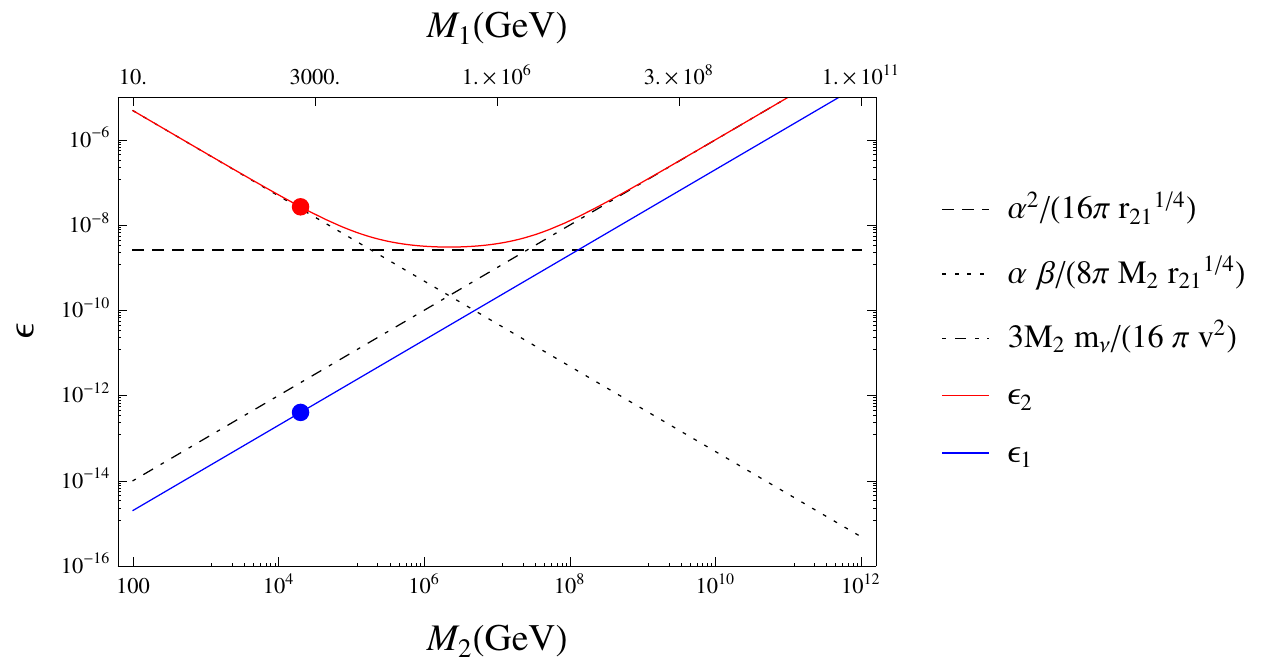}
\label{fig: LowMassExample1 asymmetry}
}\caption[]{These plots, in the same format as Fig.~\ref{fig: High Mass Example 3}, show a viable example in the low mass regime. The hidden sector-sourced 
$CP$-asymmetry from $N_2$ decays is the dominant contribution. The dominant scattering processes are also  mediated by the hidden sector. This is 
also an explicit example of a 
scenario leading to a lepton asymmetry enhancement, thanks to the evolution traversing the sphaleron freezeout temperature.}
\label{fig: Low Mass Example 2}
\end{figure}

\begin{itemize}
 \item \textbf{High-scale models}

An example of a high scale scenario is shown in Fig.~\ref{fig: HighMassExample3 Loglog}. 
The high mass regime is marked by the dominance of the Yukawa sector in contributing to the $CP$-asymmetry, so that $\epsilon_2\sim(3/16\pi v^2)\sum_\alpha 
m_\alpha 
M_2\sim\epsilon_1(M_2/2M_1)$, as displayed in Fig.~\ref{fig: HighMassExample3 asymmetry}. The general result in Eq.~\eqref{eq: general parametrization of 
the final asymmetry} reduces to
\beq\label{eq: high mass parametrization of the final asymmetry}
 Y^f_{L-\overline L}\sim\frac{9\sum_\alpha m_\alpha}{64\pi v^2}M_2\left(\kappa_2e^{-\int_{\mathcal{Z}}^\infty dz' 
(W_1/z')}+2\frac{M_1}{M_2}\kappa_1\right).
\eeq
The lepton-number changing processes mediated by the hidden sector scale as $\sim|\lambda|^2\beta^2/M^2$, and are suppressed compared to Yukawa-mediated scattering 
which scales as $\sim|\lambda|^2m_t^2/v^2$. This is shown in Fig.~\ref{fig: HighMassExample3 scattering}. Additional 
scattering processes such as $N_iS\xrightarrow[j]{}LH$ scale as $|\alpha|^2|\lambda|^2$ and are therefore suppressed if $|\alpha|^2\ll1$, which holds for all the 
viable scenarios we consider here. The lepton-number conserving scattering process $N_iN_i\rightarrow HH$, mediated by $S$ in the $s$-channel, scales as 
$|\alpha|^2\beta^2/M^2$ and is again suppressed. Finally, the lepton-mediated $t$-channel $N_iN_i\rightarrow HH$ scattering is suppressed by a factor 
$|\lambda|^2\ll1$, and has no visible effect on the neutrino abundance.

It follows that the efficiency factors $\kappa_{1,2}$ determined above can be used as a reasonably good 
approximation here, as the scattering corrections are small. Indeed, the dominant $N_i$ scattering processes are controlled by the same equilibrium parameter $K_i$, 
and the Boltzmann-suppressed number density for $T<M_i$ and the fast expansion rate at $T>M_i$ limits their range of activity. 

The possibility of having $\Delta L=1,2$ scattering processes mediated by $N_1$ that remain in equilibrium long after 
the lepton asymmetry $Y^{(2)}_{L-\overline L}$ has been generated, is an important feature of HPL. In practice, these processes need to be suppressed, and
remain out of equilibrium in viable models to avoid too much washout.  For masses chosen so that $\beta/M_1\ll m_t/v$, the washout function is dominated by 
the Yukawa processes $\int_{\mathcal{Z}}^\infty dz'(W_1/z')\simeq c K_1$, where $c\sim1.2+1.4(m_t/v)^2\sim2.5$. With $K_2\sim1-10$, satisfying the 
neutrino mass constraint, one has $\kappa_2\sim1/(cK_2)$, and taking a mass ratio large enough that the $\kappa_1$ term can be neglected, we find the 
constraint
\beq
 Y^f_{L-\overline L}\sim\frac{9\sum_\alpha m_\alpha}{64\pi v^2}M_2\frac{e^{-cK_1}}{cK_2}>\eta_L\simeq10^{-9}.
\eeq
With the neutrino mass constraint $\sum_\alpha m_\alpha\sim m_3$ as input, this implies the lower bound $M_2>(10^7\text{GeV})\cdot cK_2e^{cK_1}$. In standard 
one-flavor leptogenesis, we instead obtain the simpler bound, $M_1\gtrsim10^7$GeV. This is because the equilibrium parameter is constrained to satisfy $K_1<1$, so 
that $\kappa_1\sim1$, and because scattering processes are not in equilibrium long enough to provide any significant washout.

  \item \textbf{Low-scale models}
  
  A viable low scale example is shown in Fig.~\ref{fig: LowMassExample2 Loglog}.
 The $CP$-asymmetry $\epsilon_2$ for low values of $M_2$ is controlled by the hidden sector couplings 
$\{\alpha,\beta\}$, i.e. $\epsilon_2\propto\left(\beta/M_2+|\alpha_{21}|/2\right)|\alpha_{21}|/r_{21}^{1/4}$. At the same time, because the $CP$-asymmetry 
$\epsilon_1$ is sourced purely from the Yukawa sector, it becomes negligible at low mass, $\epsilon_1\propto\sum_\alpha m_\alpha M_1/v^2\ll\epsilon_2$, as 
exemplified by 
Fig.~\ref{fig: LowMassExample1 
asymmetry}. In this case, Eq.~\eqref{eq: 
general parametrization of the final asymmetry} reads,
\beq\label{eq: low mass parametrization of the final asymmetry}
 Y^f_{L-\overline 
L}\simeq\frac{3}{32\pi}\kappa_2\left(\frac{\beta}{M_2}+\frac{|\alpha_{21}|}{2}\right)\frac{|\alpha_{21}|}{r_{21}^{1/4}}e^{-\int_{\mathcal{Z}}^\infty dz 
(W_1/z')}.
\eeq
This relation relies on a hierarchical separation between the $N_2$ and $N_1$ phases, which is only marginally satisfied in Fig.~\ref{fig: 
LowMassExample2 Loglog} where $M_2/M_1=10$. Nevertheless, the above relation encodes the two competing effects, namely the enhancement of the low-mass 
$CP$-asymmetry, and also the increased washout. Generating a larger $CP$-asymmetry requires an increase in the ratio $\beta/M$, which at the same time increases the 
scattering rates and the washout (see the next subsection for details of the relative effects). Independent of the precise dynamics, for the relevant couplings, 
$K_2\sim1-10$,  the efficiency factor will generally lie in the range $\kappa_2\sim0.01-0.5$ (see Fig.~\ref{fig: 2 stage leptogenesis scattering less 
efficiency factors}). The washout function depends on the scattering processes, and we can approximate $\int_{\cal Z}^\infty 
dz'W_1/z'\sim(3+2\beta/M_1)K_1$, so that $\exp(-\int_{\cal Z}^\infty dz'W_1/z')\sim0.5-1$ for $K_1\sim0.01-0.1$ and $\beta/M_1\sim1$. As an example, taking 
$\kappa_2\exp(-\int_{\cal Z}^\infty dz'W_1/z')\sim0.1$ implies the following characteristic  constraint on the lepton asymmetry,
\beq 
 Y^f_{L-\overline 
L}\sim\frac{3}{32\pi}\left(\frac{\beta}{M_2}+\frac{|\alpha_{21}|}{2}\right)\frac{|\alpha_{21}|}{r_{21}^{1/4}}10^{-1}>\eta_L\simeq10^{-9},\qquad 
|\alpha_{21}|\frac{\beta}{M_2}\sqrt{\frac{M_1}{M_2}}>3\cdot10^{-7},
\eeq
where small couplings $|\alpha_{21}|\ll\beta/M_2$ have been assumed, and typically we will use $|\alpha_{21}|\sim10^{-5}$. Given the above approximations, the 
light neutrino mass scale is a subleading parameter and does not appear in this bound.

\end{itemize}

\begin{figure}[t!]
\centering
\subfigure[\footnotesize $\{M_2,M_1,K_2,K_1,K_{21},\beta\} =\{1500\gev, 155\gev, 1, 0.5, 10^5, 
20\gev\}$]{\includegraphics[width=8cm]{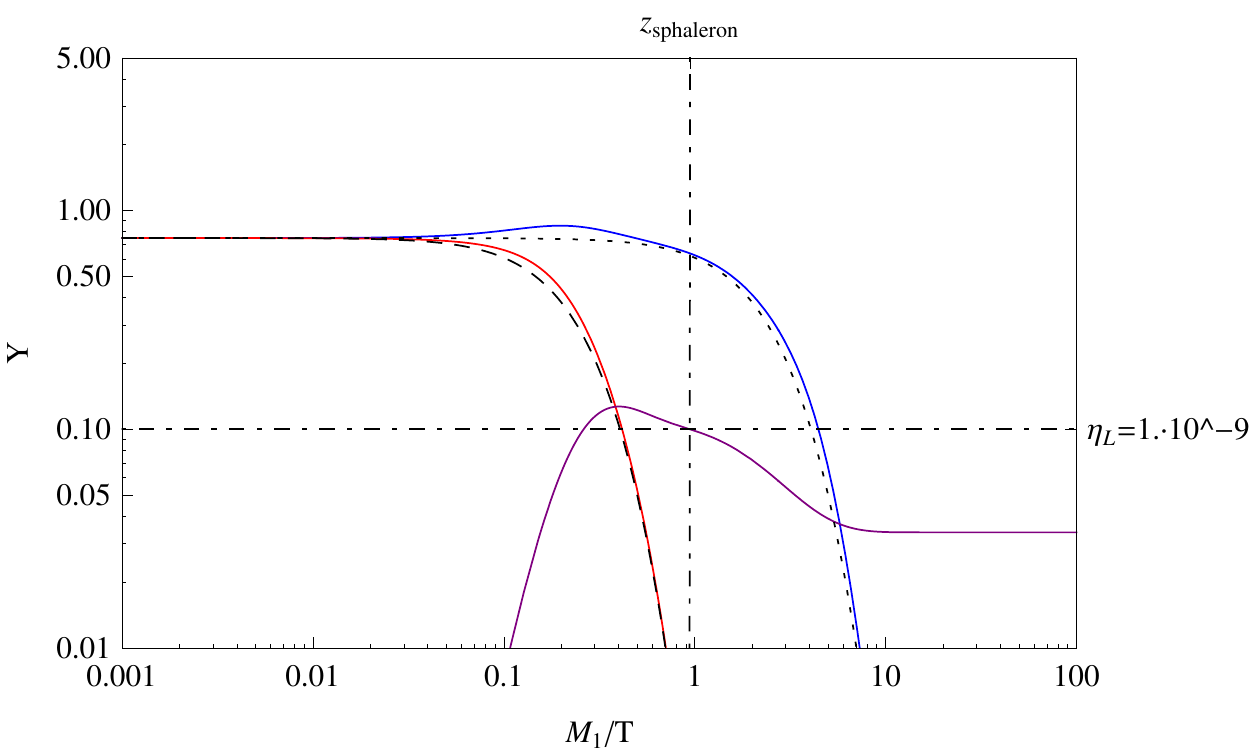}\label{fig: Low mass sphaleron effect}}
\subfigure[\footnotesize $\{M_2,M_1,K_2,K_1,K_{21},\beta\} = \newline\{10^9\gev, 130\gev, 8, 0.5, 4\cdot10^3, 
1\gev\}$]{\includegraphics[width=10cm]{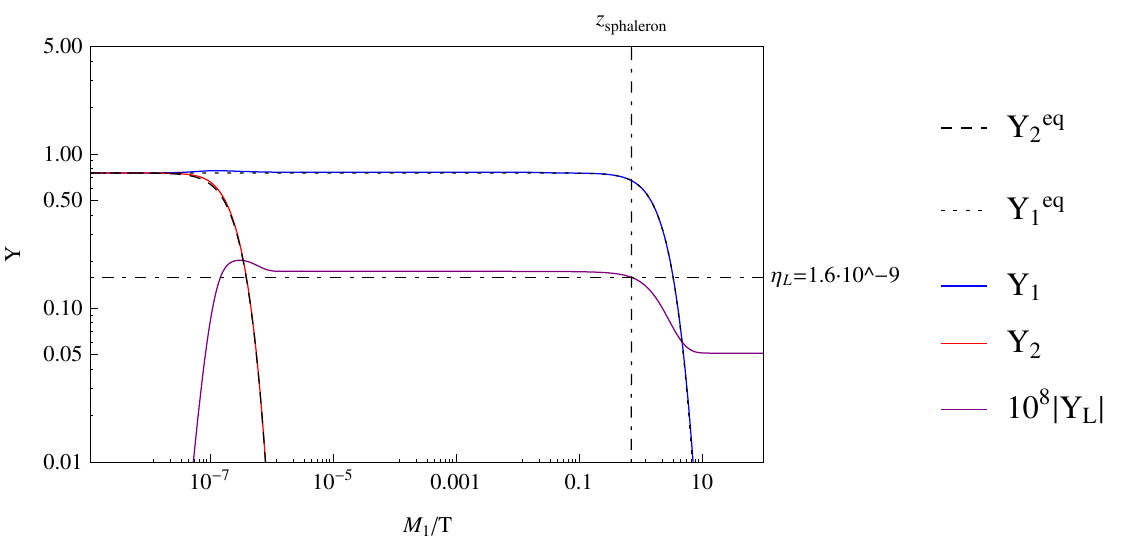}\label{fig: High mass sphaleron effect}}

\caption[]{\footnotesize Scenarios representig low (a) and high (b) scale models which take advantage of the sphaleron cutoff temperature in order to liberate the parameters 
from stringent constraints.}
\label{fig: sphaleron effect}
\end{figure}

An feature worth noting is the sensitivity to low temperature boundary conditions, namely 
the temperature at which 
$B+L$-violating sphaleron transitions go out of equilibrium, $T_{sphaleron}=159\pm1$GeV \cite{D'Onofrio:2014kta}.
The lepton asymmetry at this point effectively determines the final baryon asymmetry, while further evolution is observationally relatively 
unconstrained. This is phenomenologically interesting as it is usually quite difficult to find viable models with a large washout from $N_1$. However, if we 
choose $M_2$ above and $M_1$ well below the temperature at which sphaleron processes freeze out, the baryon asymmetry is generated right after $N_2$ falls 
out-of-equilibrium, and will not be washed out at lower scales even if the lepton asymmetry is highly suppressed through $N_1$ processes. The $N_1$ sector 
effectively decouples in this case. A low mass example is shown in Fig.~\ref{fig: Low mass sphaleron effect}. Note that 
$T_{sphaleron}>M_1>m_H$ in this case, since $M_1<m_H$ would require the inclusion of interactions such as
$N_1L\leftrightarrow H$, that have not been considered thus far.\footnote{Note that such processes may also arise on including thermal corrections to the scalar masses.}   
In the low mass regime, many interactions are relevant which cause significant washout of the lepton 
asymmetry. As a consequence, the range of parameters available for low mass scenarios is quite limited. On the other hand, an example of a high mass 
scenario which also takes advantage of the sphaleron cutoff temperature is shown in figure Fig.~\ref{fig: High mass sphaleron effect}. In that situation, all of the lepton 
asymmetry is generated out of the first phase, from $N_2$ leptonic decays. In this case, with $M_1<T_{sphaleron}$, the physics of $N_1$, including the hidden 
sector interactions, are only weakly constrained. An interesting aspect of this scenario is the possibility of exploring models where the lightest RH neutrino is so light and 
weakly coupled to leptons, that its lifetime could be long enough to play an independent cosmological role, potentially in the form of sterile neutrino dark matter 
\cite{Kusenko:2009up, *Petraki:2007gq}. In such cases, the $N_1$ abundance will have to be sufficiently depleted for consistency with constraints on the dark matter abundance. 
This can be achieved through adjusting the $N_1N_1\rightarrow HH$ rate.

\subsection{Aspects of the dynamics}\label{sec: decays vs scattering and parameter variations}

\begin{figure}[t!]
\centering
\subfigure[$\{M_2,M_1,K_2,K_1,K_{21},\beta\} =\newline\{10^8\gev, 10^4\gev, 0.8, 0.4, 4\cdot10^2, 50\gev\}$]{
\includegraphics[width=9cm]{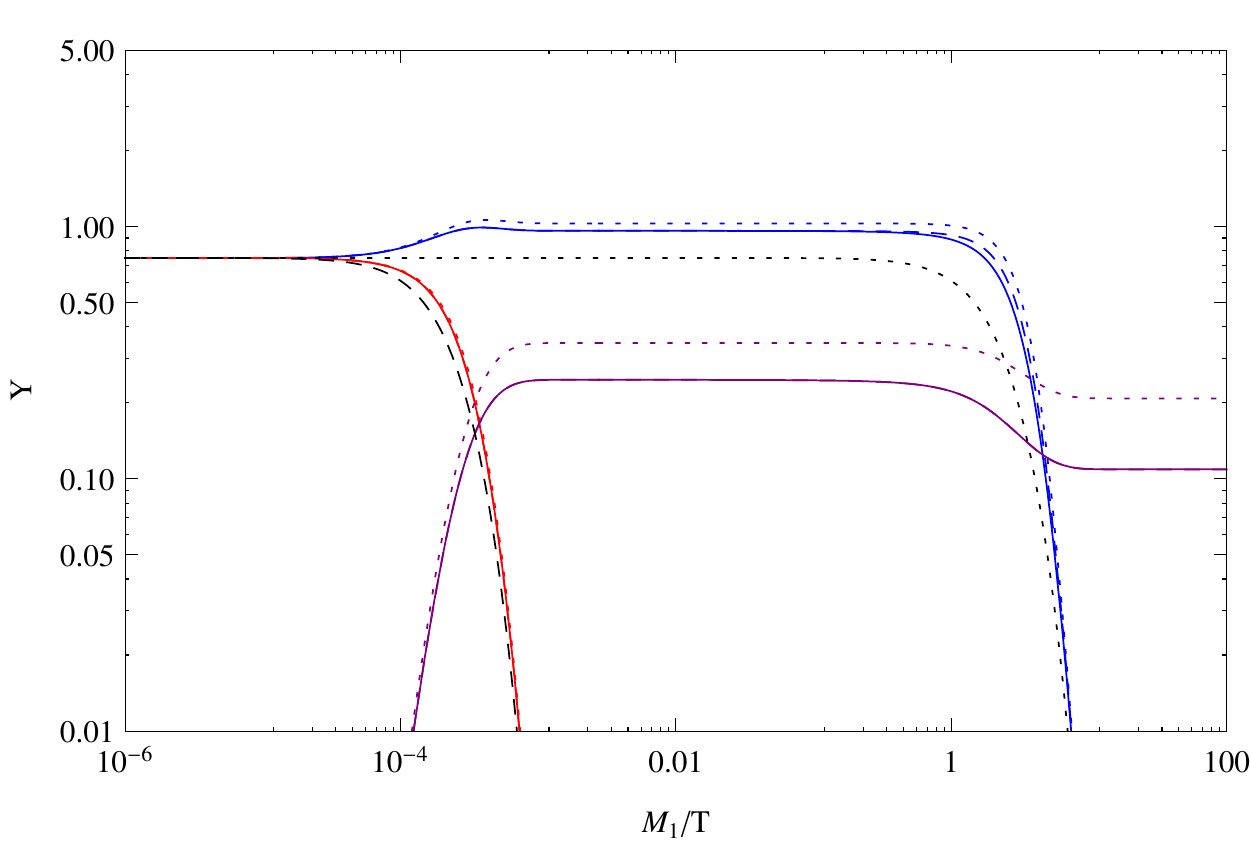}
\label{fig: Comparison_decays_scattering_high_mass Loglog}}
\subfigure[$\{M_2,M_1,K_2,K_1,K_{21},\beta\} = \newline\{2\cdot 10^3\gev, 200\gev, 0.8, 0.04, 400, 500\gev\}$]{
\includegraphics[width=9cm]{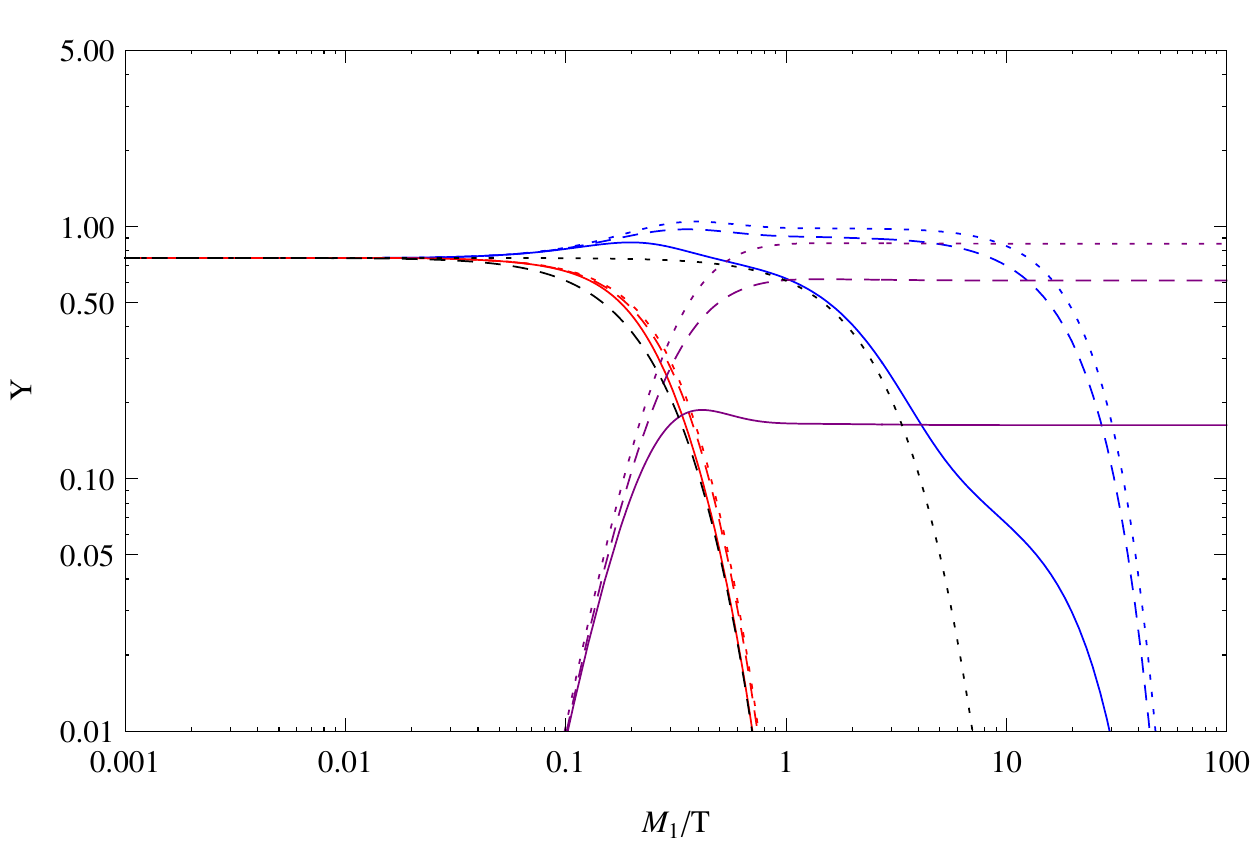}
\label{fig: LowMassExample5_withoutdecays _fordecaysvsnodecays log log}}

\caption[]{Plots showing the impact of $\Delta L=1$ and $\Delta N=2$ processes on the dynamics for high scales $\{M_2,M_1\}=\{10^8,10^4\}$GeV  (on the left) and 
low scales $\{M_2,M_1\}=\{2,0.2\}$TeV (on the right). The color coding, blue, red, purple and black, is the same as in Fig.~\ref{fig: HighMassExample3 
Loglog}~\ref{fig: 
LowMassExample2 Loglog}, ~\ref{fig: Low mass sphaleron effect}. The dotted pattern is for decays only, dashed for decays plus $\Delta L=1$ scattering, and plain for 
decays 
plus $\Delta L=1$ and $\Delta N=2$ scattering.}
\label{fig: Decays vs scattering}
\end{figure}

\begin{figure}[t!]
\centering
\subfigure[$K_2=\{0.1,5,10\}$ (plain, dashed, dotted)]{
\includegraphics[width=8.5cm]{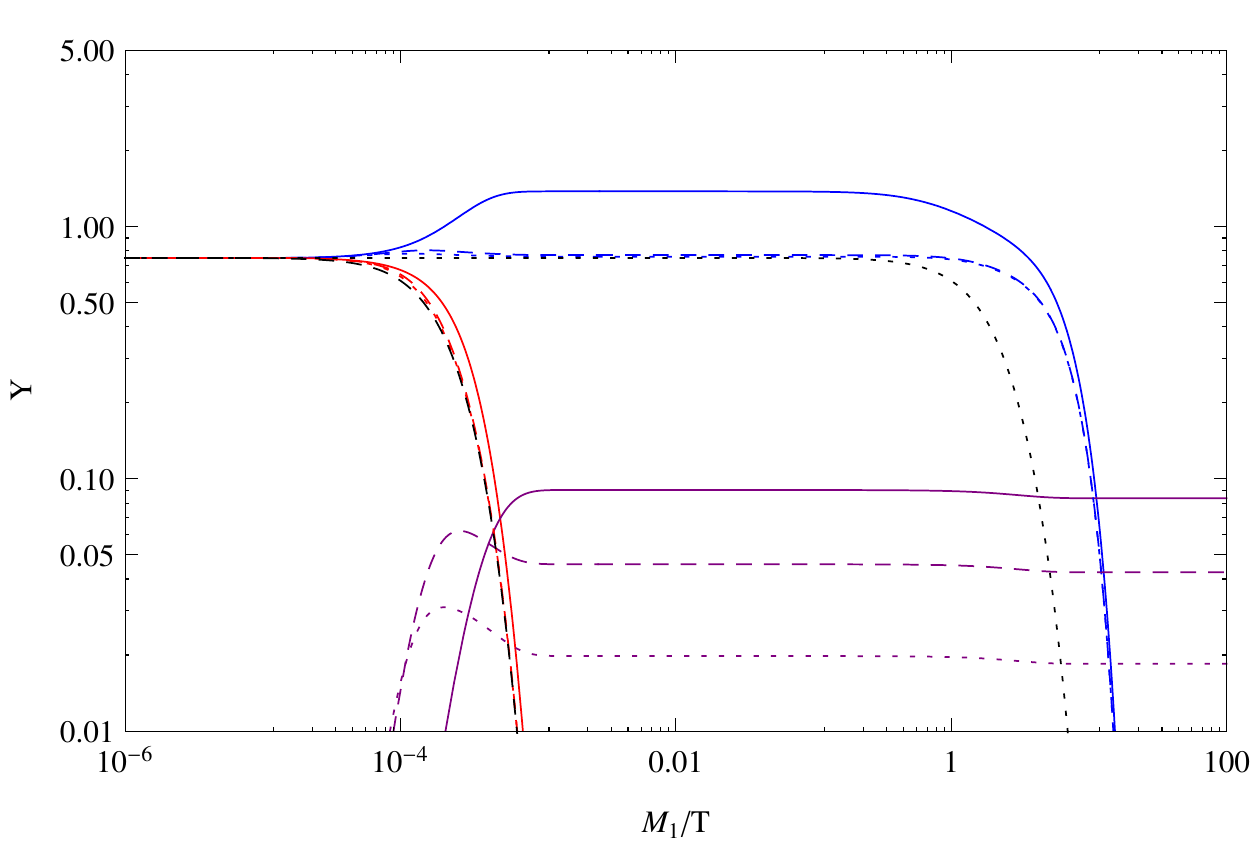}
\label{fig: Large_mass_K2effect}}
\subfigure[$K_1=\{0.1,5,10\}$ (plain, dashed, dotted)]{
\includegraphics[width=8.5cm]{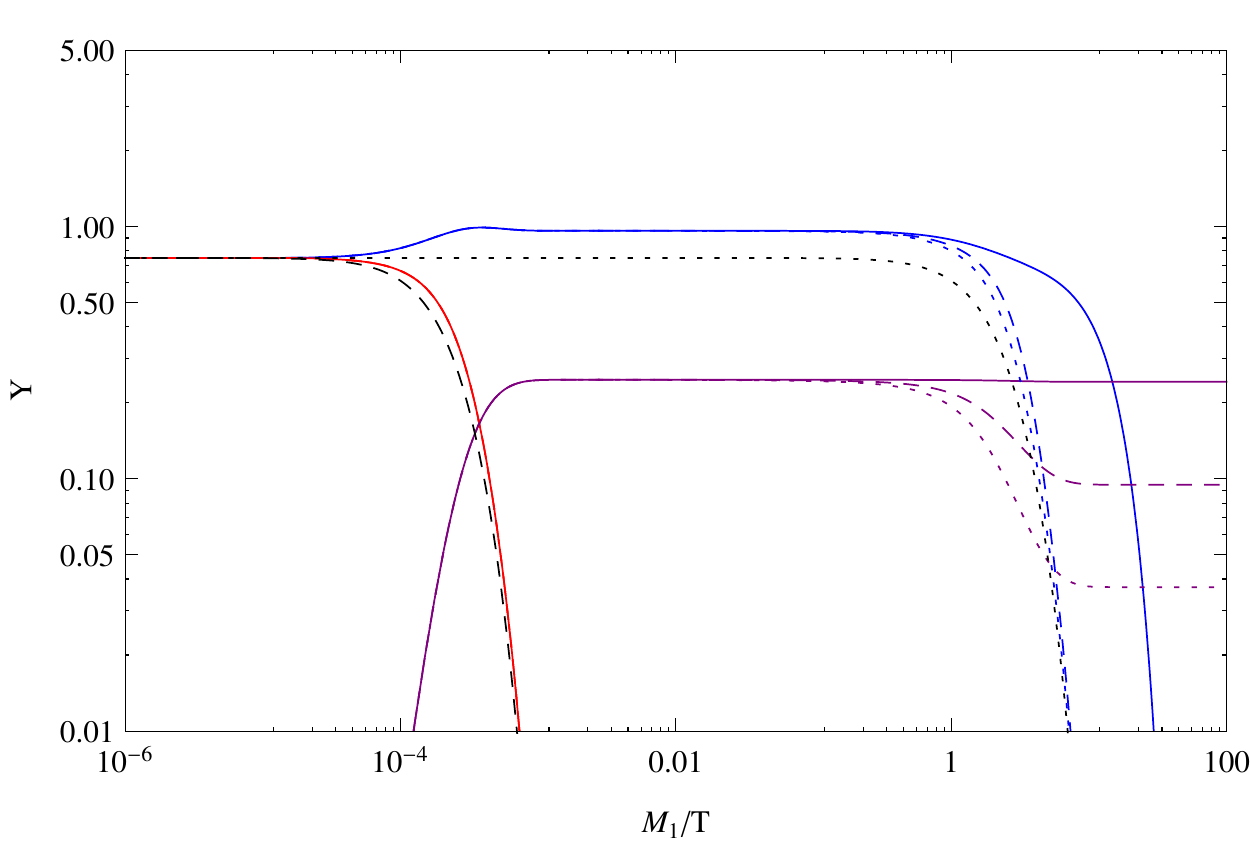}
\label{fig: Large_mass_K1effect}}
\vspace*{0.5cm}

\subfigure[$K_2=\{0.1,5,10\}$ (plain, dashed, dotted)]{
\includegraphics[width=8.5cm]{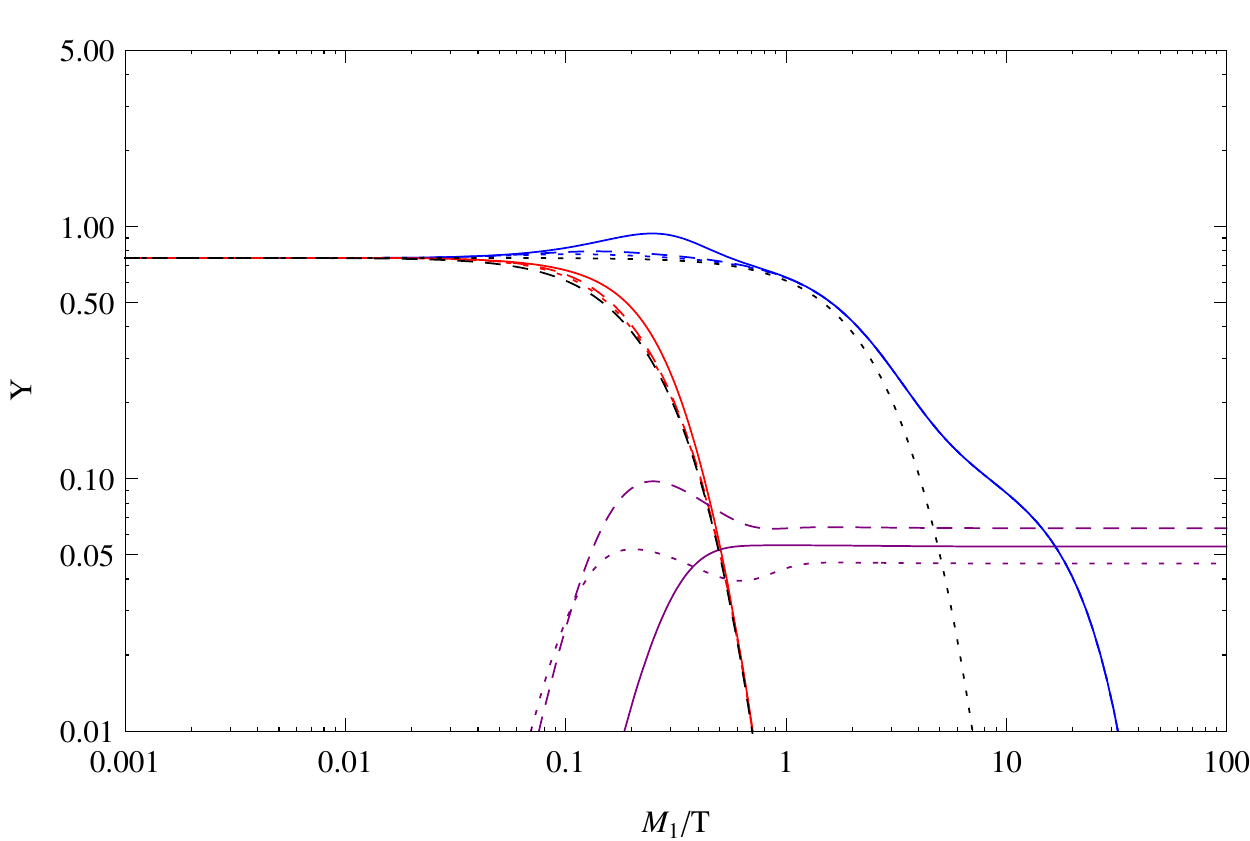}
\label{fig:  Low mass K2 effect log log}}
\subfigure[$K_1=\{0.1,5,10\}$ (plain, dashed, dotted)]{
\includegraphics[width=8.5cm]{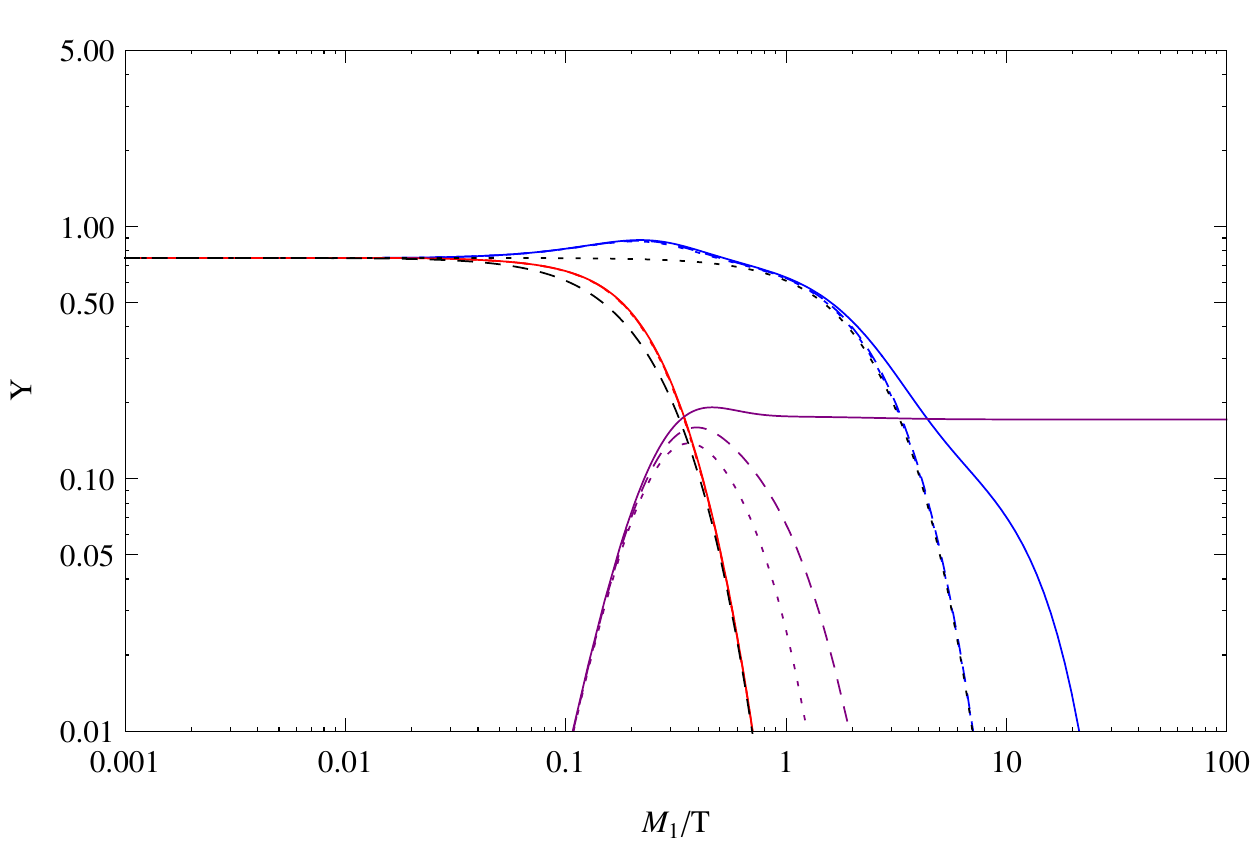}
\label{fig:  Low mass K1 effect Loglog}}
\vspace*{0.5cm}

\subfigure[$K_{21}=\{10,500,1000\}$ (plain, dashed, dotted)]{
\includegraphics[width=8.5cm]{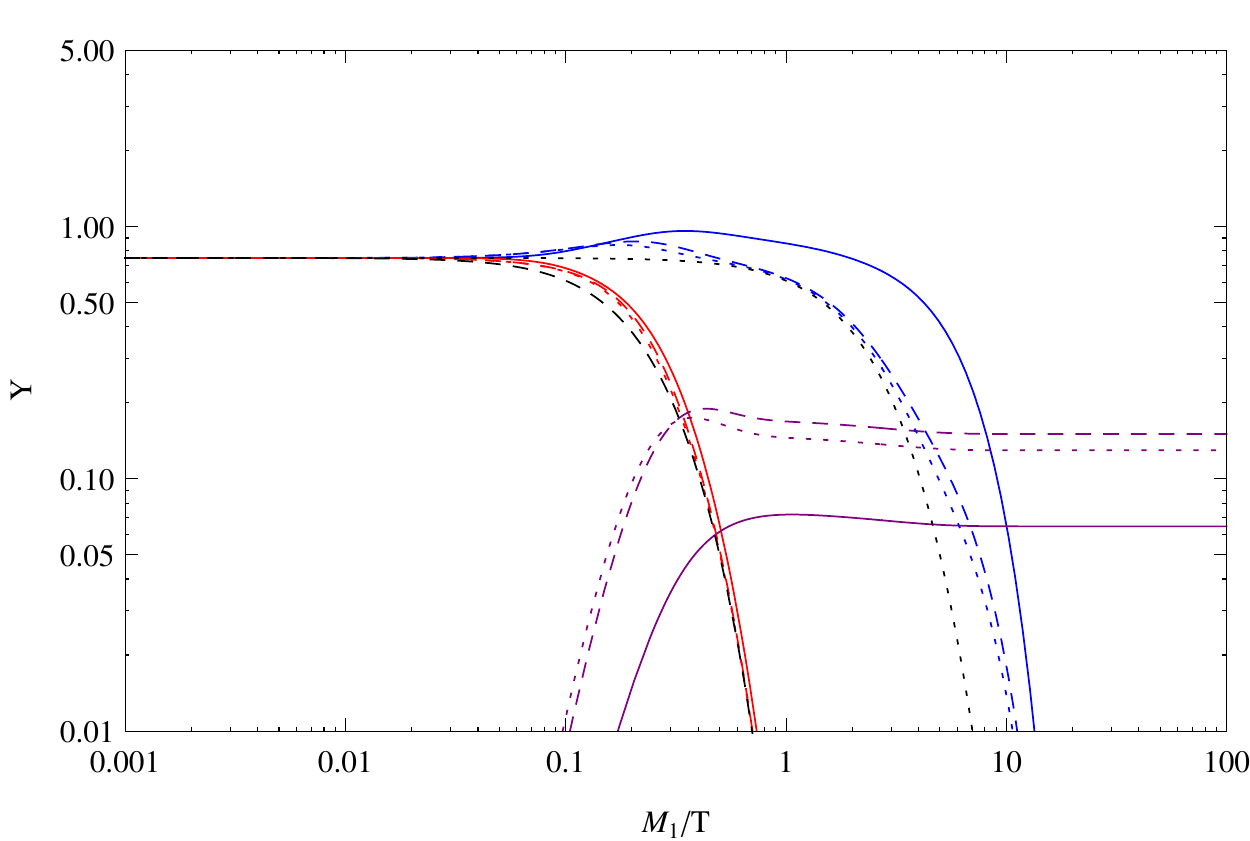}
\label{fig: Low mass K21 effect Loglog}}
\subfigure[$\beta=\{10,500,1000\}$GeV (plain, dashed, dotted)]{
\includegraphics[width=8.5cm]{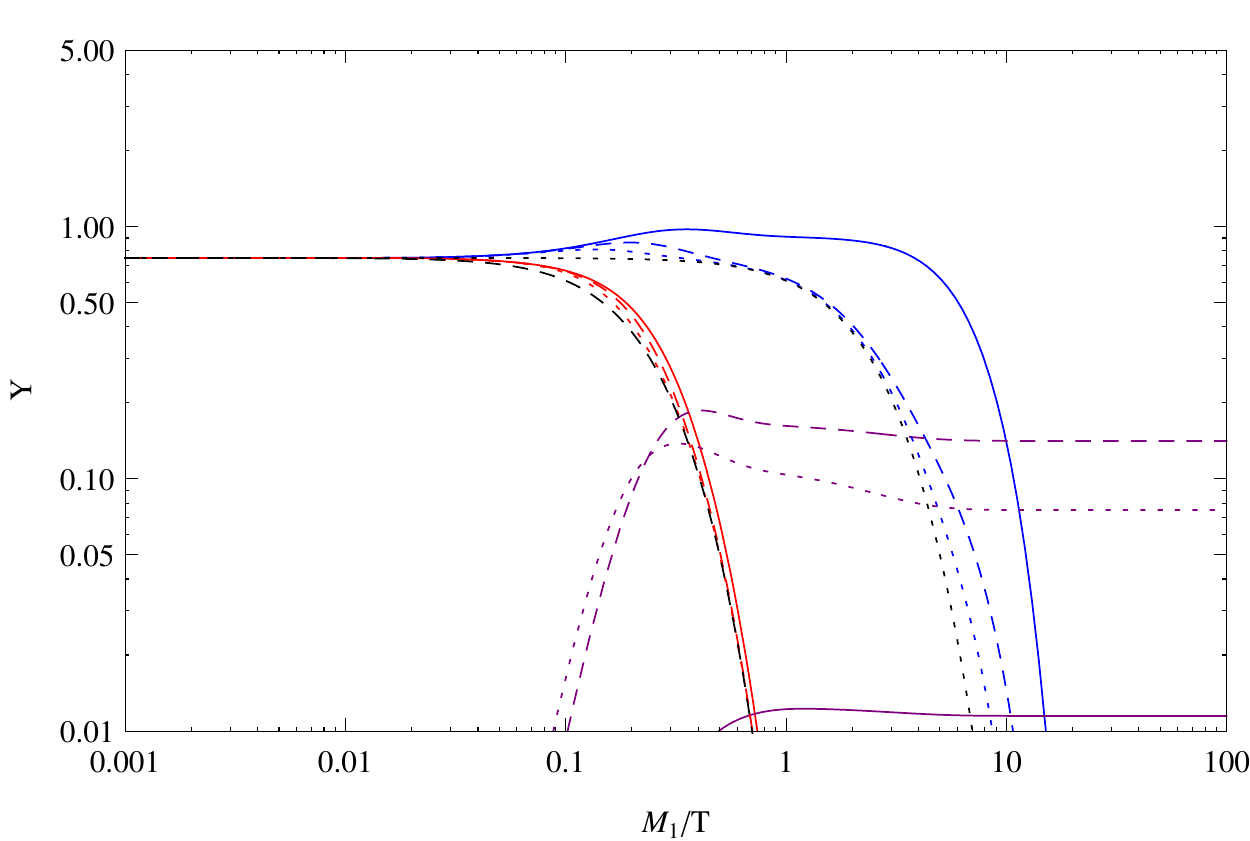}
\label{fig: Low mass beta effect log log}}

\caption[]{The effect of $\{K_2,K_1,K_{21},\beta\}$ on the dynamics for high $\{M_2,M_1\}=\{10^8,10^4\}$GeV (top two plots), and low 
$\{M_2,M_1\}=\{2,0.2\}$TeV (bottom four plots) mass regimes. The color coding, blue, red, purple and black, is the same as Fig.~\ref{fig: 
HighMassExample3 Loglog}~\ref{fig: 
LowMassExample2 Loglog}, ~\ref{fig: Low mass sphaleron effect}, while the dotted, dashed and plain lines represent the variation of each parameter.}
\label{fig: varying parameters}
\end{figure}

In this subsection, we relax some of the constraints required for physical scenarios and focus on the various dynamical components that come into play: the 
decays and scattering processes on one hand, and the various parameters on the other.

\subsubsection*{Decays and inverse decays versus scattering}

 The impact of the decays, inverse decays and scattering is summarized in Fig.~\ref{fig: Decays vs scattering}. We overlay three solutions: (i) 
just decays and inverse decays (dotted lines), (ii) decays, inverse decays and $\Delta L=1$ scattering (dashed lines), and finally (iii) decays, inverse decays 
and $\Delta L=1$  and $\Delta N=2$ scattering (plain lines). 

Starting with the high mass regime, the hidden sector scattering processes, e.g. $N_iH\rightarrow LS$, $N_iN_i\rightarrow HH$, are subdominant compared to the 
Yukawa-mediated processes since $\beta/M_i\ll m_t/v$. This can be seen from Fig.~\ref{fig: Comparison_decays_scattering_high_mass 
Loglog}, where $\Delta L=1$ scattering has a 
significant effect relative to decays, whereas the $\Delta N=2$ scatterings have no visible effect. In practice, the decays and inverse decays remain the 
dominant effect, and the results from section \ref{sec: app Toy model of the 2-level Boltzmann 
equations} can be reasonably well applied.

The low mass regime is marked instead by the significant, if not dominant, effect of hidden sector scattering processes, given $\beta/M_i\sim m_t/v$. Looking at 
the 
plot in Fig.~\ref{fig: LowMassExample5_withoutdecays _fordecaysvsnodecays log log}, the effect of adding $\Delta L=1$ scattering is similar to that in the 
high mass regime, however the impact of $\Delta N=2$ processes is greatly enhanced. Although these processes do not violate lepton number, $\Delta L=0$, they act to 
maintain the neutrino abundance closer to equilibrium. As a consequence, both the $\Delta L=1$ processes (e.g. 
decays) and the inverse processes (e.g. inverse decays) remain in equilibrium for longer, leading to an enhanced lepton washout in that regime.

\subsubsection*{Parameter Dependence}

 The dependence of the dynamics on the parameters is displayed in Fig.~\ref{fig: 
varying parameters}. The subsection above considered examples of viable models that satisfy the basic constraints, and focused on 
constraining the relevant parameters accordingly. We now ignore those constraints, and instead vary the parameters $\{K_2,K_1,K_{21},\beta\}$ to study their impact 
in both high and low mass regimes.

\begin{itemize}
 \item $\{K_2,K_1\}$:-
Figs.~\ref{fig: Large_mass_K2effect} and \ref{fig: Large_mass_K1effect} exhibit the effects of $K_2$ and $K_1$ in the large mass regime. When the $N_1$ and 
$N_2$ phases can be hierarchically separated, they both act as independent one flavor systems according to the respective efficiency 
factors $\kappa_{1,2}$. Consequently, we expect little deviation from the toy model that was studied previously.

 Figs.~\ref{fig:  Low mass K2 effect log log} and \ref{fig:  Low mass K1 effect Loglog}  exhibit the effects of $K_2$ and $K_1$ in the low mass 
regime. This regime requires a smaller mass ratio to achieve sufficient $CP$-asymmetry, therefore the overlap of the $N_2$ and $N_1$ phases induces more 
intricate dynamics, though we observe that the impact of $K_2$ and $K_1$  can still be separated.

 \item $\{K_{21},\beta\}$:-
 Figs.~\ref{fig: Low mass K21 effect Loglog} and \ref{fig: Low mass beta effect log log} exhibit the effects of varying $K_{21}$ and $\beta$ in the low 
mass scale regime. The primary effect is on the magnitude of the lepton asymmetry, via the impact on the $CP$-asymmetry of $N_2$. However, there are also effects 
due to scattering. Indeed, in both cases the maximal lepton asymmetry is achieved for mid-range values, 
$K_{21}=500$ and $\beta=500$GeV respectively. This illustrates the fact that beyond a given threshold, increasing these parameters increases the 
scattering washout more significantly which more than compensates for the increase in the $CP$-asymmetry. 

\end{itemize}

\section{Concluding Remarks}

This paper has considered a minimal extension of leptogenesis that arises by opening up the Higgs portal with a new singlet scalar. This scalar can also couple 
at the renormalizable level to the RH neutrinos which introduces a new (hidden sector) source of $CP$ violation into the theory. The new RHN decay channels that are opened allow 
Higgs portal leptogenesis to avoid the stringent constraints of the Davidson-Ibarra bound, with viable low scale scenarios that we have considered in detail. 
The new decay channels are only available for the next-to-lightest RH neutrinos, which has a number of interesting implications for phenomenology. We conclude 
in this section by mentioning a number of these as directions for future work.

\begin{itemize}

\item {\it First-order leptogenesis}: The new decay channels, e.g. $N_2 \rightarrow N_1 + S$ do not violate lepton number. Thus, this model falls into a general 
category in which the next-to-lightest RH neutrinos have both $L$-violating and $L$-conserving decays. As has recently been emphasized 
\cite{Bhattacharya:2011sy}, such models allow the original Weinberg-Nanopoulos theorem \cite{Nanopoulos:1979gx} to be evaded in that the loop-level amplitude 
can be of first-order in 
the $L$-violating vertex. This is clear from the analysis in Section~2, and thus the HPL model is a simple example illustrating this general feature.

\item {\it Light ($CP$ violating) sterile neutrinos}: Since the new sources of the $CP$-asymmetry arise from decays of the next-to-lightest RH neutrino states, 
it is possible to effectively decouple $N_1$ from leptogenesis. Indeed, since the normal heirarchy still allows one parametrically light (or massless) active 
neutrino, we can consider taking $N_1$ to be, for example, in the keV mass range for sterile neutrino dark matter. It would be interesting to explore whether 
the washout constraints on the interactions allow for viable thermal production modes in the early universe. It is notable that, since $\al_{ij}$ contains 
multiple $CP$-odd phases, this model would generically imply some new low energy contributions (albeit suppressed) to $CP$-violating observables.

\item {\it Dynamical seesaw scale}: We assumed throughout that the scalar $S$ was in a stable vacuum throughout the range of cosmological evolution relevant for 
leptogenesis. This needn't be the case, and the full scalar potential $V(H,S)$ could allow for some evolution in $\langle S\rangle$, which would in turn affect 
the RH neutrino mass scale. Some of these issues were recently considered in \cite{Sierra:2014sta}, and it would be interesting to explore the implications of 
having an early epoch where, for example, the RH neutrino mass scale were to pass through zero due to a phase transition in the scalar potential.\footnote{We 
thank Maxim Pospelov for suggesting this possibility, and related discussions.}

\end{itemize}

\section*{Acknowledgements}

We would like to thank M.~Pospelov for many helpful discussions. The work of  M.L. and A.R. is supported 
in part by NSERC, Canada.

\appendix

\bigskip
\section{Parametrization of the CP-asymmetry}\label{sec: app paramet of CP-asymmetry}

The focus of this paper is on the \textit{dynamics} of the Boltzmann evolution for HPL and the
dependence on the masses and mass ratios. Therefore, rather than working with the full $CP$-asymmetries computed in Section~\ref{sec: HPL CP-asymmetry}, in this 
Appendix we will derive an order-of-magnitude estimate that will be more convenient to use in exploring the full evolution. 
To this end, we replace the coupling constants by their magnitudes, enabling components from the different chirality chains to be combined. Doing so effectively
gives us an absolute upper bound on the asymmetries.

\subsubsection*{Hidden sector $CP$ asymmetry}\label{subsec: app hidden sector}
Starting with the hidden sector, and using the forms \eqref{eq: vertex correction correction to the hidden sector asymmetry} and \eqref{eq: wave function correction to hidden 
sector asymmetry} for the vertex and wave function 
corrections, we write

\beq
\begin{split}\label{eq: parametric size of CP-asymmetries}
 &|\epsilon_i^v|\sim\sum_{j}\frac{|(\lambda^\dagger\lambda)_{ji}|}{(\lambda^\dagger\lambda)_{ii}}\frac{|\alpha_{ij}|}{8\pi
}\frac{\beta}{M_i}\left(\mathcal{F}^v_{jLL}+\mathcal{F}^v_{jRL}\right),\\
 &|\epsilon_i^w|\sim\sum_{l,j}\frac{|(\lambda^\dagger\lambda)_{ji}|}{(\lambda^\dagger\lambda)_{ii}}\frac{|\alpha_{lj}\alpha_{il}|
}{8\pi}\left(\mathcal{F}^w_{jlLL}+\mathcal{F}^w_{jlRL}+\mathcal{F}^w_{jlLR}+\mathcal{F}^w_{
jlRR}\right).
\end{split}
\eeq
Assuming the standard see-saw mechanism, we can replace the Yukawa couplings by the light active neutrino masses through the following relations 
\cite{Casas:2001sr},
\begin{equation}\label{eq: see-saw general relation}
 m_\nu=v^2\lambda^*M^{-1}\lambda^\dagger,\qquad \lambda=\frac{1}{v}UD_{\sqrt{m}}R^\dagger D_{\sqrt{M}},
\end{equation}
where the Higgs vacuum expectation value is $v=174\,$GeV, and $R$ is a (complex) orthogonal matrix, and 
the diagonal Majorana and active neutrino mass matrices are
respectively $D_M=\text{diag}(M_1,M_2,M_3)$, and $D_m=\text{diag}(m_1,m_2,m_3)$. The matrix $U$ is the so-called unitary PMNS matrix. Using this 
notation, we can write
\begin{equation}\label{eq: general expression for Yukawa matrix elements}
 (\lambda^\dagger\lambda)_{ji}=\frac{\sqrt{M_iM_j}}{v^2}\sum_bm_bR_{jb}R_{ib}^*,\qquad(\lambda^\dagger\lambda)_{ii}=\frac{M_i}
{ v^2}\sum_bm_b|R_{ib}|^2.
\end{equation} 
The Schwartz inequality $|\sum_bm_bR_{jb}R^*_{ib}|^2\leq\sum_\alpha m_\alpha|R_{j\alpha}|^2\sum_\beta
m_\beta|R_{i\beta}|^2$, then allows the couplings to be bounded from above
\begin{equation}
 \frac{|(\lambda^\dagger\lambda)_{ji}|}{(\lambda^\dagger\lambda)_{ii}}\lesssim\sqrt{\frac{M_j\sum_bm_b|R_{jb}
|^2}{M_i\sum_bm_b|R_{ib}|^2}}\sim\sqrt{\frac{M_j}{M_i}}.
\end{equation}
The final approximation assumes all the entries of the orthogonal matrix $R$ are similar in magnitude. When $R$ is real, the orthogonality
condition $R^TR=\textbf{1}$ ensures that its elements satisfy $R_{ij}\Leq1$. When $R$ is complex, we have fewer constraints but bounding $|R_{ij}|$ by unity is a sufficient condition for orthogonality, and to obtain a characteristic estimate below we will simply assume $|R_{ij}|\sim 1$.

\begin{figure}[t!]
\centering
\subfigure[]{
\includegraphics[width=8cm]{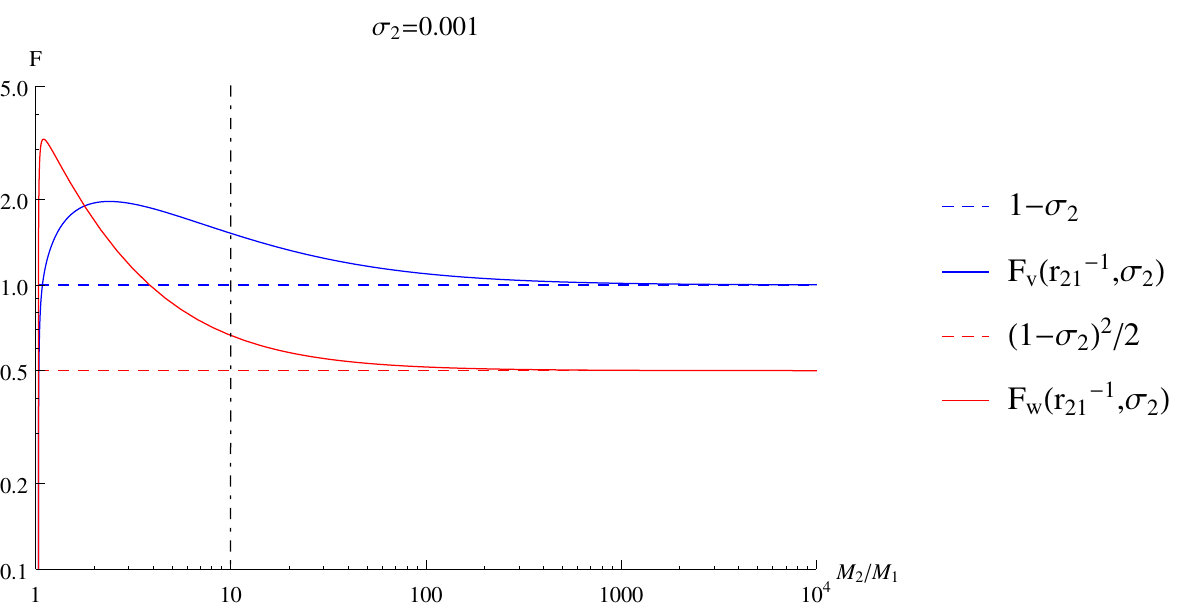}
\label{fig: Asymmetry functions and approximation}}
\subfigure[]{\includegraphics[width=9cm]{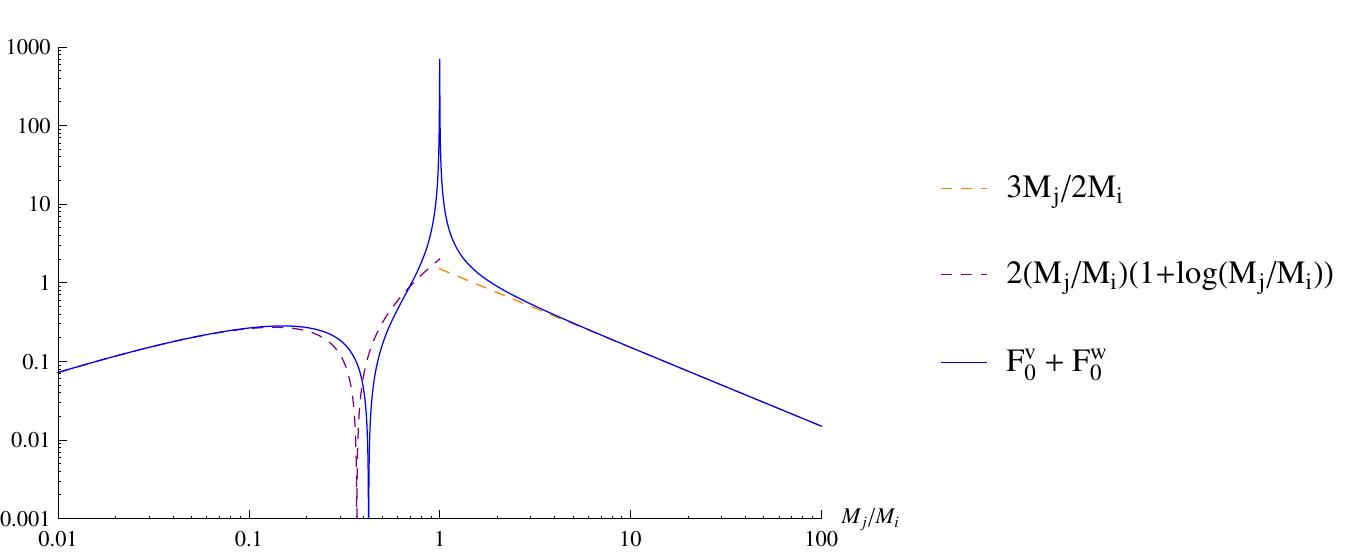}
\label{fig: standard Asymmetry functions and approximation}
}\caption[]{Plots of the full loop functions ${\cal F}$ (solid), and the approximations (dashed) discussed in this section, as a function of the RHN mass 
hierarchy.}
\end{figure}

Because of kinematic constraints, 
we have $\epsilon_1=0$. For $i=2$, the neutrino sum 
runs over $j=1,3$. However, the asymmetry vanishes when $j=3$, since $M_3>M_2$ is kinematically forbidden for the cut loop. The sum thus reduces to 
$j=1$ only. For the wave-function corrections, we have two sums over $j,l=1,3$. The index $j$ denotes the neutrino inside the loop, and is therefore
constrained to $j=1$ by the cut kinematics, but $l=3$ is not forbidden. However by assumption $r_{32}\gg1$, and in that limit the resulting 
CP-asymmetry is negligible. The sums therefore collapse to $j=l=1$. Hence, we arrive at the parametric estimate,
\beq
\begin{split}\label{eq: Cp-asymmetry function exact}
 &\epsilon_2^v\sim\frac{|\alpha_{21}|}{8\pi
}\frac{\beta}{M_2}\sqrt{\frac{M_1}{M_2}}\left(\mathcal{F}^v_{jLL}+\mathcal{F}^v_{jRL}\right),\\
 &\epsilon_2^w\sim\frac{|\alpha_{11}\alpha_{21}|
}{8\pi}\sqrt{\frac{M_1}{M_2}}\left(\mathcal{F}^w_{jlLL}+\mathcal{F}^w_{jlRL}+\mathcal{F}^w_{jlLR}+\mathcal{F}^w_{
jlRR}\right),
\end{split}
\eeq
The functions $\mathcal{F}$ depend upon the variable $1/r_{21}$, which tends to zero in the hierarchical regime, $r_{21}\gg1$. In that limit, we find the 
asymptotic behaviour
\beq
\begin{split}\label{eq: CP-asymmetries approximation}
 &\epsilon_2^v\sim\frac{|\alpha_{21}|}{8\pi
}\frac{\beta}{M_2}\sqrt{\frac{M_1}{M_2}}(1-\sigma_2),\\
 &\epsilon_2^w\sim\frac{|\alpha_{11}\alpha_{21}|}
{16\pi}\sqrt{\frac{M_1}{M_2}}(1-\sigma_2)^2.
\end{split}
\eeq
In Fig.~\ref{fig: Asymmetry functions and approximation}, we compare the above approximate functions with the exact functions calculated in the main 
text, cf. Eqs.~\eqref{eq: vertex correction correction to the hidden sector asymmetry}, \eqref{eq: wave function correction to hidden sector asymmetry}. 
The approximations are excellent for large $M_2/M_1$, but only deviate from the exact answer by a factor of 2 for $M_2/M_1\gtrsim10$.

\begin{figure}[t!]
\centering
\includegraphics[width=12cm]{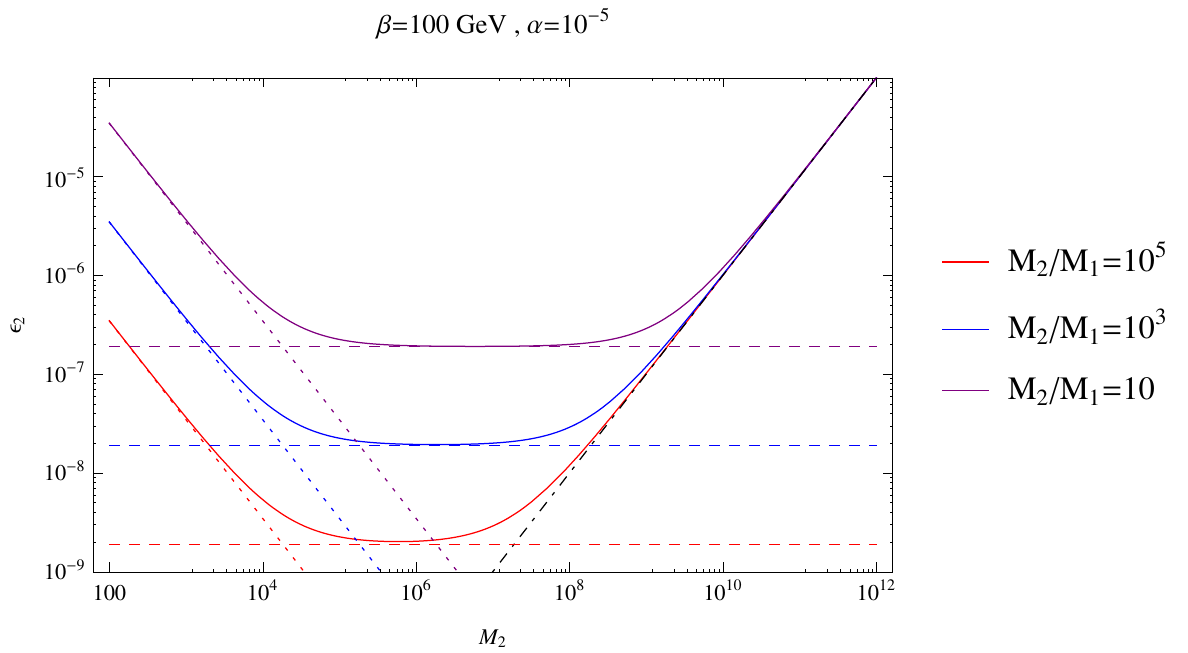}
\label{fig: Equilibrium parameters for decays and inverse decays}
\caption[]{The differing contributions to the total $CP$-asymmetry $\epsilon_2$ for various mass ratios $M_2/M_1$. The high mass ratio is dominated by the 
Yukawa-sourced $CP$-asymmetry, $\epsilon_0\propto\sum_\alpha m_\alpha M_2/v^2$ (black, dot-dashed line). The low mass $CP$-asymmetry is proportional to the 
$\beta$ coupling via the vertex correction, $\epsilon_{(\beta)}\sim\sqrt{M_1/M_2}|\alpha\beta|/M_2$ (dotted lines). In the intermediate mass range, the 
$CP$-asymmetry is dominantly generated by the wave function correction $\epsilon_{(\alpha)}\sim|\alpha|^2\sqrt{M_1/M_2}$ (dashed lines).}

\label{fig: standard and hidden cp asymmetries}
\end{figure}

\subsubsection*{Standard Yukawa $CP$ asymmetry}\label{subsec: app standard sector}

We also recall the conventional contribution to the $CP$-asymmetry, $\epsilon_0$ \cite{Covi:1996wh,Roulet:1997xa,Buchmuller:1997yu,Pilaftsis:1997jf}. In general, 
the loop-induced vertex and wave-function 
contributions from $N_i$ decays are,
\beq
 \epsilon^v_{0i}=\sum_{j\neq
i}\frac{\text{Im}\{(\lambda^\dagger\lambda)_{ji}^2\}}{8\pi(\lambda^\dagger\lambda)_{ii}}\mathcal{F}^v_{0}(r_{ji}),\qquad
 \epsilon^w_{0i}=\sum_{j\neq
i}\frac{\text{Im}\{(\lambda^\dagger\lambda)_{ji}^2\}}{8\pi(\lambda^\dagger\lambda)_{ii}}\mathcal{F}^w_{0}(r_{ji}),
\eeq
with
\beq
 \mathcal{F}^v_{0}(r_{ji})=\sqrt{r_{ji}}\left[1-(1+r_{ji})\log\left(\frac{1+r_{ji}}{r_{ji}}\right)\right],\qquad\mathcal{
F } ^w_ { 0 }(r_{ji})=\frac{\sqrt{r_{ji}}}{(1-r_{ji})}.
\eeq
Using Eq.~\eqref{eq: general expression for Yukawa matrix elements} and the Schwartz inequality, we can again relate the Yukawa coupling to the
light neutrino masses, and bound from above the magnitude of the $CP$-asymmetry. Taking $|R_{ij}|\sim1$, we find
\beq\label{eq: standard bound on CP-asymmetry}
 |\epsilon_{0i}|\sim\frac{\sum_\alpha m_\alpha}{8\pi v^2}\sum_{j\neq
i}M_j\left|\mathcal{F}^v_{0}(r_{ji})+\mathcal{F}^w_{0}(r_{ji})\right|.
\eeq
The sum over the active neutrino masses is constrained by cosmological data, with bounds in the range $\sum_{\alpha}m_\alpha<0.2-1$eV 
\cite{Ade:2013zuv,Moresco:2012by,*Xia:2012na,*Giusarma:2013pmn}. The additional assumption of a normal hierarchy leads to a stronger constraint $\sum_\alpha 
m_\alpha\gtrsim m_3\simeq\sqrt{\Delta m_{31}^2}\simeq 0.04-0.05$eV at the 3$\sigma$ level 
\cite{Capozzi:2013csa,*Fogli:2012ua,*GonzalezGarcia:2012sz,*Tortola:2012te}. Throughout this paper, we assume a normal hierarchy for the light neutrinos, taking 
$\sum_\alpha m_\alpha\sim m_3\sim0.05$eV. When the internal RHN is much heavier than the external neutrino, so that $r_{ji}\gg1$, the 
loop function has the limit $|\mathcal{F}^v_{0}+\mathcal{F}^w_{0}|\sim3/(2\sqrt{r_{ji}})$. At the other end of the spectrum, when the internal RHN is 
much lighter, $r_{ji}\ll1$, we find $|\mathcal{F}^v_{0}+\mathcal{F}^w_{0}|\sim\sqrt{r_{ji}}|2+\log(r_{ji})|$, as shown in Fig.~\ref{fig: standard Asymmetry 
functions and approximation}. The
$CP$-asymmetry from $N_1$ decays receives contributions from internal $j=2,3$ heavy neutrinos, which imply 
$\sum_{j=2,3}M_j\left|\mathcal{F}^v_{0}(r_{j1})+\mathcal{F}^w_{0}(r_{j1})\right|\sim3M_1$. The $CP$-asymmetry from $N_2$ decays receives a contribution from 
$j=3$ giving $3M_2/2$, and a contribution from $j=1$ giving $M_2(M_1^2/M_2^2)|\log(M^2_2/M^2_1)|\ll M_2$, which is neglected. In total, we obtain the standard 
parametric scaling of the $CP$-asymmetry,
\beq
 |\epsilon_{01}|\sim\frac{3M_1\sum_\alpha m_\alpha}{8\pi v^2},\qquad|\epsilon_{02}|\sim\frac{3M_2\sum_\alpha m_\alpha}{16\pi
v^2}.
\eeq
This is analogous to the Davidson-Ibarra bound \cite{Davidson:2002qv}, though somewhat less strict as it depends on $\sum_\alpha 
m_\alpha$ rather than $m_3-m_1$, which is a consequence of taking $|R_{ij}|\sim 1$. In the standard case one can go further as the $CP$-asymmetry is sensitive to 
the $(\lambda^\dagger\lambda)^2_{ij}$ elements which depend on $R_{ij}^2$, as can seen from the representation in \eqref{eq: general expression for Yukawa matrix elements}. 
Thus, the orthogonality condition $\sum_kR_{ki}^2=1$ can be used directly, leading to the conventional Davidson-Ibarra bound. The hidden sector $CP$-asymmetry on the other 
hand, only depends on $(\lambda^\dagger\lambda)_{ij}$, in which case the orthogonality condition is less constraining, and we have taken $|R_{ij}|\sim 1$ to obtain a 
characteristic magnitude. For consistency, we have also used this approach to obtain the above magnitudes for the standard $CP$-asymmetries. In practice, the scaling is very 
similar for the normal hierarchy, where $m_3-m_1\sim m_3\sim\sum_\alpha m_\alpha$.

 Combining both the standard and hidden sector contributions 
gives the 
\textit{total} $\epsilon_{1,2}$ CP-asymmetries,
\beq\label{eq: hpl D.I. bound}
\begin{split}
 &|\epsilon_1|\sim\frac{3M_1\sum_\alpha m_\alpha}{8\pi v^2},\\
 &|\epsilon_2|\sim\frac{3M_2\sum_\alpha m_\alpha}{16\pi v^2}+\left(\frac{\beta}{M_2}+\frac{|\alpha_{11}|}
{2}(1-\sigma_2)\right)\frac{|\alpha_{21}|}{8\pi
}\sqrt{\frac{M_1}{M_2}}(1-\sigma_2).
\end{split}
\eeq
The $\ep_2$ function is exhibited for a given set of parameters $\{\beta,\alpha\}$ and various mass ratios in Fig.~\ref{fig: standard and hidden cp asymmetries}.

\bigskip
\section{Boltzmann Equations and Equilibria}\label{sec: app Boltzmann Equations and Equilibrium}

\subsection*{General considerations}

For completeness, in this Appendix we review the relativistic formulation of the classical Boltzmann equations for the evolution of ensemble phase space 
densities in curved spacetime, and specifically for the FRW metric. Recall that in a thermal quantum field theoretic context, the classical Boltzmann evolution 
implicitly assumes that the effects of quantum coherence are negligible. In that case, the Boltzmann equation for the number densities $n_i$ for a particle 
species `$i$' takes the form \cite{Buchmuller:2000as},

\beq\label{eq: Boltzmann equation for number densities}
 \frac{1}{a^3}\frac{\ptl}{\ptl t}(a^3n_i)=\frac{\ptl n_i}{\ptl t}+3Hn_i=\gamma_i,\qquad 
n_i=g_i\int\frac{d^3p}{(2\pi)^3}f_i,
\eeq
and 
\beq
 n_i^{eq}=\frac{g_im_i^3}{2\pi^2}\frac{T}{m_i}K_2\left(\frac{m_i}{T}\right),\qquad 
n^{eq}_\gamma(z_i)=\frac{g_\gamma
}{\pi^2}T^3,\qquad  H(T)=T^2\sqrt{\frac{8\pi^3g_*}{90M_p^2}}\sim1.66\sqrt{g_*}\frac{T^2}{M_p}.
\eeq
In these exressions $H$ is the Hubble parameters, while the parameter $g_i$ counts the number of degrees of freedom of the particle `$i$': $g_N=2$ for the RHN, 
$g_L=2$ for components of the SU(2)
doublets, $g_{e_R}=1$ for the SU(2) singlets, and $g_H=2$ for the SU(2) Higgs doublet. The parameter $g_*$ counts the effective number of degrees of freedom in 
the 
theory, and typically, within the extension of the SM one considers for Standard Leptogenesis, $g_*\sim100$. The left-hand
side of \eqref{eq: Boltzmann equation for number densities} incorporates information about the cosmology, while the right-hand 
side is the collision term, and dictates how the interactions affect the
number densities. Boltzmann expressed the collistion term via  the \textit{Stosszahlansatz}
(collision number hypothesis),
\beq
 \gamma_i=-\sum_{m,n}\left(\gamma_{i\rightarrow mn}-\gamma_{mn\rightarrow 
i}\right)-\sum_a\sum_{m,n}\left(\gamma_{ia\rightarrow mn}-\gamma_{mn\rightarrow ia}\right)+\cdots
\eeq
For one particular interaction $ia\rightarrow mn...$, the collision term is written as
\beq
 \gamma_{ia\rightarrow mn\cdots}=\int d\Pi_{ia}d\tilde\Pi_{mn\cdots}|i\MM_{ia\rightarrow 
mn\cdots}|^2f_if_a(1\pm f_m)(1\pm
f_n)\cdots
\eeq
where $i\MM_{ia\rightarrow mn\cdots}$ is an S-matrix element, and the phase space integrals are
\beq
\begin{split}
 &\int d\Pi_{ia}\equiv g_ig_a\int\frac{d^3p_i}{(2\pi)^32E_i}\frac{d^3p_a}{(2\pi)^32E_a},\\
 &\int d\tilde\Pi_{mn\cdots}\equiv\int\frac{d^3p_m}{(2\pi)^32E_m}\frac{d^3p_n}{
(2\pi)^32E_n}
\cdots (2\pi)^4\delta(p_i+p_a-p_m-p_n-\cdots).
\end{split}
\eeq
The `+' sign in $(1\pm f)$ is for bosons, and the `-' for fermions. These are the induced emission and
Pauli blocking factors respectively \cite{HahnWoernle:2009qn}. However, we will assume that the gas of particles is dilute enough to use the classical 
Maxwell-Boltzmann approximation, $1
\pm f\approx1$. Also, under the assumption that the scattering processes are fast enough to maintain kinetic equilibrium, the
phase space densities and number densities are related by $n/n^{eq}=f/f^{eq}$ \cite{Buchmuller:2000as},
\beq
 \gamma_{ia\rightarrow mn}=\frac{n_i}{n^{eq}_i}\frac{n_a}{n^{eq}_a}\gamma^{eq}_{ij\rightarrow 
mn},
\eeq
where $\gamma^{eq}$ is given for $f_i=f^{eq}_i$. The thermal cross sections $\gamma^{eq}$ are given specifically for 2-to-2 scatterings $ia\rightarrow mn$, 
and decays $i\rightarrow mn$ by
\beq\label{eq: equilibrium thermal average cross sections}
\begin{split}
 &\gamma^{eq}_{i\rightarrow mn}(T)=n^{eq}_i\Gamma_{i\rightarrow 
mn}\left\langle\frac{M_i}{E}\right\rangle=n_i^{eq}\frac{K_1(z_i)}{K_2(z_i)}\Gamma_{i\rightarrow mn},\qquad z_i=\frac{m_i}{T},\\
 &\gamma^{eq}_{ia\rightarrow 
mn}(T)=n^{eq}_in^{eq}_a\langle v\sigma_{ia\rightarrow 
mn\cdots}\rangle=g_ig_a\frac{T^4}{32\pi^4}\int_{w_{min}}^{\infty}dw\sqrt{w}K_{1}\left(\sqrt{w}
\right)\hat\sigma_{ ij\rightarrow
mn}\left(w\frac{m_i^2}{z_i^2}\right),
\end{split}
\eeq
where $w=s/T^2$, and $K_{1,2}(z)$ are the modified Bessel functions of the second kind. The decay rate $\Gamma_{i\rightarrow mn}$ in the above 
equations is calculated in the center of mass frame of particle `$i$', while the ratio $\langle m_i/E_i\rangle=K_1(z_i)/K_2(z_i)$ is the thermal 
average of the Lorentz factor between the center of mass frame and any other frame \cite{Buchmuller:2004nz,Buchmuller:2000as,Buchmuller:2005eh}. We have also 
used above the reduced cross section, defined as 
\beq
 \hat\sigma_{ia\rightarrow mn}(s)=\frac{1}{s}\delta\left(s,m_i^2,m_a^2\right)\sigma_{ia\rightarrow
mn}(s),\qquad\delta(a,b,c)=(a-b-c)^2-4bc.
\eeq
It is convenient to switch to the comoving system of variables,
\beq
 z_i=\frac{m_i}{T},\qquad Y_i=\frac{n_i}{n^{eq}_\gamma},\qquad Y^{eq}_i=\frac{n_i^{eq}}{n^{eq}_\gamma}=\frac{3}{8}z_i^2K_2(z_i).
\eeq
With these variables, the left-hand side of the Boltzmann equation transforms into $\ptl n_i/\ptl t+3Hn_i=n^{eq}_\gamma z_iH\ptl 
Y_i/\ptl z_i$, and the full equation now reads
\beq\label{eq: Boltzmann equation for abundances general}
 n^{eq}_\gamma z_iH\frac{\ptl Y_i}{\ptl z_i}=-\sum_{m,n}\left(\frac{Y_i}{Y^{eq}_i}\gamma^{eq}_{i\rightarrow
mn}-\frac{Y_m}{Y^{eq}_m}\frac{Y_n}{Y^{eq}_n}\gamma^{eq}_{mn\rightarrow
i}\right)-\sum_{a}\sum_{m,n}\left(\frac{Y_i}{Y^{eq}_i}\frac{Y_a}{Y^{eq}_a}\gamma^{eq}_{ia\rightarrow
mn}-\frac{Y_m}{Y^{eq}_m}\frac{Y_n}{Y^{eq}_n}\gamma^{eq}_{mn\rightarrow
ia}\right),
\eeq
In addition, we can define the Decay and Scattering functions in the following way,
\beq
\begin{split}
 &D_{i\rightarrow mn}\equiv\frac{\gamma^{eq}_{i\rightarrow 
mn}}{n_\gamma^{eq}H}=z_i^2Y^{eq}_i\frac{K_1(z_i)}{K_2(z_i)}K_{i\rightarrow mn},\qquad K_{i\rightarrow mn}\equiv\frac{\Gamma_{i\rightarrow 
mn}}{H_i},\\
 &S_{ia\rightarrow mn}\equiv\frac{\gamma^{eq}_{ia\rightarrow 
mn}}{n_\gamma^{eq}H}
 =\frac{g_ig_a}{g_\gamma}\frac{m_i}{H_i}\frac{1}{32\pi^2}z_i\int_{w_{min}}^{\infty}dw\sqrt{w}K_{1}\left(\sqrt{w}
\right)\hat\sigma_{ ij\rightarrow
mn}\left(w\frac{m_i^2}{z_i^2}\right).
\end{split}
\eeq
The Hubble rate $H_i=H(T=m_i)$. Through the Hubble time $t_i=1/H_i$, we have a notion of the 
time scale before the equilibrium density of the massive particle `$i$' is Boltzmann suppressed. This is to be compared with the natural 
time scale set by the particle lifetime $\tau_i=1/\Gamma_{i\rightarrow mn}$. If the lifetime is larger than the Hubble time, $\tau_i>t_i$, 
we anticipate a number density excess relative to equilibrium. Thus the \textit{equilibrium parameter} 
$K_{i\rightarrow mn}\equiv\Gamma_{i\rightarrow mn}/H_i$, as defined above, characterizes Sakharov's non-equilibrium condition. As noted in the main text, we follow the literature in using the notation $K$ for the equilibrium parameters, to be distinguished from the modified Bessel function $K_i(z)$ by the presence in the latter of the argument $z$.

The above definition of the equilibrium parameter is not fully consistent since the decay rate is calculated at zero 
temperature. More precisely, we define \textit{thermal 
equilibrium parameters}, $\mathcal{K}_{i\rightarrow mn}=\langle\Gamma_{i\rightarrow mn}\rangle/H(T)$ for the decay rate, and similarly 
$\mathcal{K}_{ia\rightarrow mn}=n^{eq}_i\langle v\sigma_{ia\rightarrow mn}\rangle/H(T)$ for the scattering processes. It turns out that the thermal 
equilibrium parameters are related to the decay and scattering functions $D,S$, defined earlier,
\beq
\begin{split}
  &\mathcal{K}_{i\rightarrow mn}=\frac{\langle\Gamma_{i\rightarrow mn}\rangle}{H}=\frac{\gamma^{eq}_{i\rightarrow 
mn}}{n^{eq}_iH}=\frac{D_{i\rightarrow 
mn}}{Y^{eq}_i},\\
  &\mathcal{K}_{ia\rightarrow mn}=\frac{n^{eq}_i\langle v\sigma_{ia\rightarrow 
mn}\rangle}{H}=\frac{\gamma^{eq}_{ia\rightarrow mn}}{n^{eq}_aH}=\frac{S_{ia\rightarrow 
mn}}{Y^{eq}_a}.
\end{split}
\eeq
Under the assumption of $CP$ and 
$CPT$ conservation, energy conservation implies that $\gamma^{eq}_{ia\rightarrow mn}=\gamma^{eq}_{mn\rightarrow ia}$. It is 
possible to relate the above equilibrium parameters to the parameters for the inverse processes, as follows
\beq
\begin{split}
  &\mathcal{K}_{mn\rightarrow i}=\frac{\gamma^{eq}_{mn\rightarrow
i}}{n^{eq}_nH}=\frac{Y^{eq}_i}{Y^{eq}_n}\mathcal{K}_{i\rightarrow 
mn}=\frac{1}{Y^{eq}_n}D_{i\rightarrow mn},\\
  &\mathcal{K}_{mn\rightarrow ia}=\frac{\gamma^{eq}_{mn\rightarrow ia}}{n^{eq}_nH}=\frac{Y^{eq}_a}{Y^{eq}_n}\mathcal{K}_{ia\rightarrow 
mn}=\frac{1}{Y^{eq}_n}S_{ia\rightarrow mn}.
\end{split}
\eeq
Note that alternative definitions appear in the literature. For example, the decay function $D$ in 
\cite{Buchmuller:2004nz} corresponds here to $D_{i\rightarrow mn}/z_iY^{eq}_i$. The difference is a consequence of the present need to consider 2-level 
leptogenesis. In \cite{Buchmuller:2004nz}, 
 only terms of the form $1/Y^{eq}_i$ appear in the Boltzmann equations, since only one RHN flavor is 
accounted for. Instead, we take into account scattering processes with external states mixing several RHN flavors, leading to terms 
like $1/Y^{eq}_1Y^{eq}_2$ or $(1/Y^{eq}_i)^2$. In addition, the definition $z_i=m_i/T$ explicitly brings in an arbitrary choice of reference 
mass scale, $m_i$. Therefore, we choose to keep the left-hand side of 
the Boltzmann equations in the form $z_i\ptl Y_i/\ptl z_i$ in order for it to be independent of that choice. Therefore, removing the $z_iY^{eq}_i$ 
from the definition of the decay functions seems more appropriate to the present case.

\subsection*{Example with leptonic RHN decay}

\begin{figure}[t!]
\centering
\subfigure[]{
\includegraphics[width=9cm]{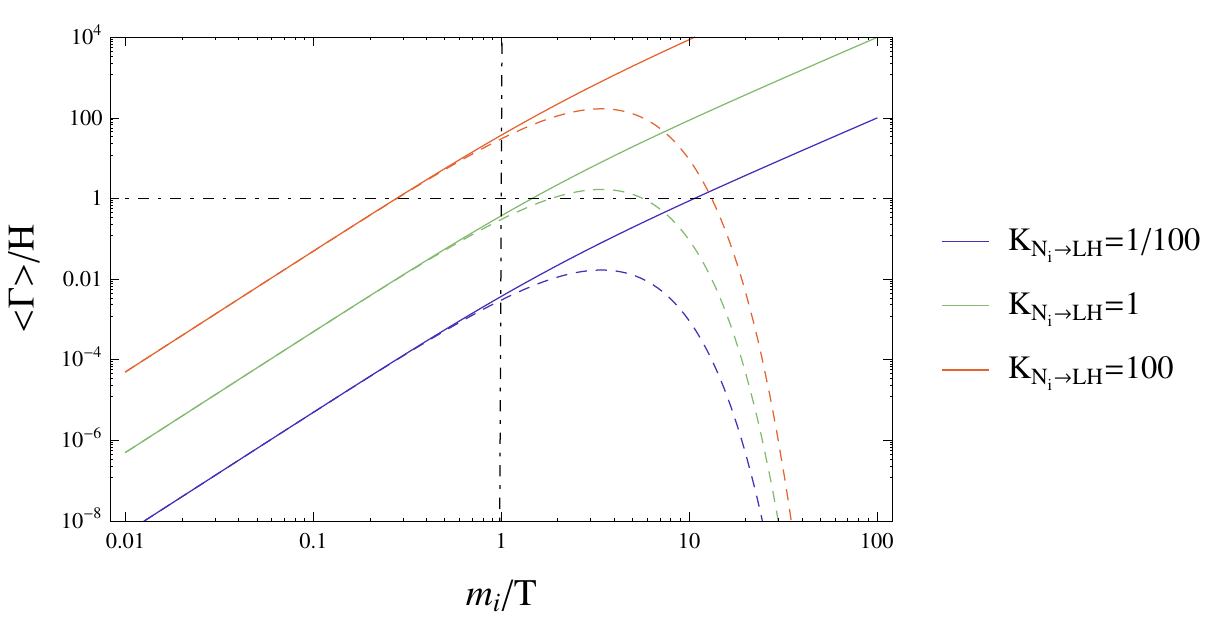}
\label{fig: Equilibrium parameters for decays and inverse decays}}
\subfigure[]{\includegraphics[width=8.5cm]{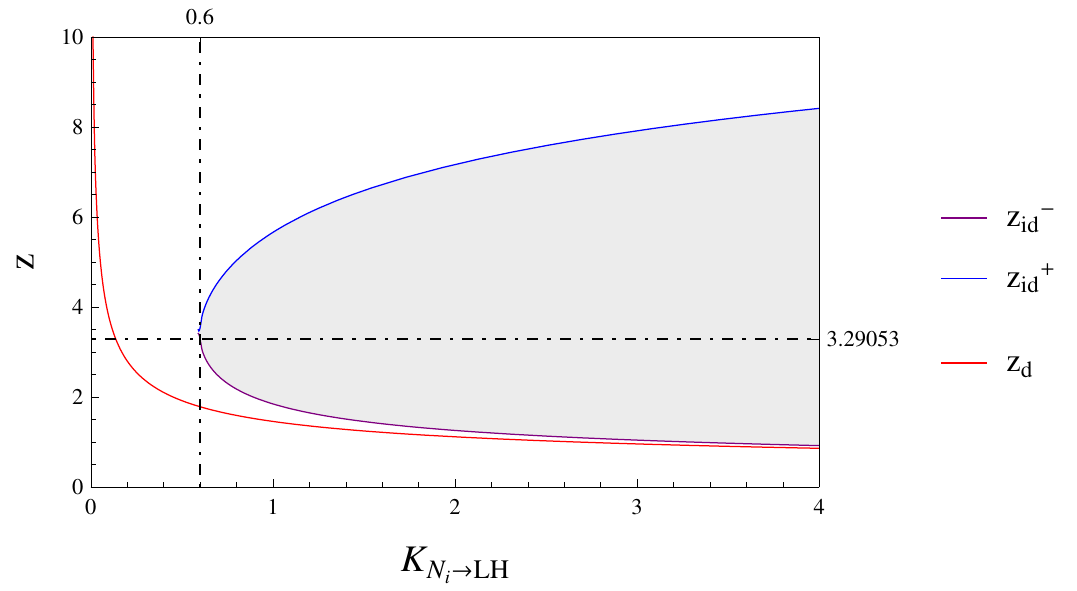}
\label{fig: Equilibrium points}}
\caption[]{The left plot shows the thermal equilibrium parameter $\mathcal{K}_{N_i\leftrightarrow LH}$ for both decays (plain) and inverse decays (dashed), 
for 
several equilibrium parameters $K_{N_i\leftrightarrow LH}$. The right plot shows the solution $z_d$ and $z_{id}^{\pm}$ to the 
equilibrium condition for
decays 
and inverse decays respectively. Decays and inverse decays are both in equilibrium in the gray area, which only happens if $K\geq0.6$.}
\label{fig: decays and inverse rates}
\end{figure}

In order to be specific, we will consider how these formulae apply to the case of leptonic RHN decays and inverse decays, $N_i\leftrightarrow LH$, 
directly relevant for leptogenesis. In equilibrium, we have
\beq
 Y^{eq}_i=\frac{n_i^{eq}}{n^{eq}_\gamma}=\frac{3}{8}z_i^2K_2(z_i),\quad Y^{eq}_L=\frac{3}{4}.
\eeq
Note that the derivation of the Boltzmann equations in the form \eqref{eq: Boltzmann equation for abundances general} assumed 
Maxwell-Boltzmann statistics for all the particles. This assumption should technically lead to $Y^{eq}_i=z_i^2K_2(z_i)/2$. However, we 
know that relativistic fermions have an abundance $Y^{eq}_L=3/4$. In order to reproduce the relativistic RHN abundance at 
$T\gg M_i$, we multiply by an overall $3/4$, so that $Y^{eq}_i(z_i\rightarrow0)=3/4$ \cite{HahnWoernle:2009qn,Davidson:2008bu}. The thermal equilibrium 
parameters 
take the form
\beq
\begin{split}
 &\mathcal{K}_{N_i\rightarrow LH}=K_{N_i\rightarrow LH}z_i^2\frac{K_1(z_i)}{K_2(z_i)},\\
 &\mathcal{K}_{LH\rightarrow N_i}=K_{N_i\rightarrow LH}z_i^2\frac{K_1(z_i)}{K_2(z_i)}\frac{Y^{eq}_i}{Y^{eq}_L}.
\end{split}
\eeq
Of course, it is a matter of convention to use $1/Y^{eq}_L$ rather than $1/Y^{eq}_H$ in the definition of the  thermal 
equilibrium parameter for inverse decays. These two functions have been 
plotted in figure \ref{fig: Equilibrium parameters for decays and inverse decays} for various parameters $K_{N_i\rightarrow LH}$. 
The equilibrium conditions for the decays and inverse decays are given as $\mathcal{K}_{N_i\rightarrow LH}>1$ and $\mathcal{K}_{LH\rightarrow N_i}>1$ 
respectively. The first condition translates to a restriction on $z_i>z_d$, while the second translates to the condition $z^+_{id}>z_i>z^-_{id}$. The solutions 
$z_d$, and $z^\pm_{id}$ have been numerically solved and plotted in Fig.~\ref{fig: Equilibrium points}.

\bigskip
\section{Decay rates and scattering cross sections}\label{sec: app scattering cross sections}

\begin{figure}[!t]
\centerline{\includegraphics[width=15cm]{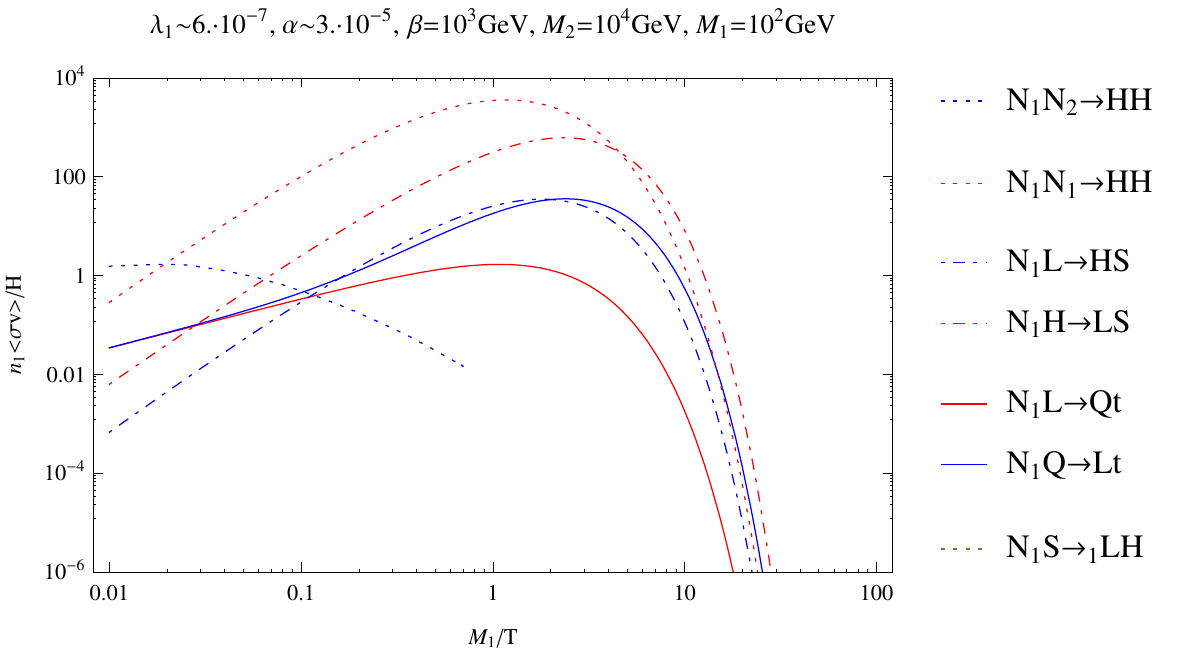}}
\caption{This plot is a summary of the important scattering processes to account for, in order to properly manage the effect of $N_1$ on the lepton asymmetry. 
The corresponding Feynman diagrams are shown in the main text, in Fig.~\ref{fig: scattering cross sections summary}.}
\label{scat}
\end{figure}

This appendix compiles the relevant decay rates and scattering cross sections used in the HPL Boltzmann equations. We summarize the relevant scattering processes in 
Fig.~\ref{scat}.

\subsubsection*{Decays and inverse decays}

At tree level, one has the RHN decay rates to leptons,
\beq
\begin{split}
 &\Gamma(N_i\rightarrow 
LH)=\Gamma(N_i\rightarrow\overline{LH})=\sum_{k,\alpha,\beta}\Gamma(N_i\rightarrow L_k^\alpha 
H^\beta)=\frac{(\lambda^\dagger\lambda)_{ii}}{16\pi}M_i,\\
 &\Gamma_i=\Gamma(N_i\rightarrow LH)+\Gamma(N_i\rightarrow\overline{LH})=\frac{(\lambda^\dagger\lambda)_{ii}}{8\pi}M_i.
\end{split}
\eeq
The indices $\alpha,\beta=1,2$ refer to the SU(2) doublets, and enumerate the electron-type and the neutrino-type leptons. The 
index $k=1,2,3$ runs through the three families, electron, muon, and tau. One of the hidden sector decays we encounter is $N_2\rightarrow N_1S$,
\beq
 \Gamma(N_2\rightarrow 
N_1S)=\frac{|\alpha_{12}|^2M_2}{16\pi}\left[\left(1+\frac{M_1}{M_2}\right)^2-\frac{m_S^2}{M_2^2}\right]\sqrt{\left(1-\frac{M_1^2}{M_2^2}
-\frac{m_S^2}{M_2^2}\right)^2-4\frac{M_1^2}{M_2^2}\frac{m_S^2}{M_2^2}},
\eeq
where we used the approximation that $\alpha_{ij}=\text{Re}\{\alpha_{ij}\}$, which is not generally true because of the complex 
phases contained in $\alpha$, but the $CP$-odd contributions are not relevant here as the decay is $L$-conserving.

\subsubsection*{Visible sector scattering}

These processes involve the SM quarks and leptons in external states. 

\begin{itemize}
 \item  {\it $s$-channel}: the $N_iL\leftrightarrow Qt$ cross section and reduced cross section read
\beq
\begin{split}
 &\sigma(N_iL\rightarrow Q\overline t)=\sum_{\alpha,k}\sigma(N_iL_k^\alpha\rightarrow Q\overline 
t)=\frac{(\lambda^\dagger\lambda)_{ii}}{8\pi}\frac{m_t^2}{v^2}\frac{1}{s},\\
 &\hat\sigma(N_iL\rightarrow Q\overline t)=\frac{(\lambda^\dagger\lambda)_{ii}}{8\pi}\frac{m_t^2}{v^2}\frac{(s-M_i^2)^2}{s^2}.
\end{split}
\eeq
\end{itemize}

\begin{itemize}
 \item {\it $t$-channel}: We consider $N_iQ\leftrightarrow Lt$ and $N_it\leftrightarrow LQ$. Because the 
leptons and quarks are assumed massless, these two channels are in fact equal. We have the cross section and reduced cross 
sections
\beq
\begin{split}
 &\sigma(N_iQ\rightarrow 
Lt)=\frac{(\lambda^\dagger\lambda)_{ii}}{8\pi}\frac{m_t^2}{v^2}\frac{1}{s}\left[\frac{s-2M_i^2+2m_h^2}{s-M_i^2+m_h^2}+2\frac{M_i^2-m_h^2}{s-M_i^2
} \ln\left(\frac{s-M_i^2+m_h^2}{m_h^2}
\right)\right],\\
 &\hat\sigma(N_iQ\rightarrow 
Lt)=\frac{(\lambda^\dagger\lambda)_{ii}}{8\pi}\frac{m_t^2}{v^2}\frac{s-M_i^2}{s}\left[\frac{s-2M_i^2+2m_h^2}{s-M_i^2+m_h^2}+2\frac{M_i^2-m_h^2}{s-M_i^2
} \ln\left(\frac{s-M_i^2+m_h^2}{m_h^2}
\right)\right ].
\end{split}
\eeq
Throughout this paper, we take the approximation that the zero-temperature Higgs mass $m_h=0$, but in this limit, these cross sections are infrared 
divergent. The regulator to use, however, is the thermal mass which can potentially be quite large at leptogenesis temperatures. In practice, the cross section 
is only logarithmically sensitive to the regulator, and we therefore make the conventional choice $m_h/M_i=10^{-5}$.

\end{itemize}

\subsubsection*{Hidden sector scattering}

\begin{itemize}
 \item {\it $s$-channel}: $N_iL\leftrightarrow HS$ through the $\beta HHS$ vertex. The cross section and reduced cross section are
\beq
\begin{split}
 &\sigma(N_iL\rightarrow HS)=\sum_{\alpha,k}\sigma(N_iL_k^\alpha\rightarrow 
HS)=\frac{(\lambda^\dagger\lambda)_{ii}\beta^2}{8\pi}\frac{s-m_S^2}{s^3},\\
 &\hat\sigma(N_iL\rightarrow HS)=\frac{(\lambda^\dagger\lambda)_{ii}\beta^2}{8\pi}\frac{(s-m_S^2)(s-M_i^2)^2}{s^4}.
\end{split}
\eeq

 \item {\it $s$-channel}: $N_iS\leftrightarrow LH$, through the hidden sector vertex $\alpha_{ij}N_iN_jS$. This process is mediated by $N_j$, and the amplitude 
should thus be summed over 
all flavors. However, we shall simplify the discussion by considering only one internal flavor. In the the limit of massless $S$, the cross section takes the 
form
\beq
\begin{split}
 &\sigma(N_iS\rightarrow LH)=\sum_j\sum_{\alpha,k}\sigma(N_iS\rightarrow L_k^\alpha 
H)=\sum_j\frac{(\lambda^\dagger\lambda)_{jj}|\alpha_{ij}|^2}{16\pi}\frac{(s+M_i^2)(s+M_j^2)-4sM_iM_j}{\sqrt{\delta(s,M_i^2,m_s^2)}
((s-M_j^2)^2+{\cal E}_j^2)},\\
 &\hat\sigma(N_iS\rightarrow LH)=\sum_j\frac{(\lambda^\dagger\lambda)_{jj}|\alpha_{ij}|^2}{16\pi 
}\frac{(s-M_i^2)}{s}\frac{(s+M_i^2)(s+M_j^2)-4sM_iM_j}{(s-M_j^2)^2+{\cal E}_j^2},\qquad{\cal E}_j=M_j\Gamma_j.
\end{split}
\eeq

 \item {\it $s$-channel}: $N_iN_j\leftrightarrow HH$ mediated by $S$. Taking the notation, $\delta(s,M_i,M_j)=(s-M_i^2-M_j^2)^2-4M_i^2M_j^2$, the cross section and reduced cross 
section are given by
\beq
 \begin{split}
  &\sigma(N_iN_j\rightarrow HH)=\frac{|\alpha_{ij}|^2\beta^2}{32\pi}\frac{s-(M_i+M_j)^2}{\sqrt{\delta(s,M_i,M_j)}(s-m_S^2)^2},\\
  &\hat\sigma(N_iN_j\rightarrow 
HH)=\frac{|\alpha_{ij}|^2\beta^2}{32\pi}\sqrt{\delta(s,M_i,M_j)}\frac{s-(M_i+M_j)^2}{s(s-m_S^2)^2}.
 \end{split}
\eeq
\end{itemize}

\begin{itemize}
 \item {\it $t$-channel}: $N_iH\leftrightarrow LS$ and $N_iS\leftrightarrow LH$, both mediated by a Higgs. Care is needed in computing these cross sections, because the Higgs 
mediator can be produced on-shell. This is true even 
if the Higgs has a small but finite mass. Thus, these processes will almost always be divergent, as they include the kinematic regime where the RH neutrino 
decays on-shell to $LH$. 
It is therefore important to note that the cross section is regulated by the external neutrino decay width. This subtlety has been noted previously in a different context
\cite{Ginzburg:1995js,*Melnikov:1996iu}. Starting with $N_iH\rightarrow LS$, 
the amplitude squared takes the form
\beq
 \sum_{\alpha,\beta,k}\sum_{spins}|i\MM(N_iH^\beta\rightarrow 
L^\alpha_kS)|^2=2\beta^2\left[\frac{(\lambda^\dagger\lambda)_{ii}M_i^2}{t^2+{\cal E}_i^2}-(\lambda^\dagger\lambda)_
{ii}\frac{t}{t^2+{\cal E}_i^2 } \right].
\eeq
Upon integration over the transfer momentum, the second term leads to a logarithmic divergence. The first term is naively more problematic
because it leads to a linear divergence, $1/{\cal E}_i$. However, in the narrow width limit,  upon integration this term gives the delta function $\delta(t)$. 
This is the signature 
of an on-shell mediator, which splits the scattering into the two on-shell subprocesses $N\rightarrow LH$ followed by $HH\rightarrow S$. The first part is 
already accounted for in the Boltzmann equations, and should be subtracted in order to avoid double-counting \cite{Giudice:2003jh}. In effect, this is a 
$t$-channel RIS. The subtracted scattering cross section we use is then
\beq
\begin{split}
 &\sigma(N_iH\rightarrow 
LS)=\frac{(\lambda^\dagger\lambda)_{ii}\beta^2}{16\pi(s-M_i^2)^2}\log\left(\frac{
s^2(s-M_i^2-m_S^2)^2 +s^2{\cal E}_i^2} {M_i^4m_S^4+s^2{\cal E}_i^2} \right),\\
 &\hat\sigma(N_iH\rightarrow 
LS)=\frac{(\lambda^\dagger\lambda)_{ii}\beta^2}{16\pi s}\log\left(\frac{s^2(s-M_i^2-m_S^2)^2+s^2{\cal E}_i^2}{m_S^4+s^2{\cal E}_i^2}
\right),\quad{\cal E}_i=M_i\Gamma_i.
\end{split}
\eeq
The decay rate to account for in ${\cal E}_i$ is the \textit{total} rate, that is ${\cal E}_1=M_i\Gamma_1$ for $N_1$ but 
${\cal E}_2=M_2\left(\Gamma_2+\Gamma_{21}\right)$ for $N_2$. This is model dependent, though we can safely assume that 
$\lambda^2,\alpha^2\Leq10^{-5}$, which inspires our choice ${\cal E}_i/M_i^2=10^{-5}$. The cross section is only weakly dependent on the prescription, as the 
residual divergence is logarithmic.
\end{itemize}

\bibliography{hpl2_bib}
\end{document}